\documentclass[aps,groupedaddress,nofootinbib,11pt]{revtex4}

\usepackage[dvips]{epsfig}
\usepackage{amsfonts}
\usepackage{amsmath}
\usepackage{}

\newcommand{\Eqref}[1]{(\ref{#1})}
\newcommand{\ket}[1] {\mbox{$ \vert #1 \rangle $}}
\newcommand{\kett}[1] {\mbox{$ \vert #1 \rangle_2 $}}
\newcommand{\bra}[1] {\mbox{$ \langle #1 \vert $}}
\newcommand{\abs}[1] {\mbox{$ \vert #1 \vert $}}

\newcommand{\inv}[1]{\frac{1}{#1}}

\newcommand{\dpartial}{\stackrel{\leftrightarrow}{\partial}}
\newcommand{\dpartialeta}{\stackrel{\leftrightarrow}{\partial_{\eta}}}

\newcommand{\tprod}{\widetilde{\prod\limits_{\bold k}}}
\newcommand{\tprodl}{\widetilde{\prod\limits_{\lambda}}}

\newcounter{subequation}[equation] \makeatletter
\expandafter\let\expandafter\reset@font\csname reset@font\endcsname
\newenvironment{subeqnarray}
  {\arraycolsep1pt
    \def\@eqnnum\stepcounter##1{\stepcounter{subequation}{\reset@font\rm
      (\theequation\alph{subequation})}}\eqnarray}
  {\endeqnarray\stepcounter{equation}}
\makeatother

\newcommand{\ba}{\begin{eqnarray}}
\newcommand{\ea}{\end{eqnarray}}
\newcommand{\sba}{\begin{subeqnarray}}
\newcommand{\sea}{\end{subeqnarray}}

\def\th{\mbox{th}}
\def\ch{\mbox{ch}}

\begin{document}

\vskip 1truecm
%D25/11/03
\title{Space-time correlations in inflationary 
%fluctuation 
spectra, \\
a wave-packet analysis} 
%of primordial fluctuations in inflation}
%\title{Space-time correlations in inflationary 
%%primordial 
%spectra}
%%\title{Primordial spectrum correlations in inflationary models}
%%resulting from the detection of a Post selection of a classical wave.}
\vskip 1truecm
\author{David Campo}
\email[]{campo@phys.univ-tours.fr}
\affiliation{Laboratoire de Math\'{e}matiques et Physique
Th\'{e}orique, CNRS UMR 6083,
Universit\'{e} de Tours, 37200 Tours, France}
\author{Renaud Parentani}
%\email[]{parenta@celfi.phys.univ-tours.fr}
\email[]{Renaud.Parentani@th.u-psud.fr}
%\affiliation{Laboratoire de Math\'{e}matiques et Physique
%Th\'{e}orique, CNRS UMR 6083,
%Universit\'{e} de Tours, 37200 Tours, France}
\affiliation{Laboratoire de Physique Th\'{e}orique, CNRS UMR 8627,
B\^atiment 210,
Universit\'{e} Paris XI, 91405 Orsay Cedex, France}
\date{December 8, 2003}
\maketitle

\vskip 2truecm

\centerline{{\bf Abstract }}
\vskip 0.5truecm

%Inflation predicts the properties of the
%primordial density fluctuations. In particular t
The inflationary mechanism
of mode amplification predicts that the state of each mode
with a given wave vector is correlated to that of
its partner mode with the opposite vector.
This implies nonlocal correlations which leave their imprint 
on temperature anisotropies in the cosmic microwave background.
%primordial fluctuations, which might be observable. In this article we show that t
Their spatial properties are best revealed by using local wave packets. 
%This can be done
%can be revealed 
% to extract from the mean 
%the contribution of the configurationsapp forming a wave packet.  
%These %space-time correlations 
This analysis
shows that all density fluctuations giving rise
the large scale structures originate in pairs which are
born near the reheating. 
%have a dipolar structure, in which
In fact each local density fluctuation is paired with
an oppositely moving partner 
%which has an
%same amplitude, which is in phase opposition, 
with opposite amplitude. 
%The spatial separation
%is always twice the Hubble time multiplied by the speed of sound. 
To obtain these results we first 
%These can be inequitroduced by
apply
%ing 
a ``wave packet transformation'' 
with respect to one argument of the two point correlation function.
A finer understanding of the correlations is then reached 
by making use of coherent states.
%Indeed, when applying the wave packet transformation one obtains 
%a structure consisting of 
%three aligned wave packets.
%aling with a snapshot of fluctuations.
%This follows from a two-folded degeneracy associated with
%the transformation.
%the fact that one deals with a snapshot of fluctuations.
%The degeneracy can be lifted if one makes use of t
The knowledge of the velocity field 
%is stressed as  its knowledge is necessary 
is required to extract the contribution 
of a single pair of wave packets.
Otherwise, there is a two-folded degeneracy which gives
%, one obtains 
three aligned wave packets arising from two pairs.
%obtain this two-folded degeneracy which is present
%when applying the wave packet transformation and which gives 
%rise to a structure consisting of three aligned wave packets. 
%dealing with a snapshot of fluctuations. 
The applicability of these methods to observational data
is briefly discussed. 
%s described by coherent states 
%or by applying a ``wave packet transformation'' 
%with respect to one argument of the two point function.

\newpage
\section{Introduction} \label{sec:intro}

In inflationary models, primordial density fluctuations and
primordial gravitational waves are described by 
Gaussian ensembles with well defined correlations
in wave-vector space $\bold k$.
These correlations lead to the 
%D25/11/03
temporal coherence
of the modes when re-entering
the Hubble horizon in the adiabatic era
\cite{Mukhanov81,Staro79,MBF92,Albrecht00}.
As far as the primordial density fluctuations are concerned, 
the temporal coherence can be now considered as an observational fact 
since it is necessary 
to obtain multiple acoustic peaks in the spectrum of CMB anisotropies 
\cite{wmap,Albrecht00}.

So far the analysis of the correlations have been mostly performed in 
Fourier $\bold k$-space, 
simply because they are diagonal in this representation. 
Nevertheless, it is of value to also analyse the two-point 
correlation function in the position 
%$\bold x$-
representation. 
Indeed, as shown in \cite{BB}, this analysis displays 
the space-time causality of the mode amplification process. 
In this paper,
we shall use a mixed representation based on local wave packets.
%to re-analyse the two-point function.
%$G_{in}$. 
%As such there is no new phenomena. 
%Nevertheless t
This third analysis possesses its own virtues, 
and should be thought as complementary to the 
two other representations.
%usual Fourier ${\bf k}$-space analysis (with diagonal correlations 
%and sine functions representing growing modes) 
%and the less usual analysis in ${\bf x}$-space.
%(wherein space-time causality is clearly displayed). 
In fact, we found 
%the mixed representation
it the most appropriate
when focusing on the spatial correlations 
on a given time slice.
%, we found the mixed representation the most appropriate. 
This is because the use of local wave packets 
introduces spatial correlations by coupling different
${\bf k}$-modes which were so far independent. 
Notice also that we shall work with the three dimensional
Green function and not with its restriction to the Last 
Scattering Surface. This choice gives simpler expressions
unencumbered by the projection on a 2-sphere.

Wave packets are introduced by applying a  
``wave packet transformation'' to the two-point 
correlation function with respect to one of its argument.
Doing so one obtains a one-point function which displays a 
%well defined
universal  structure consisting of three local wave packets on a line. 
(To our knowledge this has not been noticed before).
The central wave has relative amplitude two and corresponds to the chosen 
wave. The two others have amplitude minus one and correspond
to two partners.
% possiblity of recovering these specific correlations 
%rests on the possibility of identifying the primordial velocity field.
As we shall later see, the reason for this three-folded structure follows
from the fact that one deals with a snapshot of field configurations
when using the two-point function on a given time slice. 
%on the last scattering surface, 
%one deals with a snapshot of field configurations. 
%with a two-folded degeneracy. 
This implies that one cannot distinguish
``left'' moving from ``right'' moving configurations,
% described by the same Fourier components. 
thereby inducing 
%This leads to 
a doubling of the partners. To raise the degeneracy, one needs to 
take into account the velocity of the waves, or equivalently 
%and therefore 
%%R% and not only it Fourier content. This amounts  
distinguish positive from negative frequencies.
% for a given wave vector.
Mathematically this raises no difficulty and can be obtained 
by using the Klein-Gordon product when applying the wave packet transform. 
Doing so, only two local wave packets of opposite amplitude are found,
as one would have expected.  

There is a complementary way to interprete 
the use of wave packets: 
%David07/07/04
%which furthermore introduces the 
%more elaborate procedure we shall present in the last two Sections. 
%Wave packets 
they can be viewed as introducing a filter in ${\bf k}$-space. 
This opens the way to refine the procedure of filtering by working
with the distribution of field configurations rather than with mean values. 
Indeed the Gaussian ensemble of field configurations 
determines, on one hand, the {mean} properties
%of the primordial fluctuations 
such as the two-point correlation function. 
These, together with Boltzmann equations \cite{priorM,Mukhanov03},
%in turn 
determine the $C_l$ the power of the temperature anisotropy multipoles.
On the other hand, the knowledge of the ensemble also 
gives the probability to find a particular set of configurations,
i.e. a particular realization of the ensemble. 

In the last two Sections, we exploit this second aspect in order to determine 
%present a method which allows to study 
the spatial properties of the correlations associated with the realization of
configurations described by a {local wave packet}.
The procedure \cite{CP1} consists in isolating these field configurations 
%D25/11/03
during the adiabatic era
%at late time 
and to compute the correlations within that 
restricted set. 
%The motivation for 
These correlations show up specific spatial 
properties which are some how smeared when 
dealing with the entire ensemble, i.e. with the mean values. 
These correlations have a double origin. 
First, their localization and their Fourier content
depend 
%are specific to 
on the chosen set. Second, their space-time
structure is independent of this set and directly follows from
%is intrinsic to 
the amplification process (or equivalently, the neglect of the decaying mode).

It should be noticed that
the space-time properties of these correlations coincide with those
obtained by having applied a wave-packet transform to the two-point function.
However, this second procedure 
is more general in that it gives also rise to
correlations in amplitude. These cannot be obtained by working 
with the two-point function because in that case 
the mean has already been taken. 
Finally, even though the first procedure
%of wave-packet transforming the two-point function 
is simpler, 
%since we work with the mean of a statistical ensemble,
% of pairs of modes,   
%its probabilistic
the physical interpretation of the correlations it displays
is unclear, at least to us. 
On the contrary, the interpretation of the analysis performed
in configuration space is unambiguous 
and reached on a more fundamental level. 
%Up to now correlations have mostly been studied 
%in the mean by analysing  
%David07/07/04
The question of whether our procedures
can be implemented to observational data 
%(given their present status) 
is addressed at the end of the paper.

%%%%%%%%%%%%%%%%%%%%%%%%%%%%%%%%%%%%%%%%%%%%%%%%%%%%%%%%%%%%%%%%%%%%%%%%%%%%
%%%%%%%%%%%%%%%%%%%%%%%%%%%%%%%%%%%%%%%%%%%%%%%%%%%%%%%%%%%%%%%%%%%%%%%%%%%%
%%%%%%%%%%%%%%%%%%%%%%%%%%%%%%%%%%%%%%%%%%%%%%%%%%%%%%%%%%%%%%%%%%%%%%%%%%%%
%%%%%%%%%%%%%%%%%%%%%%%%%%%%%%%%%%%%%%%%%%%%%%%%%%%%%%%%%%%%%%%%%%%%%%%%%%%%

\section{Bogoliubov transformation and two-mode states} \label{sec:settings}

%The first and second parts of this section present no new material. 
In this section we recall the basic elements which 
define Bogoliubov transformations in a 
%its probabilistic
the physical interpretation of the correlations it displays
is unclear, at least to us. 
On the contrary, the interpretation of the analysis performed
in configuration space is unambiguous 
and reached on a more fundamental level. 
%Up to now correlations have mostly been studied 
%in the mean by analysing  
%David07/07/04
The question of whether our procedures
can be implemented to observational data 
%(given their present status) 
is addressed at the mological context. 
%We also introduce 
In Eq. \Eqref{tildeP} we introduce the notion of two-mode states
which will play a central role in encoding 
the correlations we shall focus on.
%We conclude this Section by two remarks which relate our work
%to more conventional treatments.
%induced by pair creation in this context.

It has been shown that the evolution of linearized cosmological perturbations
(primordial gravitational waves and density perturbations) reduces to the
propagation of real, massless, minimally coupled scalar fields in FRW spacetimes 
\cite{MBF92}. 
For simplicity, in this article, we shall study the fluctuating properties
of a test scalar field $\xi $ in a homogeneous background. 
The translation of the results 
to physical fields represents no difficulty.

We work with a line element with flat spacial surfaces:
\ba \label{metric}
 ds^2 &=& a(\eta)^2 \left[ -d\eta^2 +  \delta_{ij}dx^{i}dx^{j} \right] \, .
\ea
The field $\xi(\eta,{\bf x})$ obeys the d'Alembertian equation:
\ba \label{PropPhi}
 \partial_{\eta}^2 \xi + 2 {\cal{H}} \partial_{\eta} \xi - 
 \inv{a^2}\bold{\nabla}^2 \xi = 0  \, , 
%\nonumber 
\ea
where ${\cal{H}} ={\partial_{\eta} a}/{a}$ is the conformal Hubble parameter and
$\bold{\nabla}$ is the gradient with respect to the comoving coordinates
$\bf x$.
It is convenient to introduce the rescaled field $\phi = a  
\xi $ and
to decompose it into Fourier modes
\ba
% a(\eta) \label{rescalledfield} \xi(\eta, \bold x) &=&  
\phi(\eta, \bold x)
&=& \int\!\!d^3k \, \frac{e^{i \bold k \bold x}}{(2\pi)^{3/2}} 
  \phi_{\bf k}(\eta) \, .
% \, \nonumber\\&=& \int\!\!d^3k \, (A_{\bf k}\tilde \phi_{\bf{k}} + c.c.) \, .
\ea
%R%When dealing with the rescalled field $\phi$ and 
%Using conformal coordinates,
%the scale factor $a$ drops from the space-time modes
% $\tilde \phi_{\bf{k}}$. 
%Hence they reduce to the usual Minkowski expression:
%\ba \label{stmodes}
% \tilde \phi_{\bb{k}}(\eta,\bold x ) = 
%\frac{e^{i \bold k \bold x}}{(2\pi)^{3/2}} \phi_{k}(\eta) \, .\ea
%Moreover they 
%They form a complete orthonormal basis with respect to the Klein-Gordon scalar
%product.
The time dependent mode $\phi_{\bold k}$
% = A_{\bf k} \phi_{k}+ A_{\bf -k}^* \phi_{k}^*$ 
obeys 
%follow the equation of a parametrized harmonic oscillator
\ba \label{PropPhi_k}
 \left(\partial_{\eta}^2  + \omega_{k}^2
%(\eta)
 \right) \phi_{\bf k} = 0 \, ,
\ea
where $k=\abs{\bold k}$ and where the time-dependent frequency is given 
by\footnote{The $\bold k$-mode of the gravitational waves obeys 
Eq. \Eqref{PropPhi_k} equation whereas the
%is the mode equation of motion for gravitational waves. For the 
density fluctuation mode has a frequency given by 
%perturbations, the dispersion law is
\ba \label{omega_k2}
 \omega_{k}^2(\eta) = c_s^2{\bold k}^2 - \frac{z''}{z} \, .
\nonumber 
\ea
%where t
The function $z(\eta)$ is determined by the 
% of the a combination of 
the background evolution:
\ba
 z= a\frac{({\cal{H}}^2 - {\cal{H}}')^{1/2}}{{\cal{H}}c_s} \, .
\nonumber 
\ea
%where 
%$c_s$ is the sound velocity: 
During inflation $c_s=1$ and during the adiabatic era it is given by the 
sound velocity $c_s^2=\delta p / \delta\rho$ \cite{MBF92}.}
\ba \label{omega_k}
 \omega_{k}^2(\eta) = {\bold k}^2 - \frac{a''}{a} \, .
\ea

In second quantization, these modes are decomposed as
%operators:
%R% Fourier components of the rescalled field operators are 
\ba \label{hatphik}
 \hat \phi_{\bold k}(\eta) = 
 \hat a_{\bold k}  \phi_{k}(\eta) + 
 \hat a_{-\bold k}^{\dagger} \phi_{k}^{*}(\eta) \, ,
\ea
where the 'hat' characterizes operators.
In this decomposition, 
%the c-number mode 
$\phi_k$ is a solution of 
Eq. \Eqref{PropPhi_k} with unit positive Wronskian \cite{B&D}. 
%\ba \label{unitwronskian}
%\left(\phi_k,\, \phi_k \right) = \phi_k^* i\!\!\dpartialeta \phi_k = 1 \, .
%\ea
%Here $\phi_k$ 
It depends on
the norm of $\bf{k}$ only since we work
%the field propagates 
in an isotropic background.
The operators $\hat a_{\bold k}$ and $\hat a_{\bold k}^{\dagger}$ 
are the creation and
annihilation operators of a quantum of comoving momentum $\bf{k}$.
%They are defined by the Wronskians
%\ba \label{defcreationop}
% \hat a_{\bf{k}} =  \left( \phi_k,\hat \phi_{\bf{k}} \right)  \mbox{, \ \ \ \ } 
% \hat a_{\bf{-k}}^{\dagger} = - \left( \phi_k^*,\hat \phi_{\bf{k}} \right) \, ,
%\ea 
%or equivalently by the overlaps based on the Klein-Gordon scalar product 
%\ba \label{defcreationop2}
% \hat a_{\bf{k}} =  \langle \tilde \phi_{\bf{k}},\hat \phi  \rangle_{KG}  
% \mbox{, \ \ \ \ } 
% \hat a_{\bf{-k}}^{\dagger} = 
% - \langle \tilde \phi_{\bf{-k}}^*,\hat \phi \rangle_{KG}  \, ,
%\ea 
%where $\hat \phi(\eta, \bf{x})$ is the field operator.
% of the Hilbert space with respect to the Klein-Gordon product. 
%The completeness of the modes Eq. \Eqref{stmodes} 
%guarantees that the canonical equal time commutation relations
%\ba \label{ETCR}
%[\hat \phi(t,\mathbf{x}')
%%_{\mathbf k}(t), 
%\partial_t \hat \phi(t,\mathbf{x})]=
%%_{\mathbf{-k'}}(t)]=
% i\delta^3(\mathbf{x}-\mathbf{x}')  \, , \ea
%are equivalent to 
%\ba \label{Commut}
%\, [\hat a_{\mathbf{k}}, \hat a_{\mathbf{k'}}^{\dagger}]
%&=&\delta^3(\mathbf{k}-\mathbf{k'}) \, , 
%\nonumber \\
%\, [\hat a_{\mathbf{k}}, \hat a_{\mathbf{k'}}] &=& 0  \,  .
%\ea
The ground state of each mode $\bf{k}$ is defined by
\ba
 \hat a_{\mathbf{k}} \ket{0, \, \bf{k}} = 0 \, .
% 0 \qquad \mbox{for \ all \ } \bf{k}.
\ea
Since annihilation operators of different momenta commute, the ground state
of the field is 
%factorises as 
the tensorial product over all $\bf{k}$:
\ba
 \ket{0} = 
%{\bigotimes_{\bold k}} 
\prod_{\bold k} \otimes 
 \ket{0,\,\bold k} \, .
\ea

When studying correlations due to pair creation, it is appropriate
to re-write the vacuum in terms of two-mode states:
\ba \label{tildeP}
 \ket{0} = \widetilde{\prod_{\bold k}}
%\bigotimes_{\bold k}} 
\otimes \kett{0,\,\bold k} \, .
\ea
%In the second line we have introduced the notation for a 
The tilde tensorial product takes into account only half of the modes 
and the $\bold k$-th two-mode vacuum state is defined by
\ba
  \kett{0,\,\bold k} = \ket{0,\,\bold k} \otimes \ket{0,\,-\bold k} \, .
\ea
%and the tilde tensorial product which takes into account half of the modes.
%This rewritting 
%in terms of two-mode states is 
%particuliarly convenient for two reasons. .........
%because, in homogeneous background, 
%each created pair is made of quanta with opposite momentum. 
%since in expanding backgrounds, one gets pair creation
%of quanta of opposite momentum.
%Under the evolution, the state of the
%field developes EPR correlations. In order to exhibe these correlations
%amongst quanta of a same pair, 
(Notice  that this notion 
can be generalized to other states
%, like occupation number states: 
%\ba \label{2modestate}
%\kett{\psi, \, \bold k } = 
%\ket{\psi, \, \bold k } \otimes \ket{\psi, \, -\bold k } \, .\ea
whenever both modes are 
in the same 1-mode state.)
% $\psi$.) 
When using this writing, one must pay attention not to
count modes twice. To this end, for a real scalar field, 
one needs to separate (arbitrarily) the momentum space in two. 
For definiteness, we choose the separation according to the sign
of $k_x$, the $x$-component of the momentum.
Then the product in Eq. (\ref{tildeP}) is performed
over momenta with  positive $k_x$ only.
To emphasize this 
%the partnership bewteen $\bold k$ and $-\bold k$ modes, 
we shall call modes, states, and operators
%separate modes according to the sign of $k_x$ and call the 
right ($R$) or left ($L$) according to the sign of $k_x$.
Hence we write $\hat a_{\bold k}^{R}= \hat a_{\bold k}$
and $\hat a_{\bold k}^{L}= \hat a_{-\bold k}$.
%For instance
%the states are now rewritten as 
%%modes respectively and note them:
%\ba \label{twomodevac}
% \ket{\psi,\, {\bold k}, R} = \ket{\psi,\, {\bold k}}
%k_x > 0} 
%\, , \qquad 
% \ket{\psi,\, {\bold k}, L} = \ket{\psi,\, - {\bold k}}
%%k_x < 0} 
%\, .
%\ea
%Likewise, sub- or super-scripts $R$ (resp. $L$) 
%designate states and operators 
%which contain only $k_x > 0$ (resp. $k_x < 0$) componants.
%For instance, a two-mode vacuum state is defined by
Using this notation, the two-mode vacuum state obeys
\ba
\label{twomodevac2}
 \hat a_{\bold k}^{R} \kett{0,\,\bold k} = 
 \hat a_{\bold k}^{L} \kett{0,\,\bold k} = 0 
 \, .
% \qquad  \mbox{for \ all \ } {\bf k} \mbox{\, with \, } k_x > 0 \, ,
\ea 
 
%The vacuum of the field is the tensorial product
%\ba
% \ket{0} =  \widetilde{\prod\limits_{\bold k}}  \kett{0,\,\bold k} \, , 
%\ea
%where the tilde product $\widetilde{\prod}=\prod_{k_x >0}$ is the product over 
%half momentum space.
%The field is decomposed into
%\ba \label{fielddecomposed}
% \hat \phi(\eta, \bold x) 
% &=& \hat \phi_{R} + \hat \phi_{L} \nonumber \\
% &=&  \tint \frac{e^{i \bold k \bold x}}{(2\pi)^{3/2}}
% \left(\hat a_{\bold k}^{R} \phi_{k} + 
% \hat a_{-\bold k}^{R\, \dagger} \phi_{k}^{*} \right)  
% + \tint \frac{e^{-i \bold k \bold x}}{(2\pi)^{3/2}} 
% \left(\hat a_{\bold k}^{L} \phi_{k} + 
% \hat a_{-\bold k}^{L\, \dagger} \phi_{k}^{*}\right) \, ,
%\ea
%where the measure $d^3\tilde k$ 
%indicates that the sum extends to half momentum space.

In non-stationary backgrounds, the frequency Eq. (\ref{omega_k}) depends
on time. Hence the non-adiabaticity of the propagation
leads to spontaneous excitations of the various modes. 
To characterize these transitions, it is appropriate to introduce two 
sets of modes. These 
are positive frequency solutions of Eq. \Eqref{PropPhi_k} at 
early and late time. 
In Appendix A, they are explicitly given when considering a 
cosmological evolution which starts with an inflationary phase and ends
by a matter dominated period after having experienced a radiation
dominated period.
% which asymptotically have positive frequency.  
%of the equation \Eqref{unitwronskian} does not define 
%positive frequency modes. 
%Therefore one can define several vacua.
As usual, we shall use the labels 'in' and 'out' to designate 
%creation and
states and operators 
%(as well as states)
which are defined with respect to the corresponding modes. 
% during the matter domination.three periods relevant for cosmology, namely, 
%inflation, radiation and matter domination.
%a or an explicite
%time and at a later one.  
Since Eq. \Eqref{PropPhi_k} is homogeneous and  linear,
%, second order ordinary differential 
 in and out modes 
%solutions 
are related by a 
%linear transformation, called a 
Bogoliubov transformation 
%in this context:
\ba \label{inoutmodes}
 \phi_k^{in}(\eta) = \alpha_k \phi_k^{out}(\eta) + 
 		     \beta_k^* \phi_k^{out\, *}(\eta) \, .
\ea
The corresponding 
%Bogoliubov 
transformation between in and out operators is
%can be found:
\ba \label{inoutoperator}
 \hat a_{\bold k}^{R, \, in} = \alpha_k^{*} \, \hat a_{\bold k}^{R,\, out} -
		      \beta_k \, \hat a_{\bold k}^{L,\, out \, \dagger} \, .
%, \qquad      \mbox{for\, all\, } {\bf k} \, .
\ea 
Because of the homogeneity of the background,
this transformation 
%couples two-mode of opposite momenta. Therefore the Bogolibov transformation 
is $2\times 2$ block-diagonal as it couples ${\bf k}$ to $-{\bf k}$ only.
Hence every produced ${\bf k}$-particle will be accompanied by a partner 
of momentum $-{\bf k}$. Moreover particles characterized 
by different momenta are incoherent in the in vacuum
(in the sense that in the expectation value of any
product of annihilation and creation operators of different momenta 
will factorize).

These two properties are made explicit when  
expressing the in vacuum in terms of out states (i.e.
states with a definite out particle content). From  
Eq. \Eqref{inoutoperator}, using the notations
of Eq. \Eqref{tildeP}, one gets (see App. B in \cite{CP1}, \cite{PhysRep95})
\ba \label{inoutvact}
 \ket{0,\, in} &=&  \widetilde{\prod\limits_{\bold k}} \otimes  
 \kett{0,\, \bold k, in} 
\nonumber\\ 
&=& 
\widetilde{\prod_{\bold k}}  \otimes 
%\bigotimes_{\bold k}} \otimes 
\left(  \inv{\vert \alpha_{k} \vert} 
   \exp \left( z_{k}\, 
  \hat  a_{\bold k}^{R,\, out \, \dagger}  \, 
  \hat a_{\bold k}^{L, \, out \, \dagger} \right)
  \kett{0,\, \bold k, out} \right) \, ,
\ea
where $z_{k}=\beta_{k}/\alpha_{k}^{*}$. From this writing we see that
the in vacuum factorizes into a product over half the momenta
of sums of two-mode out states. 
%R%
It has to be emphasized that these out states carry
no 3-momentum since they contain exactly the same number of $R$ and $L$ 
out ${\bold k}$-particles. 
%Having so defined the in anihilation operators, the $\bold k$-th 
%two-mode in-vacuum obeys,
%as in Eq. \Eqref{twomodevac2}, 
%%two-mode in-vacuum obeys
%%is defined by $\kett{0,\,\bold k, in}$ by
%\ba \label{a^ina^out}
% \hat a_{\bold k}^{R,\,in} \kett{0,\,\bold k, in} = 
%%%%%,R and ,L supressed
% \hat a_{\bold k}^{L,\,in} \kett{0,\,\bold k, in} = 0 \, .
%\ea 
%Similarly, the out vacuum is defined with the out annihilation operators. 

Our aim is to analyze how the these properties
determine the {\it space-time structure} of the correlations
of the $\phi$ field. 
%R%
%We shall study the mean description in
%Sections III and IV. 
%two different cases: in the mean,
%when 
%re the only knowledge on 
%In Sections V and VI we shall study 
%the field state is the in vacuum to 
%the situation obtained
%by adding the information that classical waves have been detected.
We shall use two different approaches.
%The correlations present in the in vacuum
%can be analyzed in two manners. 
In Section III and IV we shall work directly
%present the first one which  
%simplest and usually adopted approach is based on 
with the two-point correlation function
\ba \label{2ptf}
 G_{in}(\eta, {\bf x}; \eta' ,{\bf x}') = 
 \bra{0\, in}\hat \phi(\eta, {\bf x}) \,  \hat \phi  (\eta' ,{\bf x}') 
 \ket{0\, in} \, .
\ea

In Section V and VI, we develop an alternative 
%and more decotailed way 
%to study the correlations 
approach, based on \cite{CP1}, which is more fundamental as it is based on 
%the analysis of 
the correlations in configuration space encoded in
 Eq. \Eqref{inoutvact}.
%exist before having taken the mean. 

\section{Spatial correlations induced by pair-creation} 
\label{sec:EPR}

%Let us present these two approaches in turn.

%\subsection{Two-point correlation functions in the large occupation number limit} 
%\label{sec:EPR,one}

Since we are dealing with a free field, 
all (in-in) expectation values of products of the field operator 
can be decomposed in terms of the two-point function $G_{in}$. 
Its late time properties are best revealed by decomposing the field operator 
into out modes. One gets
\ba \label{2ptf2}
 G_{in}(\eta, {\bf x}; \eta' ,{\bf x}') &=& 
 G_{out}(\eta, {\bf x}; \eta' ,{\bf x}') \nonumber\\
 &&+ 2 
 \int\!\!d^3k \, \frac{e^{i {\bf k}( {\bf x} -  {\bf x}' )}}{(2 \pi)^3} \,
 %\int d^3k' {e^{i {\bf k}( {\bf x} -  {\bf x}' )} \over (2 \pi)^3} \left[ 
 n_k \, Re\left\{ \phi^{out}_{k}(\eta) \phi^{out\, *}_{k}(\eta' )\, 
 %P({\bf k},{\bf  k}') \tilde \phi^{out}_{\bf k}(\eta, {\bf x})  \tilde \phi_{\bf k'}^{out\, *}(\eta', {\bf x'}) 
 %\right.
 %\nonumber\\
 %&&\quad \quad \left.
 + 
 %2 Re \{  
 \frac{c_k}{n_k} \phi^{out}_{k}(\eta) \phi^{out}_{k}(\eta' )\right\}
 %C({\bf k},{\bf  k}') \tilde \phi_{\bf k}^{out}(\eta, {\bf x})  \phi_{\bf k'}^{out}(\eta', {\bf x'}) \} 
 %\right] 
 \, . \quad \quad  
% \label{2ptf2}
\ea
In the first line, $G_{out}$ is the Wightman function evaluated 
in the out vacuum.
This quantum contribution is $O(1)$ whereas 
the second term is proportional to the occupation number $n_k$. 
Hence when $n_k \gg 1$
the vacuum contribution can be neglected (unless one computes 
operators containing commutators, because in certain cases,
the second term might not contribute since it is symmetric in $x, x'$).  
Notice also that we could have split $G_{in}$ into a commutator and an 
anti-commutator.
In the large occupation number limit, the dominant terms coincide.

The second line of Eq. \Eqref{2ptf2}
is governed by two quantities. First one has a diagonal term 
%(in occupation number): 
\ba \label{occup} 
 % P({\bf k},{\bf  k}') = 
 \langle \hat a_{\bold k,\, R}^{out\, \dagger} \,  
 \hat  a_{\bold k',\, R}^{out} \rangle_{\rm{in}}
 =   \langle \hat a_{\bold k,\, L}^{out\, \dagger}  \, 
 \hat  a_{\bold k',\, L}^{out} \rangle_{\rm{in}}  
 = n_k  \, \delta^3(\bold k - \bold k' )
 = \abs{\alpha_k}^2 \abs{z_k}^2 \, \delta^3(\bold k - \bold k')  \, ,
 %= \abs{z_{k}}^2 \abs{\alpha_k}^2   \, .
\ea 
which fixes the mean number  $n_k$. 
(The symbol $\langle \, \cdot \, \rangle_{\rm{in}}$ designates
the in vacuum expectation value: 
$\langle 0 in \vert  \,  \cdot \, \vert 0 in \rangle$.)
Second one has an interfering term
\ba \label{cross} 
 %C({\bf k},{\bf  k}')&=& 
 \langle \hat a_{\bold k,\, R}^{out}  \,  
 \hat a_{\bold k',\, L}^{out} \rangle_{\rm{in}}
 = c_k \, \delta^3(\bold k - \bold k' ) =
  \abs{\alpha_k}^2 z_{k} \, \delta^3(\bold k - \bold k' )  \, ,
 %\nonumber \\ &&
 %\langle \hat a_{\bold k,\, R}^{out}  \,  \hat a_{\bold k,\, R}^{out} \rangle_{\rm{in}}
 %=\langle \hat a_{\bold k,\, L}^{out}  \,  \hat a_{\bold k,\, L}^{out} \rangle_{\rm{in}}= 0
\ea 
which governs the {\it coherence} of the distribution.  
By coherence we mean that the expectation value of a product
does not factorize. In the present case, it is  
$\langle \hat a_{\bold k,\, R}^{out}  \,  
\hat a_{\bold k,\, L}^{out} \rangle_{\rm{in}}
\neq 
\langle \hat a_{\bold k,\, R}^{out}  \rangle_{\rm{in}}
\langle \hat a_{\bold k,\, R}^{out}   \rangle_{\rm{in}} = 0$
which thus expresses the coherence. For incoherent distributions,
such as thermal baths, one would get $c_k = 0$ for all ${\bf k}$.
Notice also that the expectation values  
which differ from the above ones by one additional $\dagger$ on 
an operator $\hat a$ all vanish. This last property is valid for all
homogeneous and isotropic 
distributions (and not only those resulting 
from pair creation).
%Eq. \Eqref{cross} tells us that every $R$ $\bold k$-particle 
%is accompagned by one partner which is a $L$ $\bold k$-particle. 
The {degree of coherence} of the distribution
is given\footnote{
In this definition, $n_k+1/2$ and $c_k$ 
are the non-vanishing elements of the covariance matrix in the two-mode
state $\kett{0,\, {\bf k}, in}$, defined by the expectation values of the
anticommutators of $\hat a_{R}$, $\hat a_{L}$, $\hat a^{\dagger}_{R}$ 
and $\hat a^{\dagger}_{L}$. 
This matrix is also the 
covariance matrix of the corresponding classical distribution 
\cite{cpPeyresq}.} by $\abs{c_k}/ (n_k + 1/2)< 1$.
For pair creation from vacuum, one has
$\abs{c_k}/ (n_k + 1/2) = 2 \abs{z_k} / (1 + \abs{z_k}^2)$.

For macroscopic occupation numbers, 
the norms of the diagonal and interfering terms 
%Eqs. \Eqref{occup,cross}
coincide since $\vert z_k \vert= \vert \beta_k /\alpha_k\vert \to 1$. 
In this limit the interfering term can thus be written as
%Eq. \Eqref{cross} gives
%implies that 
\ba \label{coherenceterm}
c_k = - n_k  \, e^{i 2 \psi_k} \, .
\ea 
%which means that the distribution is maximally coherent.
%every $R$ $\bold k$-particle 
%is accompanied by one partner, a $L$ $\bold k$-particle. 
Taking into account the isotropy of the distribution,
the dominant part of the two-point function simplifies and reads
\ba \label{infl2ptf1}
G_{in}(\eta, {\bf x}; \eta' ,{\bf x}') 
%&=& 
%\int_0^\infty \frac{dk k^2}{\pi^2} 
%\frac{\sin(k\vert {\bf x} - {\bf x}'\vert)}{k\vert {\bf x} - {\bf x}'\vert} 
%\,  n_k \, Re\left\{ \phi^{out}_{k}
%\phi^{out\, *}_{k} - e^{2 i\psi _k} \phi^{out}_{k} \phi^{out}_{k}\right\}
%\,
%\nonumber \\
&=& \int_0^\infty \frac{dk k^2}{\pi^2} 
\frac{\sin(k\vert {\bf x} - {\bf x}'\vert)}{k\vert {\bf x} - {\bf x}'\vert} 
\, 4 n_k \, Im\left\{ e^{ i\psi _k} \phi^{out}_{k}(\eta) \right\} 
Im\left\{ e^{ i\psi _k} \phi^{out}_{k}(\eta')\right\} \, .\quad 
\ea
%D25/11/03
%As noted in \cite{GuthPi85}, t
%The dominant part of the two-point function
The integrand is a product of two classical waves 
($= Im\left\{ e^{ i\psi _k} \phi^{out}_{k}\right\}$). 
Three remarks are in order.
First, this factorization could not have been performed
if the distribution did not obey 
%was not fully coherent, i.e. if 
$\vert c_k \vert = n_k$. 
%R%
Therefore the fact that it can be done
is an expression of the two-mode coherence of the underlying
distribution.

Second, when working with a (two-mode)
coherent state,  
%(see Eq. \Eqref{Impeq}),
the Green function is also a product of two classical waves,
% when the state of the field is a coherent state,
see Eq. %\Eqref{cohgreen} ?ou bien C4??. \Eqref{phiintopsi} and 
\Eqref{greenforcomplexocs}. 
%with $w= -v$.
This therefore suggests to view Eq. \Eqref{infl2ptf1}
as resulting from an ensemble of two-mode coherent states,
see Section V. 
% and \cite{cpPeyresq} for more details.

Third, 
%it is appropriate to relate this
the two-mode coherence giving rise to the classical waves 
$Im\left\{ e^{ i\psi _k} \phi^{out}_{k}\right\}$
%is closely related to 
yields the usual 
description \cite{GuthPi85,PolStar96,Matacz94,KieferPol98}
based on the neglect of the decaying mode,
see \cite{cpPeyresq} for more details. 
For simplicity, let us consider 
a radiation dominated universe which follows a period of inflation. 
In this case the classical wave with $k\eta_r \ll 1$, 
where $\eta_r$ is the time of reheating defined in Eq. \Eqref{aeta},
are proportional to
%Im\left\{ e^{ i\psi _k} \phi^{out}_{k}(\eta)\right\}= 
$\sin(k\Delta \eta)$
%\eta - \psi_k) /(2k)^{1/2}$.  
%Therefore, for modes such that $k\eta_r \ll 1$ 
%where $\eta_r$ is defined in Eq. \Eqref{aeta}, 
%the value of $\psi_k$ is such that The argument of the sine is equal to $k 
%where 
up to a correction term of the 
order of $(k \eta_r)^3 \simeq 10^{-75}$ when inflation lasts 
for about 60 e-folds. 
They correspond to growing modes since 
the conformal time-lapse $\Delta \eta $ is proportional to $a(\eta)$.
%the conformal time 
%evaluated since the reheating. 
Given this strict correspondence
between $Im\left\{ e^{ i\psi _k} \phi^{out}_{k}\right\}$
and growing modes, one can abandon the quantum settings
and proceed with the effective description based on growing modes 
with stochastic amplitudes.
% \cite{GuthPi85,PolStar96,Matacz94,KieferPol98}. 
In this paper we shall nevertheless use the quantum formalism 
in Sections V and VI for the following reason.
%s.First it requires no approximation. Second, i
It allows to treat separately right and left moving wave packets, 
a feature which is useful and which leads to
a transparent interpretation of the results. 
The next Section however
is based on the two-point function and can therefore be interpreted
in either formalism. 

We now wish to illustrate how the above coherence 
(or equivalently the neglect of the decaying mode)
induces spatial structures
on a given time slice, e.g. on the Last Scattering Surface (LSS). 
%with this As a concrete example,
% followed by 
%When applying this to the case of in a radiation dominated universe.
%In this case, for modes
%such  that $k\eta_r \ll 1$, 
%where $\eta_r$ is the time of reheating defined in Eq. \Eqref{aeta},
%one verifies that the 
%D25/11/03 degree of coherence satisfies 
%norms of $|c_k|/n_k = 1 + O(1/n)$. Since$\psi_k = 4 k \eta_r$, 
%For these modes,
%In this case,
%the classical waves ($= Im\left\{ e^{ i\psi _k} \phi^{out}_{k}\right\}$)
%are proportional to
%Im\left\{ e^{ i\psi _k} \phi^{out}_{k}(\eta)\right\}= $\sin(k\Delta \eta)$
%\eta - \psi_k) /(2k)^{1/2}$.  
%Therefore, for modes such that $k\eta_r \ll 1$ 
%where $\eta_r$ is defined in Eq. \Eqref{aeta}, 
%the value of $\psi_k$ is such that The argument of the sine is equal to $k 
The lapse $\Delta \eta $ appearing in the classical 
waves determines the characteristic size of the structures on the LSS. 
%locus of constructive interferences.
Indeed when $\eta' = \eta$, the stationary phase
condition applied to the integrand of Eq. \Eqref{infl2ptf1}
gives two solutions. 
First one gets $\vert {\bf x} - {\bf x}'\vert= 0 $ 
which is responsible for the usual divergence 
in coincidence point limit. More importantly, there also 
exists a
%is another
non-trivial solution:
\ba
\label{distc}
\vert {\bf x} - {\bf x}'\vert = 2 \Delta \eta
= 2 (\eta -  \partial_k \psi_k ) = 
2 (\eta -  2 \eta_r ) \, .
\ea
This {\it only} results from the interference term 
$c_k$. The lapse $2 \Delta \eta$
designates the mean separation reached by the particles and 
their partners from their birth near $a=0$, see Appendix A.
This interpretation will become clear in the sequel.

It should be also noticed that the contribution of this second
term is negative, thereby causing a {\it dip} in the two-point 
function \cite{BB}. 
The origin of this dip can be traced to Eq. \Eqref{coherenceterm}
and the fact that $\psi_k \ll 1$.
It tells us that only the growing mode has been kept.
As we shall demonstrate in Section VI, 
this implies that the partner of {\it any} local
over-density is a local under-density. 
Notice finally that the relative weight of the usual solution
and that of Eq. \Eqref{distc} is two. In the next Section
this factor shall be recovered and explained 
in terms of local wave packets.

%Notice also that in a radiation dominated universe, the distance 
%Eq. \Eqref{distc} is equal to twice the Hubble radius (times $c_s/c$ 
%if the field propagates at the speed $c_s$).
%Therefore the particle and its partner are flying on the (particle) horizon
%when $c_s=c$. 

%%%%%%%%%%%%%%%%%%%%%%%%%%%%%%%%%%%%%%%%%%%%%%%%%%%%%%%%%%%%%%%%%%%%%%%%%%%%%
%%%%%%%%%%%%%%%%%%%%%%%%%%%%%%%%%%%%%%%%%%%%%%%%%%%%%%%%%%%%%%%%%%%%%%%%%%%%%
%%%%%%%%%%%%%%%%%%%%%%%%%%%%%%%%%%%%%%%%%%%%%%%%%%%%%%%%%%%%%%%%%%%%%%%%%%%%%%%
\section{wave packet transform and space-time correlations} 
\label{sec:wpt}

In this section we analyze the space-time correlations 
obtained 
%two-point function evaluated in the in vacuum, without performing
%any projection in configuration space. To this end, let us perform 
by wave packet transforming the two-point function with respect 
to a $R$-moving wave packet. We shall consider two different 
scalar products: one based on the usual overlap and the
other based on the Klein-Gordon product. The first product leads
to a spatial structures containing three packets whereas the second leads
to two packets only. The reason comes from the fact that 
$R$-moving and $L$-moving modes equally contribute 
in the first case, thereby leading to a doubling of the 
partners. The knowledge of the velocity field is required
to lift this degeneracy.

% is role of the velocity field will be revealed 

%This wave packet 
A $R$-moving wave packet can be written as 
%recuperate the set of Fourier coefficients $v_{\bf k}$ of Eq. \Eqref{wp}
%defining the classical R-wave given in Eq. \Eqref{clRwave}.
%We then  use the 
\ba \label{condvalues}
 \bar \phi_{R, \, {\cal V}_R}(\eta, {\bf x}) &=& 
 \int\!\!\widetilde{d^3 k} \, 
     \left( v_{\bold k} \, \frac{e^{i \bold k \bold x}}{(2\pi)^{3/2}} 
  \phi_{k}^{out}(\eta)
  %\tilde \phi_{\bold k}^{out}(\eta, {\bf x}) 
  + c.c.
     %v_{\bold k}^{*}\,  \tilde \phi_{\bold k}^{out \, *} (\eta,{\bf x}) 
     \right) \, ,
    \nonumber\\
     &=& \bar \phi_{R, \, {\cal V}_R}^{(+)}
     %(\eta, {\bf x}) 
     + \bar \phi_{R, \, {\cal V}_R}^{(-)} \, .
\ea     
where the tilde integral means that only positive values of $k_x$ are
considered. 
%pHence the wave packet is purely $R$-moving.
The symbol ${\cal V}_R$ designates the set of Fourier
amplitudes $v_{\bold k}$. It is used to remind that the specification
of the wave packet has been made in the $R$-sector. 
%Later on we shall designate its $L$-moving partner by 
%$\bar \phi_{L, \, {\cal V}_R}$. 
In the second line, we have decomposed
the wave into its positive and negative frequency content.
This will be needed when considering the Klein-Gordon product.

To be specific we shall consider 
%an explicit example. 
%Suppose that the classical configuration is composed of 
a single Gaussian wave packet.
For clarity of the equations, we write 
\ba
 v_\bold k = \bar v \, f_R(\bold k) \, , 
\ea
where $\bar v$ is real and positive and 
where the function $f_R$
is normalized to unity:
\ba \label{normv_k}
\int\!\!\widetilde{d^3k} \, \abs{f_R}^2 = 1 \, .
\ea 
%The function $f_R(\bold k)$
%characterizes the Fourier content of the classical configurations.
We shall use the following Gaussian wave
\ba \label{wp}
% v_{\bold k} =  \bar v\,  
f_R(\bold k) = 
%\bar v 
e^{i\phi} \, N\, 
 e^{-\frac{(\bf k - \bar{\bf k})^2}{4\sigma^2}} 
 e^{-i\bf k\bf x_0} e^{i k \eta_0} \, ,
\ea
where $N>0$ is a constant such that Eq. \Eqref{normv_k} is satisfied. 
%The amplitude and the phase of 
%detected classical wave are given by $\bar v$ and $\phi$ respectively
%(see Eq. \Eqref{cltrajincohstate}).
The mean momentum of the wave packet is $\bar{\bf k}$ and 
its mean position at $\eta_0$
is ${\bf x}_0$. $\phi$ is the phase of the positive frequency part of the wave 
evaluated at ${\bf x}_0, \eta_0$.
%the center.
%D25/11/03
% of the wave packet.
%R%
(Notice that when considering only the 1-particle sector, 
this phase would be inaccessible. 
Instead when dealing with coherent states, it is an observable). 
Suppose also that 
we are working in
%the detection occurs during the
radiation dominated era, so that positive frequency modes are
\ba \label{clRwave}
\phi_{k}^{out}(\eta) = 
 \inv{\sqrt{2k}} e^{-ik \eta} 
 %e^{i\bf k\bf x} 
 \, .
\ea
%With our conventions, the complex function $z_k$ is
%Therefore the conditional amplitude for  t
The $R$-wave is thus
\ba \label{barphiR}
 \bar \phi_{R, \, {\cal V}_R} = \bar v N \int \widetilde{d^3k} \inv{\sqrt{2k}} 
 	e^{-\frac{({\bf k - \bar{\bf k}})^2}{4\sigma^2}} 
	%e^{-ik \eta} e^{i\bf k\bf x}
 \left(e^{i\bf k(\bf x - \bf x_0)}e^{-i k (\eta-\eta_0)}e^{i \phi} + c.c.
 \right) \, .
\ea
By making use of the stationary phase condition, 
one finds, as expected, that this wave-packet is maximum along the 
classical light-like trajectory 
\ba \label{maxR}
{\bf x}_R(\eta, \bar{\bf k}) = {\bf x}_0 +
 (\eta- \eta_0) {\bf 1}_{\bar{\bf k}} \, ,
\ea
which passes through $ {\bf x_0} $ at $\eta_0$ with a 
momentum $\bar{\bf k}$. (The vector
${\bf 1}_{\bar{\bf k}} $ is the unit vector in the direction of the
velocity of the chosen wave.) 
%Therefore, as expected, the 
%$R$ component of the field describes the wave packet which has been detected
%at $\eta_0$. 

%this wave 
\subsection{Wave packet transform based on the usual product}

We use the above $R$-moving wave to ``wave-packet transform'' the 
two-point function with respect to one of its arguments:
\ba \label{Theta}
 \Theta({\bf x}, \eta \, ; {\cal V}_R) 
 &=& \int\!\!d^3x' \, 
 \bar \phi_R (\eta_{0}, {\bf x}')\, 
 G_{in}(\eta, {\bf x}\, ; \eta_{0}, {\bf x}')  \, .
 \ea
% \nonumber \\
In Fourier transform we get
\ba \label{Theta1}
 \Theta({\bf x}, \eta \, ; {\cal V}_R)  
 &=& \int\!\!\frac{\widetilde{d^3k}}{(2 \pi)^{3/2}} \, 
 \left( v_{\bf k}{e^{i\bf k \bf x}} \phi_{k}^{out}(\eta_{0}) + c.c.\right) \, 
 \phi_k^{in}(\eta) \phi_k^{in\, *}(\eta_{0}) \, .
\ea
By 
%decomposing the two in modes (which are two 
keeping only the growing mode, 
%into out modes, 
see Eq. \Eqref{neglectdecay}, and using Eq. \Eqref{clRwave}, one obtains 
%They give rise to the following expression
%four-fold structure:
\ba \label{Gbar}
 \Theta({\bf x}, \eta \, ; {\cal V}_R) 
 &=&  \Psi_{{\cal V}_R}({\bf x}-{\bf x}_{0}, \eta-\eta_{0})
               + \Psi_{{\cal V}_R}({\bf x}-{\bf x}_{0}, -(\eta-\eta_{0}))
 \nonumber\\  
 &&  \quad\,\,
 - \Psi_{{\cal V}_R}({\bf x}-{\bf x}_{0},-(\eta+\eta_{0}-4\eta_r))
 -\Psi_{{\cal V}_R}({\bf x}-{\bf x}_{0},\eta+\eta_{0}-4\eta_r) \, .
\ea
The function $\Psi_{{\cal V}_R}$ is defined by 
\ba \label{psiVR}
 \Psi_{{\cal V}_R}({\bf x}, \eta) &=&
 2 Re \left\{ \int\!\!\frac{\widetilde{d^3k}}{(4 \pi k)^{3/2}} \,\, n_k
 %\frac{n_k}{(2k)^{3/2}} 
 \vert v_{\bf k} \vert \, e^{i \phi}
 e^{i {\bf k} {\bf x}} e^{-ik \eta}
 \right\} 
 \nonumber\\
 &=&\Psi_{{\cal V}_R}^{(+)}
 %({\bf x}, \eta) 
 + \Psi_{{\cal V}_R}^{(-)}  \, ,
\ea
where we have decomposed the amplitude $v_{\bf k}$ 
of Eq. \Eqref{wp} as its norm times 
its phase in order to exhibit the linear phases in ${\bf x}$ and $\eta$.
%%%%%%%%%%% D %%%%%%%%%%%%%%
%Notice that we have introduced a shift in time $\widetilde\eta_0$. By
%fine-tunning this phase, one is able to adjust the spreads of the waves in
%Eq. (\ref{Gbar}), see below.
In Eq. (\ref{Gbar}) it is the same 
function $\Psi_{{\cal V}_R}$ which governs the four contributions.
The reason for this simplicity arises from the fact that the differences 
between the integrands in Eq. \Eqref{Theta1} 
are given by $z_k = \beta_k/\alpha_k^*$
whose phase is linear in $k$, see Eq. \Eqref{z}. 
Hence the four contributions only differ by their temporal argument
and their relative sign. This relative sign guarantees that the 
integral over  $\bf x$ of $\Theta$ vanishes irrespectively of 
the chosen wave packet.
%Because of this symmetry, 
%all the pair with the same many pairs are taken into account within 
%a single wave packet transform. 
%One also notices that the function $\psi_{{\cal V}_R}$ 
%differs from $\bar \phi_R^{(+)}$ by its amplitude which is 
%amplified by $n_{k} \simeq \abs{\alpha_{k}}^2$. 
%one can evaluated the integral with a saddle-point approximation. 
%The result factorizes into
%In this approximation, the amplitude reduces to an overall factor given by 
%the occupation number evaluated for the
%mean momentum $\bar {\bf k}$. 

%It is of interest to analyze the limit of 
%narrow wave packets in $\bf k$ such as Eq. \Eqref{wp}.
%Moreover,
%%hen considering such wave packets,
%one can make a saddle-point approximation to evaluate the integral
%defining $\psi_{{\cal V}_R}$ in Eq. \Eqref{phiVR}. 
% which are well peaked around
%their mean momentum $\bar {\bf k}$, 

It is of interest to analyze the case where the integral 
Eq. \Eqref{psiVR} can be 
%In this case, one can 
evaluated by %the integrals with 
a saddle-point approximation.
%In this approximation, 
Then the shifts in $\eta$ translates
into shifts in $\bf x$. 
%the $v_{\bf k}$ given in Eq. \Eqref{wp},
% with 
%$\eta_{0}= \eta_{eq}$,
%}}the notations of the former Section, 
Indeed, given a wave-packet of mean momentum $\bar {\bf k}$,
Eq. \Eqref{Gbar} becomes 
%in this approximation
%one gets 
%at recombination
\ba 
\label{36a}
 \Theta({\bf x}, \eta;\, {\cal V}_R)  &=& 
 \Phi_{{\cal V}_R}\left({\bf x}-{\bf x}_R(\eta, \bar {\bf k}) ; \,
 \Sigma_{\bot,\, R}\right)
       + \Phi_{{\cal V}_R}\left({\bf x}-{\bf x}_R(\eta, -\bar {\bf k}); \,
 \Sigma_{\bot,\, R}\right)
 \nonumber\\ 
 && \, \,  
 - \Phi_{{\cal V}_R}\left({\bf x}-{\bf x}_L(\eta, \bar {\bf k} ); \,
 \Sigma_{\bot,\, L}\right)
 - \Phi_{{\cal V}_R}\left({\bf x}-{\bf x}_L(\eta, - \bar {\bf k}); \,
 \Sigma_{\bot,\, L}\right) \, .
\ea
The spreads 
%functions 
$\Sigma_{\bot,\, R}$ and $\Sigma_{\bot,\, L}$
are rather complicate matrices (in 3D). 
We postpone the analysis
of their interesting properties in a separate subsection.
The function ${\bf x}_R$ has already been defined in Eq. (\ref{maxR}).
Notice that we have enlarged its definition to ``negative'' wave vector 
%${\bf k} = - \bar {\bf k}$ 
so as to describe the second term of Eq. \Eqref{36a} which is a 
L-moving wave packet. 
The new function ${\bf x}_L$ is defined by
\ba \label{maxL}
{\bf x}_L(\eta, \bar{\bf k}) &=& {\bf x}_R(-\eta+ 4 \eta_r, \bar{\bf k}) \, ,
\nonumber \\
&=&
{\bf x}_0 - 
(\eta + \eta_0 - 4 \eta_r) {\bf 1}_{\bar{\bf k}} \, .
\ea
It corresponds to the trajectory of the partner of 
the wave of momentum $\bar {\bf k}$. This will be clearly
established in Section VI when dealing with coherent states.
We can already notice that   
%More importantly, it is separated from the detected wave by
the separation between the two waves is
\ba \label{maxL2}
 {\bf x}_R(\eta, \bar{\bf k})-{ \bf x}_L(\eta, \bar{\bf k}) = 
 2 (\eta - 2\eta_r)\, {\bf 1}_{\bar{\bf k}}\, .
\ea
The interpretation of this result is clear: it is the separation reached 
by the two waves since their creation (amplification)
near the big bang.
Indeed, 
%with our conventions, 
$\eta- 2\eta_r  \propto a(\eta)$ 
%after the reheating, 
during the radiation dominated era, see Appendix A. 

%Notice that t
The distance $\vert {\bf x}_R -{ \bf x}_L \vert$
is universal in the following sense. 
First, as expected, it is independent of both ${\bf x}_0$ (because
of the homogeneity of the process) 
and the direction specified by ${\bf 1}_{\bar{\bf k}}$ (because of isotropy).
Somehow more surprisingly\footnote{This is at least not usual
in that this is not what is found when considering pair production 
of massive particles in de Sitter space, eletro-production 
in a constant electric field, 
or pair production giving rise to Hawking flux
in quantum black hole physics. 
In all those cases, the norm of the
wave vector does characterize pairs produced at different times,
see \cite{CP1,PhysRep95}.}
it is also independent of the norm of ${\bar{\bf k}}$.
This results from the conformal (or scale) invariance of the theory. 
This independence of the traveled distance is therefore 
complementary to the well known fact that 
the power spectrum is (nearly) scale invariant. 
When considering the field at the origin of primordial density fluctuations,
see e.g. \cite{Albrecht}, 
this result implies that all wave packets at the origin of the large scale
structures are born near the reheating time, see Figure 3.
The same remark also applies to primordial gravity waves. 
%In that case, as for the massless test field
%we are considering, 
%the two waves travel on the would be particle horizon,
%i.e. the particle horizon if there were no inflation before $\eta_r$,
%see Figure 3.  

It should also be remarked that Eq. \Eqref{maxL2} is the version
of Eq. \Eqref{distc} wherein one has fixed the direction 
specified by the mean momentum $\bar{\bf k}$. The role of the 
wave packet transform is therefore to isolate from the mean 
the contribution specified by $\bar \phi_R$, i.e. by the set 
of Fourier components $v_{\bf k}$. 

%have been defined in 
%Eqs. \Eqref{maxR} and \Eqref{maxL}. 

%such filters the two-point function with a broad wave-packet 
%(in position space) whose partner will be very peaked around the semi-classical
%position. 
%$\eta_0 = -(\eta_{LSS} - 4\eta_r)$.  
%if the spread of the wave $\bar \phi_R$ is narrow in 
%$\bf k$ at recombination, the spread of its partner at that time 
%is narrow in $\bf x$. This adjustment of the spread of the R-wave 
%can be achieved by fine-tunning the phase shift 
%$\widetilde \eta_0$. 
%while the functions $\widetilde{\bf x}_R$ and $\widetilde{\bf x}_L$ are
%respectively the symmetric points with respect to ${\bf x}_0$.
%The interpretation of the four wave packets is now clear.
We now proceed with the description of the 
%To understand the 
space-time properties of four waves
in $\Theta({\bf x}, \eta;\, {\cal V}_R)$.
To this end we first 
%In Figure 1 we have 
represent in Figure 1 the function $\Theta$ 
for a one dimensional wave packet.
By construction the two wave packets governed by $x_R(\pm \bar {\bf k})$ 
merge at $\eta= \eta_0$,
see Eq. \Eqref{Gbar} and Eq. \Eqref{36a}.
%We have chosen the phase of $v_{\bf{k}}$ such that the positions of the two
%R wave packets merge at $\eta= \eta_0$
Thus, at that time we have a three-folded picture: 
a central wave with weight $2$ 
(since both right and left movers equally contribute) surrounded by the two
partner waves of weight $1$ which are respectively a $L$ and a $R$ moving
waves. 
These are shifted to the left and to the right with respect to 
the central wave in the direction specified by $\bar {\bf k}$. Their 
maxima are located along the classical trajectories Eq. \Eqref{maxL}.
 One clearly sees the factor 2 in amplitude and the phase opposition
 of the partner waves. One also verifies that the integral $\int d{\bf x}
 \Theta(\eta,{\bf x})$ identically vanishes for all wave packets.

We thus see that the wave packet transform with respect to 
a $R$-moving wave has isolated {\it two} pairs of wave packets. 
The doubling arises from the fact that the Fourier transform in 
Eq. \Eqref{Theta1} is insensitive to the velocity of the wave.
Hence the wave packets with mean momentum $\pm \bar {\bf k}$
equally contribute to $\Theta(\eta,{\bf x})$. 
This  two-folded degeneracy can be lifted if one works with a 
product which is sensitive to the velocity of the waves. 
To this end we shall use the Klein-Gordon product
in the next subsection.

%and its symmetrical. 
%This is clearly seen on Figure 2.
%according to the classical 
%trajectories Eqs. \Eqref{maxR} and  \Eqref{maxL}. 
%The two other waves are the partners of these two contributions. 
%They are therefore shifted to the left and to the right of the central wave 
%in the direction specified by $\bar {\bf k}$ according to the classical 
%trajectories Eqs. \Eqref{maxR} and  \Eqref{maxL}. 
%It is of interest to consider the
%more general case where $\eta_0 \neq \eta$. In this case,
%the central wave splits into its two R and L contributions,
%as clearly seen in Figure . 

\begin{figure}[h]
\vbox{
\hbox to\linewidth{\hss
    \resizebox{7.5cm}{6.5cm}{\includegraphics{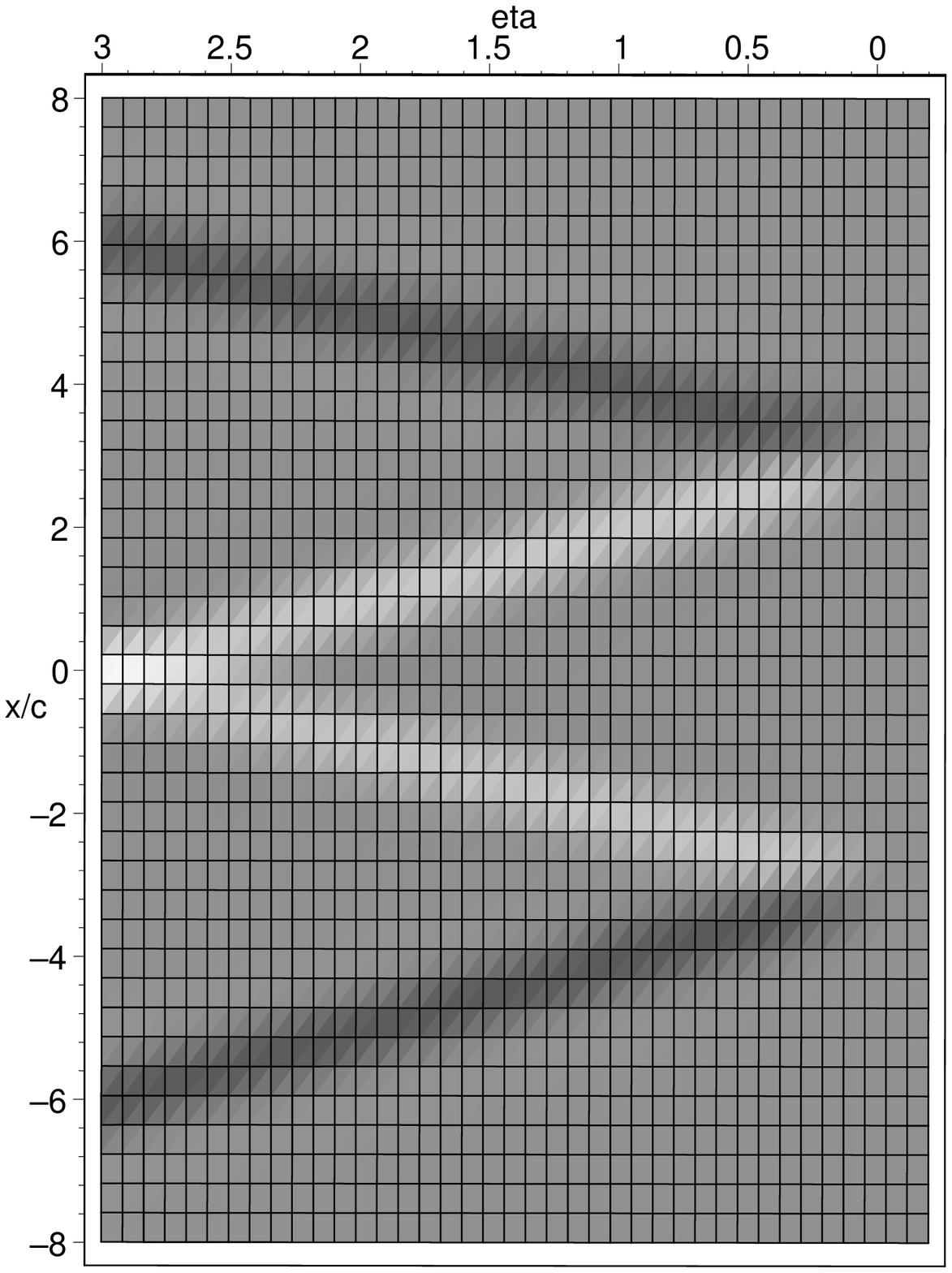}}
\hspace{5mm}
        \resizebox{7.5cm}{6.5cm}{\includegraphics{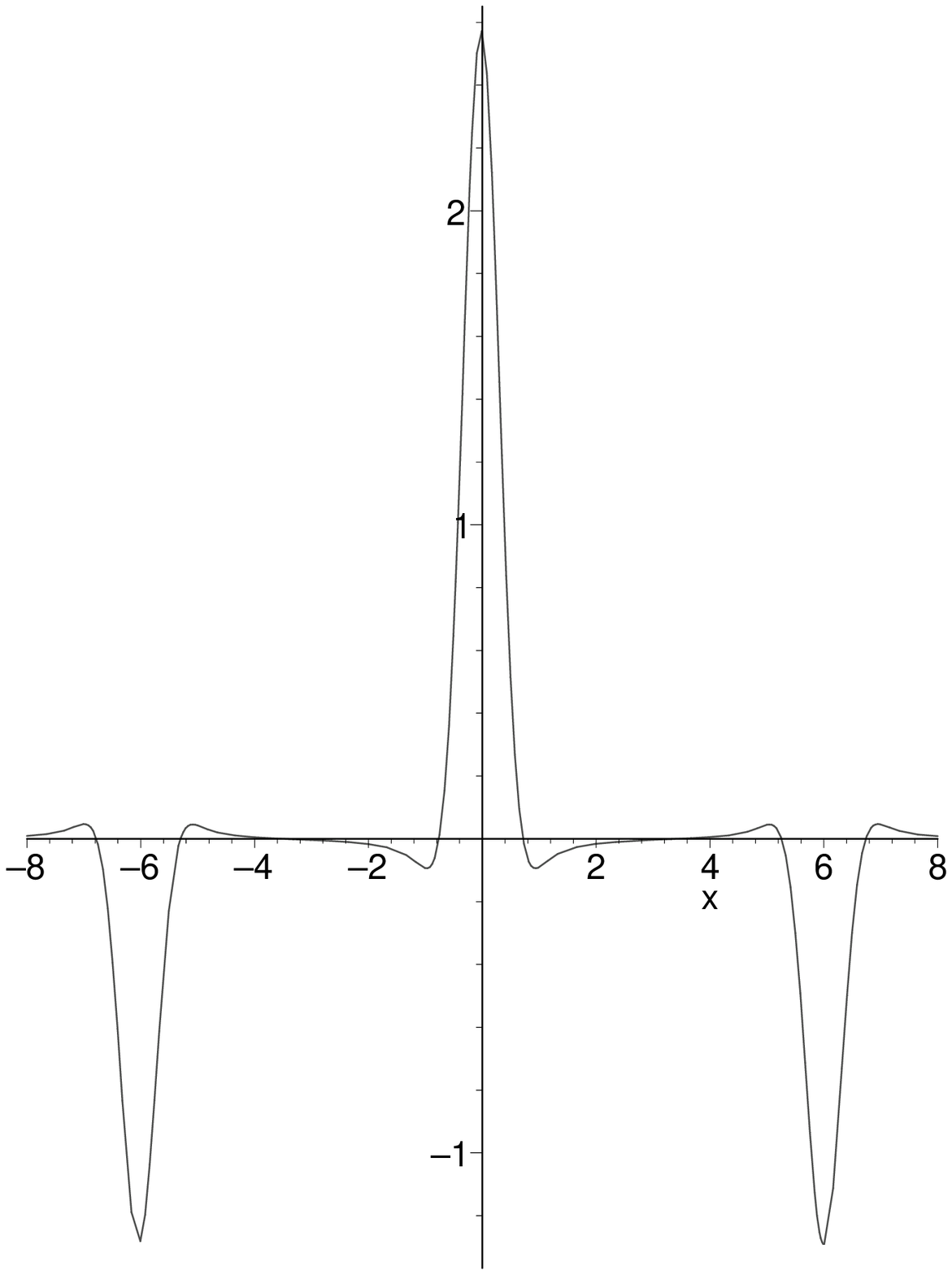}}}
\hss}
\caption{On the left, we have plotted $\Theta(\eta,{\bf x})$ of 
Eq. \Eqref{Theta} in a 1 dimensional case.
%when the wave-packet is transform of the two-point function with
%respect to one of its arguments. 
We have rescalled the spatial direction
by $c_s$ the speed of the wave.
The wave-packets have a mean momentum 
%%%R$\bar {\bf k} = \pm \pi/3$
$c_s \bar {k} =  \pi/3$, a spread
$\sigma=3$ and a phase $\phi=0$. The chosen R-wave is centered on
$x=0$ at $\eta_0=3$. Reheating is at $\eta_r=-10^{-2}$. 
(Hence one has $c_s \bar {k}\eta_0 =  \pi$, i.e. $ \bar {k}$ corresponds
to the first peak in the CMB if $\eta_0$ corresponds to the recombination.)
%The wave-packet $\bar \phi_R$ has a mean momentum $c_s\bar k=\pi/3$, 
%a spread $\sigma =2$ and a phase $\phi=0$. It is centered at $x_0 = 0, \,
%\eta_0=3$. 
The conformal time ranges from reheating to 
 %recombination at 
 $\eta = \eta_0$. 
 %The slight oscillations at the extremas are caused by the
 %pixellisation. 
 On the right, we present a spatial section at $\eta=\eta_0$.}
\end{figure}

%\begin{figure}[ht] \label{wptranform}
% \epsfxsize=8.0truecm
% \epsfysize=6.0truecm
% \centerline{{\epsfbox{wptransform.eps}}}
% \caption{The result of a wave-packet transform of the two-point function with
% respect to one of its arguments. The characteristics of the wave-packet are the
% same as in Fig. 1. The conformal time ranges from reheating to combination 
% at $\eta_{LSS} = 4$.}
%\end{figure}

%The general case where $\eta \neq \eta_{eq}$ has been represented on Figure XXX. 
%The Fig. XXX shows the result of such a wave-packet transformation. 
%As in Fig XXX, the two partners are perfectely anti-correlated to the central 
%temperature fluctuation. 
%wave. The relative heigh of these symmetric fluctuations with respect to
%the central one is $1/2$. This is because
%this three-fold structure is the superposition of two identical
%patterns, symmetric with respect to the the center ${\bf x}_R(\eta_{LSS})$ of
%the wave-packet $\Phi_{\lambda,R}^{out}$. The two patterns correspond to
%wave-packets of a same mean momentum $\bar k_{\lambda}$ but with opposite
%velocities.

\subsection{The Klein-Gordon product}

%Mathematically, this can be achieved by 
When using the 
Klein-Gordon scalar product in the place of the simple 
product of Eq. \Eqref{Theta},
% ``current''
%between the classical R-wave of Eq. \Eqref{clRwave} and the two-point function.
%In this case, 
%in the large occupation number limit, 
one gets
\ba
\label{ThetaKG}
\Theta_{KG}({\bf x}, \eta;\, {\cal V}_R)
&=& \int\!\!{d^3x'} \, G_{in}(\eta, {\bf x},\eta_0, {\bf x}') \,
 i\!\!\dpartial_{\eta_0} (\bar \phi_{R}^{(+)}(\eta_0, {\bf x}') - c.c. ) =            
 \nonumber\\
&=& 
%\quad \quad
\int\!\!\widetilde{d^3k} \, \, n_k \, 
v_{\bf k} {\frac{e^{i{\bf k \bf x}}}{(2\pi)^{3/2}}}\, 
 \left( 
 \phi_{k}^{out} +
%D25/11/03
%+  c.c. +
 z_k^* 
 \phi_{k}^{out\, *} + c.c. \right) 
  \nonumber\\
&=& 
2Re \left( i \partial_\eta \Psi_{{\cal V}_R}^{(+)} ({\bf x}- {\bf x}_0, 
\eta- \eta_0) 
\right.
\nonumber\\
&&\quad \quad \quad\left. 
- i \partial_\eta \Psi_{{\cal V}_R}^{(+)} ({\bf x}- {\bf x}_0, -(\eta
+ \eta_0 - 4\eta_r) \right)
 \, ,
\ea
where the positive frequency wave
$\Psi^{(+)}_{{\cal V}_R}$ is defined in Eq. \Eqref{psiVR}.
On the first line,
we have used the difference between 
the positive and negative
%substract the negative 
frequency components 
%took the imaginary part of the positive frequency complex
of $\bar \phi_{R}$ 
in order to cancel the minus sign which appears in the Klein Gordon
product. 
%for negative frequencies.
%cancels the one appearing after taking the time derivative. 
On the second line we thus get the real part of the integrand. 
It differs from that of Eqs. (\ref{Gbar},\ref{psiVR}) in two respects.
First there is an extra factor of $k$ which arises from the 
derivative with respect to $\eta$. Second and more importantly,
 the KG product of wave with opposite frequencies 
(i.e. velocities for a given $\bf k$) vanishes. Hence 
this leads to a reduction of the four contributions 
of Eq. \Eqref{Gbar} to two waves forming a single pair. 
%the four contributions of Eq. \Eqref{Gbar}.

When using the KG product between the two-point function $G_{in}$ 
and a $R$-moving wave, one thus correctly isolates {\it the} 
pair whose $R$-moving mode corresponds to the chosen
wave. Together with Eq. \Eqref{Gbar}, 
this is the main result of this Section. 

The extraction of the contribution of a single pair  
is clearly displayed in Figure 2.
%wherein $\Tetha_{KG}$ is represented. 
%It is helpful to visualize the situation. In  Figure 1
%To this end 
%we present the time evolution of the amplitude
%$\langle \hat \phi \rangle_{\cal{V_R}}$ of a one-dimensional
%wave packet. 
%We used Eq. \Eqref{wp}
%and got an analytic result using Eq. $7.1.12$ in \cite{Grad}. 
%Notice that propagated backward in time, the two partner waves form a wave
%packet of in R-modes only damped by a factor 
%$\alpha_{\bar k} \simeq \bar k^2 \eta_r^2 \ll 1$.
%in $R$-modes given at Eq. (\ref{modes}a). 
%At late times, the amplitude is given by
%Eqs. \Eqref{barphiR} and \Eqref{barphiL}. 
%The result  is
%\ba
%  \bar \phi_R = 
%\ea
%Because there is no dispersion, 
%this picture represents a (conformal) familly of wave-packets, 
%Indeed, the wave function is unchanged under the rescalling 
%$\bar k \to \lambda \bar k, \, x \to \lambda^{-1} x, \,  
%\eta \to \lambda^{-1} \eta, \, \sigma \to \lambda sigma$.
%%a rescalling in
%%$\bar k$ is compensated by rescalling the spread in momentum $\sigma$ 
%%by the inverse rescalling of $\eta$, $\eta_r$, $x$.
%The equivalent classes are characterized by two parameters. One is 
%$\bar k / \sigma$, that is the number of nodes of the wave packet. The other is
%the spread. The latter governs the size $1/{\sigma}^2$ 
%of the space-time region where the two wave packets overlap.
The two wave packets originate near the reheating 
from a small patch which is centered at $({\bf x}_R + {\bf x}_L )/2$. 
In fact the two waves travel on the would be particle horizon,
i.e. the particle horizon if there were no inflation before $\eta_r$. 
This can be seen from the fact that the solution of  ${\bf x}_R -
{\bf x}_L  = 0$, i.e. the tip of the lightcone,
 gives $\eta= -2 \eta_r$ which would correspond to $a =0$. 
The spatial extention of the patch at the reheating is governed
by the spread of the waves: $1/\sigma$. 
%In that case, as for the massless test field
%we are considering, 
%the two waves travel on the would be particle horizon,
%i.e. the particle horizon if there were no inflation before $\eta_r$,
%see Figure 3.  
%They thus propagate along the particle horizon  
From these results, 
one finds that in a radiation dominated era, the partner of a gravitational 
wave is always
outside the Hubble radius centered on the other wave. This means that  
the coherence will never be detectable by any measurement performed within a 
Hubble radius. (In this one gets a situation similar to that 
of Hawking radiation since the partners of Hawking quanta are all 
inside the horizon\cite{PhysRep95}).
%, the coherence will never be seen
However, thanks to the imprint they leave on the LSS, both members can now
be seen by us (when they are properly aligned, see Figure 3).

As previously discussed, the two waves are in phase opposition. 
Moreover, because of scale
invariance, the above properties are valid for 
%D%ALL 
all wave packets.

\begin{figure}[h]
\includegraphics[width=8cm,height=7.5cm]{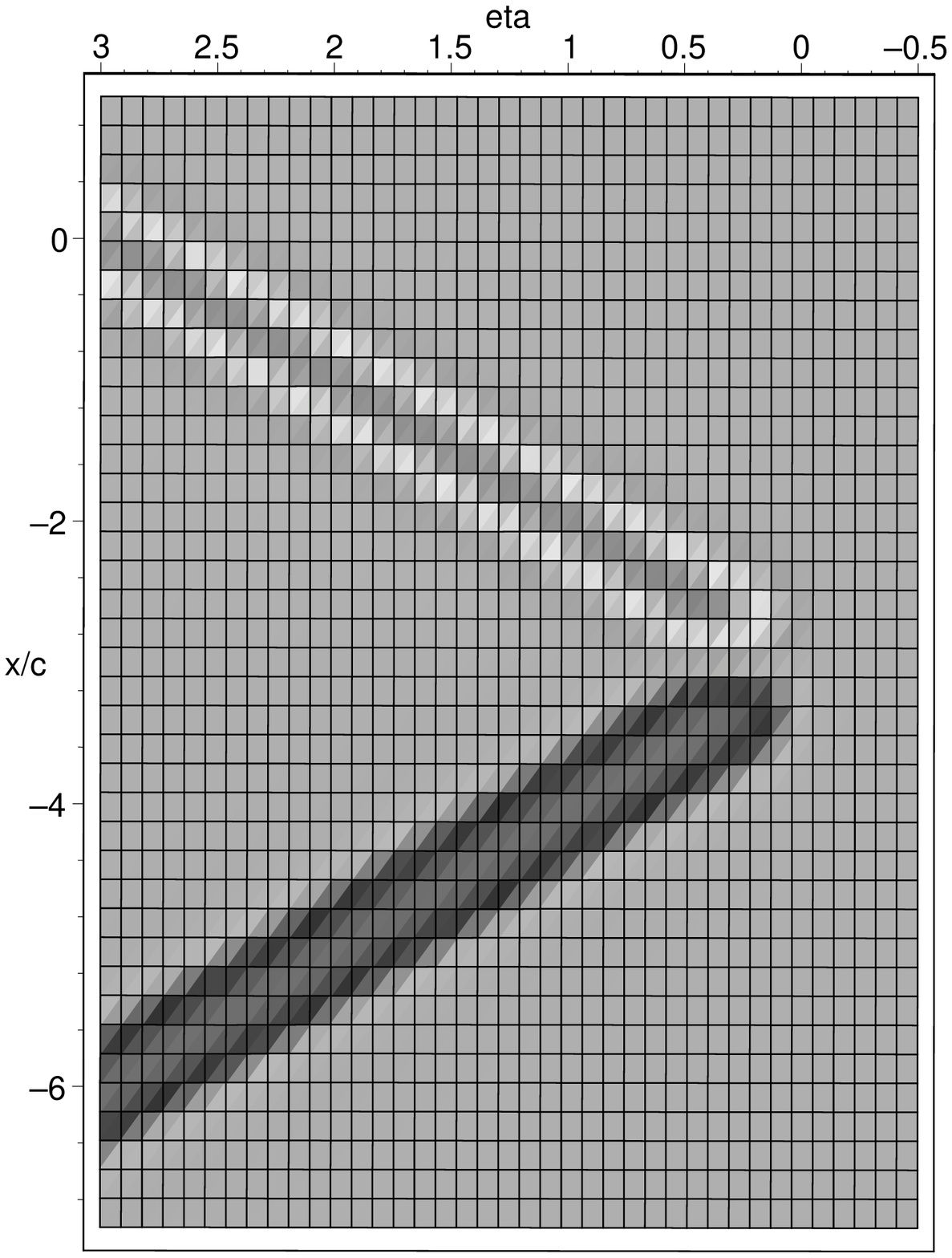}\hfill
\parbox[b]{6cm}{
\includegraphics[width=6cm,height=3.5cm]{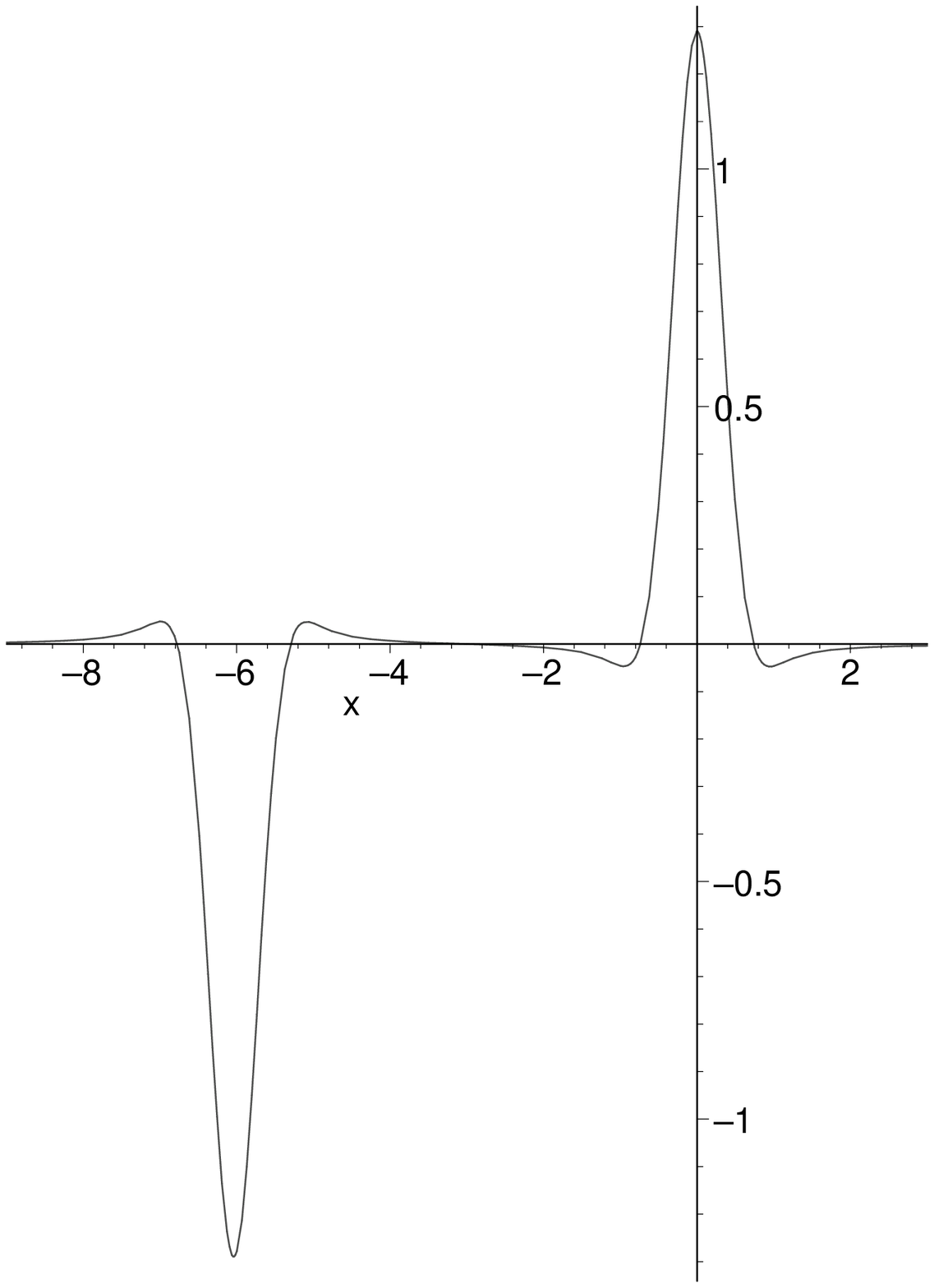}\\[0.5cm]
\includegraphics[width=6cm,height=3.5cm]{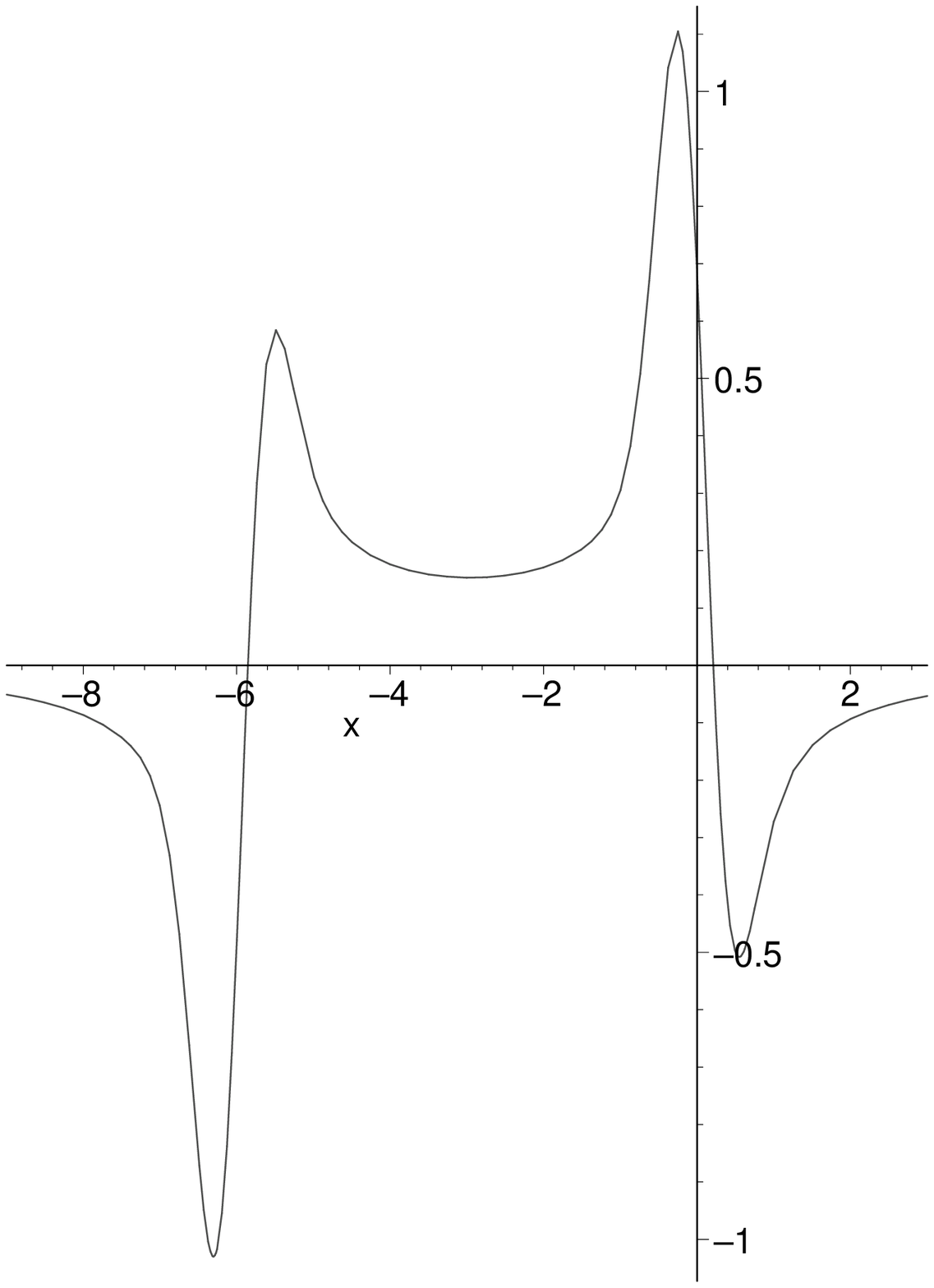}}
\caption{On the left, we have plotted $\Theta_{KG}$
%the amplitude of 
%$\langle \hat \phi \rangle$ of Eq. \Eqref{condvalues}
in conformal coordinates. 
%The darker squares (with respect to the gray background) 
%represent positive amplitude and the brighter one correspond to negative a 
%amplide. 
%We have chosen the phases so that the wave-packet is centered at 
%$x_0=0$ at $\eta=\eta_r$.
The conventions are identical to those of Figure 1.
On the right we have presented sections at $\eta=\eta_0=3$ of 
this amplitude for two different choices of the 
phase $\phi$: $\phi=0, \, \pi/3$. 
In each case, the partner wave is in phase opposition in the sense that
Eq. \Eqref{phaseop}.
%s$\bar \phi_{L}(x - x_L ; \phi) = \bar \phi_{R}(x - x_R ; \phi + \pi)$. 
The two waves are separated by 
$2(\eta_0-2\eta_r) \simeq 6$. 
This filtering giving rise to the contribution
of a single pair will be recovered in Section VI 
%e recover a
%situation similar to that of Eq. \Eqref{condvalues} which was obtained
by performing a projection in configuration space
rather than a product in space-time as in Eq. \Eqref{ThetaKG}. 
}
\end{figure}

\subsection{Spreads}

It is of great interest to analyze the spreads
of the various contributions in Eq. \Eqref{36a} or Eq. \Eqref{ThetaKG}.
Indeed their properties can be exploited {\it to reveal}
the presence of the correlations between $R$ and $L$ sectors. 
%In particular, we shall see that the coherence between these sectors
%can lead to 

The result of the saddle-point evaluation of
Eq. \Eqref{psiVR} is given by, see \cite{CP1}:
\ba \label{phiVR}
 \Phi_{{\cal V}_R}({\bf x}) \sim Re\left[
 \frac{1}{\sqrt{{\rm {det} \Sigma\vert_{\bar{\bf k}}}}} 
 e^{\chi({\bf \bar k})} 
 \exp\left(- \frac{1}{2}{\bf x}_i (\Sigma^{-1})_{ij} {\bf x}_j \right) \right]
 \, ,
\ea
where the function $\chi({\bf k})$ regroups the terms in the 
exponential of the integrand in
Eq. \Eqref{psiVR}. The matrix $\Sigma$ is defined by 
$2\Sigma_{ij} = -\partial_i\partial_j \chi \vert_{{\bf \bar k}}$
where $i,j= 1,2,3$ label the three conformal coordinates. 
The three eigenvalues of $\Sigma_{ij}$ give the spreads in 
position of the Gaussian
wave-packet Eq. \Eqref{phiVR}.
In the case of a single wave-packet with axial symmetry around $\bf \bar k$, 
%such as Eq. \Eqref{}
the three eigenvalues reduce to two scalars. The first 
one is the spread in direction
longitudinal to $\bf \bar k$. It is independent of time and equal to $1/|\sigma|$,
as in one dimension. The other governs the spread in the directions 
orthogonal to $\bf \bar k$. It  grows
linearly with time, as in the case of a non-relativistic wave-packet.
Its value therefore differs for the $R$ and $L$ wave:
%Straightforward algebra gives 
%More precisely, for an axisymmetrical wave-packet with a maximum at 
%$(\eta_0, {\bf x}_0)$, one can write
%\ba\chi({\bf k}) = i {\bf k} ({\bf x} - {\bf x}_0) - i k (\eta - \eta_0) + 
%\psi({\bf k}) \, ,
 %v_{\bf k} \propto e^{-\frac{(\bf k-\bf \bar k)^2}{4\sigma^2}}
 %e^{i\bf k (\bf x - {\bf x}_0)}e^{-i k (\eta - \eta_0)} \, , \ea
%where by definition the first derivative of the function $\psi$  vanishes at 
%$\bar {\bf k}$. This implies that near $\bar {\bf k}$, the function behaves as
%\ba \psi({\bf k}) = \psi(\bar {\bf k}) -  
%\frac{({\bf k} - \bar {\bf k})^2}{4\widetilde \sigma^2} \, , \ea
%where $\widetilde \sigma$ is a complex number 
%\footnote{Then, the saddle-point definition 
%${\bf \nabla} \chi \vert_{{\bf k}^*} = 0$ gives the semi-classical trajectory
%\ba \partial_j \chi\vert_{\bar {\bf k}} = i x_j(\eta; \, \bar {\bf k}) \, ,
%\ea
%and one gets the equations \Eqref{maxR} and \Eqref{maxL}.}.
%The spreads in the directions orthogonal to $\bar {\bf k}$ 
%are respectively for the R and L-waves :
\sba \label{spread}
 2 \Sigma_{\bot, \, R}^{2}(\eta) &=& %\frac{1}{2\sigma^2} + 
 %-\partial_{22}^2 \psi \vert_{\bar {\bf k}} + 
 \frac{1}{2\sigma^2} +
 i \frac{\eta - \eta_0}{\bar k} \, , \\
 2 \Sigma_{\bot, \, L}^{2}(\eta) &=& %\frac{1}{2\sigma^2} + 
 %-\partial_{22}^2 \psi \vert_{\bar {\bf k}} +
 \frac{1}{2 \sigma^2} +
 i \frac{\eta + \eta_0 - 4\eta_r}{\bar k} \, .
\sea
When $\sigma$ is taken real, the spread of the R-wave is minimal
%%%R
at $\eta= \eta_0$, as expected.
(Remember that the spread in the perpendicular directions
is given by the square root of the norm of $\Sigma$). 
%whereas
% when $\sigma$ is taken real.
The spread of the partner L-wave is thus larger 
as it is governed by the accumulated conformal time 
from the detection of the R wave at $\eta_0$,
%till $\eta_0$ 
{\it back} to 
%and forth to 
$\eta_r$ where the pair emerges,
and then forward to  $\eta_0$, for more details see \cite{CP1}. 
% D21/11/03
%for more details see \cite{CP1}.
The relative increase of L-wave spread becomes large 
for short wave lengths modes.
% as these have reentered the horizon at an early time. 
As an example, for a wave-packet of
mean momentum 
$\bar k(\eta_0-2\eta_r)=10$ 
(this would correspond to the third peak in the 
CMB anisotropies spectrum)
%that is corresponding to a multipole $l\sim ?$,
and spread $\sigma=\bar k$, one has  
%D25/11/03
$\vert \Sigma_L(\eta_0) / \Sigma_R(\eta_0) \vert \simeq 6$.

It is now crucial to notice that one can fine tune 
the imaginary part of $\sigma$
to obtain a partner wave which is {\it more peaked} 
than the wave which is chosen in Eq. (\ref{Theta}). 
Indeed, one can exploit the 
coherence of the wave packet {\it and} that of the vacuum, i.e.
the ${\bf k}$ dependence of 
%the EPR correlations (which are diagonal in ${\bf k}$),i.e. 
Eqs. (\ref{reducedstate}, \ref{z}), 
to obtain constructive interferences around ${\bf x}_L$ 
by {\it canceling} the accumulated effect due to the total time of flight.
The maximal effect, i.e. the minimal norm of $\Sigma_L(\eta_0)$,
% contend of the second
%derivative $\partial^2_{ij}\psi \vert_{\bar {\bf k}}$, one
%fine-tunning the parameter $\eta_0$, 
is reached when taking $Im(\sigma^{-2}) = 
- 4(\eta_0 - 2 \eta_r)/\bar k$.
When considering the above example with $\bar k(\eta_0-2\eta_r)=10$,
one gets
%the ratio of the spreads is now
%D25/11/03
\ba
\abs{{\Sigma_L(\eta_0)}/{\Sigma_R(\eta_0)}} \simeq \frac{1}{6} \, ,
\label{neq}
\ea
that is, the perpendicular 
spread in space of the partner wave is about six times smaller
than that of the wave which is used in the integration of Eq. (\ref{ThetaKG}). 
This reduction of the partner wave spread is a direct
consequence of the two-mode $R-L$ coherence. 
Hence it can be used as a check to quantify the degree of coherence 
given some observational data. 
It should be also noticed that it is not necessary to identify the velocity
in order to bring this effect into evidence. 
Indeed this reduction 
%of the spread of the partner wave 
equally applies to the spreads of the last two 
waves in Eq. \Eqref{Gbar}. (In that case, the above symbols 
$R$ and $L$ should be interpreted in their generalized sense
given after Eq. \Eqref{36a}.)

%such filters the two-point function with a broad wave-packet 
%(in position space) whose partner will be very peaked around the semi-classical
%position. 
%$\eta_0 = -(\eta_{LSS} - 4\eta_r)$.  
%if the spread of the wave $\bar \phi_R$ is narrow in 
%$\bf k$ at recombination, the spread of its partner at that time 
%is narrow in $\bf x$. This adjustment of the spread of the R-wave 
%can be achieved by fine-tunning the phase shift 
%$\widetilde \eta_0$. 
%while the functions $\widetilde{\bf x}_R$ and $\widetilde{\bf x}_L$ are
%respectively the symmetric points with respect to ${\bf x}_0$.
%The interpretation of the four wave packets is now clear.

In the next Sections, we present a more sophisticated 
procedure to extract space-time correlations.
It is based on the introduction of a filter in configuration space 
before computing expectation values.
%and not in Fourier space
Being not based on mean values,
it allows to generalyse the former analysis. In particular it reveals 
correlations in the {\it amplitude} of wave packets.  
%and to interprete the results in terms of conditional probabilities.
%It is based on the introduction of a filter in 
%configuration space which isolates a set of final configurations.
The reader not interested by these developments can proceed
directly to the Conclusions where he will find a resume of the 
results.

%%%%%%%%%%%%%%%%%%%%%%%%%%%%%%%%%%%%%%%%%%%%%%%%%%%%%%%%%%%%%%%%%%%%%%%%%%%%%%%
%%%%%%%%%%%%%%%%%%%%%%%%%%%%%%%%%%%%%%%%%%%%%%%%%%%%%%%%%%%%%%%%%%%%%%%%%%%%%%%
%%%%%%%%%%%%%%%%%%%%%%%%%%%%%%%%%%%%%%%%%%%%%%%%%%%%%%%%%%%%%%%%%%%%%%%%%%%%%%%
%%%%%%%%%%%%%%%%%%%%%%%%%%%%%%%%%%%%%%%%%%%%%%%%%%%%%%%%%%%%%%%%%%%%%%%%%%%%%%%

%%%%%%%%%%%%%%%%%%%%%%%%%%%%%%%%%%%%%%%%%%%%%%%%%%%%%%%%%%%%%%%%%%%%%%%%%%%%%

\section{Coherent states, classical waves and random processes} 
\label{sec:cohstates}

We introduce coherent states of the field $\phi$.
They are the key elements for our analysis of 
the correlations 
%R%
associated with a realization of a particular set
of classical configurations. In quantum terms such a realization 
can be described by the detection of the corresponding states,
namely coherent states.  
%Then the proper way to implement the detection relies on
%making use of 
%coherent states, see Appendix B for a brief review of their properties.
%(see \cite{Glauber63a} for 
%the use of coherent states for
%calculation of Green functions in quantum optics).
%Indeed, sSince 
%This is no surprise since the occupation number $n_k$ is macroscopic. 
%We already saw in Eq. \Eqref{infl2ptf2}
%that the dominant part of $G_{in}$
%can be written in terms of two classical waves, 
%a result valid when dealing with coherent states.
%In the following we shall see that the same is true for expectation values 
%in coherent states.
Moreover, since we want to select localized waves, we need to form 
wave packets by summing  different $\bold k$-modes.
Therefore, we shall work with coherent states of these wave 
packets.\footnote{The use of coherent 
states can be conceived from several point of views: 
either as a 
%well defined 
mathematical way to introduce 
%the concept of
%local wave packets which behaves like 
classical waves in quantum terms, or more physically, as resulting from 
a detection of such waves, or even more intrinsically, 
through decoherence induced by interactions.
Indeed oscillators weakly coupled to an environment 
%(i.e. to other oscillators) 
evolve into coherent states \cite{Zurek93,Hu97}.
% are preferably selected as a pointer basis, which
%reenforces one's conviction that they are the closest counterparts 
%of classical points in phase-space \cite{Zurek93}.
Therefore it is to be expected that in cosmology the weak non-linearities which 
are generally ignored will replace the pure (two-mode squeezed) state
by a mixture of (two-mode) coherent states \cite{cpPeyresq}.
%, i.e. will
%R%provide the required decoherence to legitimize the classical description. 
However, to our knowledge, the properties of this decoherence process
%\cite{Hu97} 
(when it occurs, how it modifies the state of the $\bold k$-modes, 
and how to put it into evidence)  have not been
fully derived.
%%%R
%As we shall later see, 
Finally, the use of
% approach based on 
coherent states provides an interesting alternative 
to \cite{GuthPi85,Matacz94,PolStar96,KieferPol98}
when analysing
%understand
the emergence of classical and stochastic properties in 
inflation.
Indeed the detection of coherent states 
%of classical configurations described by coherent states 
is a Gaussian random process. 
%This provides 
%an alternative way to understand how classical properties emerge
%from the amplification of vacuum fluctuations. 
Notice also that coherent states have been already used
in \cite{Albrecht94,Albrecht} to study the semi-classical limit.}
%Notice that the usual approach also rests 
%on the assumption that some decoherence occurs 
%D25/11/03
%\cite{GuthPi85,Matacz94,PolStar96,KieferPol98}.}
%%%%R which allows 
%thereby allowing to use
%abandoning the quantum description and 
%a classical distribution in the place of the 
%replacing 
%R%that this effectively replaces quantum  one.}
%R%Moreover, we shall see that the use of coherent states in the 
%present context naturaly leads
%to the notion of stochastic processes governed by classical statistics.
%Therefore this approach based on coherent states 
%provides an interesting alternative to understand
%the emergence of classical and stochastic properties in inflation.
%(The usual
%ly adopted procedure 
%%%%%%%%%%%%%%%%%%%%%%%%%%%%%%%%%%%
%%%%%Irrespectively of the point of view one adopts, the following propositions
%%are valid.

The filtering of a set of classical waves is implemented by
introducing a projector on the corresponding coherent state.
The introduction of this projector modifies the correlations which 
existed in the ``in vacuum'', i.e. the state of the system 
which encodes the pair creation process. 
%We shall see that 
New correlations are introduced 
at the expense of reducing the pre-existing ones.
In particular 
%%%R
%specific space-time 
{\it spatial} correlations are generated 
%through the coherence introduced by 
%%%R
by detecting a local wave.
%classicity emerges from investigating the out configuration content
%of the in vacuum.

\subsection{EPR correlations and conditional values} \label{sec:EPR,two}

To describe the correlations which exist in the 
in vacuum, we analyze 
%is \`a la Schr\"{o}dinger, through 
correlations amongst out states. To justify this analysis 
in the present context, 
let's consider the following gedanken experiment.
Suppose that $n$ $R$-out-particles of momentum $\bold k$ have been detected
and that nothing is known for the other modes except that, before this 
detection, the field was in the in vacuum. Given that detection,
one can ask what are the probabilities to 
find $m$ particles of momentum $\bold k'$. 
The answers to this type of questions
are governed by the partially `reduced' state obtained by projecting 
the in vacuum onto the state characterizing the (partial) detection, 
namely $\ket{n, \, \bold k, R, out}$:
\ba \label{reducedstate}
\vert {\rm red}_{n, \, \bold k, R} \rangle =   
 \langle n, \, \bold k, R, out \vert 0,  in \rangle_2 = 
 \frac{ ( z_k)^{n}}{\vert \alpha_{k} \vert} \,
 \ket{n, \, \bold k, L, out } 
 \widetilde{\prod\limits_{\bold k' \neq \bold k}}\otimes  \vert 0, \, \bold k', in 
 \rangle_2 \, .
\ea
In the $\bold k$-th two-mode sector, one finds the one-mode $L$-state entangled 
to $\ket{n, \, \bold k, R, out}$. 
It is a pure state with the same occupation number. 
This results from the EPR-type correlations present in the in vacuum, see 
Eq. \Eqref{inoutvact}. 
(These correlations deserve the label `EPR' since they are of the
same character as those encountered with spins: 
the spin projection on a axis is here replaced by the occupation number.
Notice also that in both cases, a symmetry is at the origin of the entanglement,
rotation invariance there, translation invariance here.) 
The two-mode sectors with $\bold k' \neq \bold k$ are all unaffected by 
the detection.
Therefore the probabilities to find particles of momentum 
$\bold k' \neq \bold k$ are unchanged. 

More generally, according to the rules of quantum mechanics, 
the ``conditional'' expectation values which result from the above 
detection 
%of the above out-particles 
can be written as
\ba
\label{condv}
\langle \hat {\cal O}  \rangle_{cond}  =  
 \frac{ \langle \hat \Pi \, \hat {\cal O} \,\hat \Pi \rangle_{\rm in}}{ 
 \langle \hat \Pi \rangle_{\rm in}}\, , 
\ea
where the projector onto the detected state $\ket{n, \, \bold k, R, out}$
is
\ba \label{proj1}
 \hat \Pi_{{n, \, \bold k, R, out}} = \ket{n, \, \bold k, R, out} 
 \langle n, \, \bold k, R, out \vert 
 \otimes {\bold 1}_{\bold k, L}
 \otimes  {\bold 1}_{\bold k' \neq \bold k}\, .
\ea

In Eq. \Eqref{condv}
we have used $\hat \Pi^2 = \hat \Pi$ to simplify 
the denominator.\footnote{A technical 
comment is in order for the readers familiar with the work of 
Aharonov {\it et al.} \cite{Aharonov64&90} 
%and collaborators 
or with its applications to pair
creation \cite{PhysRep95,MaPa98,CP1}.
In these works, 
%a different expression was used: instead of Eq. \Eqref{condv}
a different conditional value of $\hat {\cal O} $, called 
weak value, was used. It is given by 
$ \langle \hat \Pi \hat {\cal O} \rangle/\langle \hat \Pi \rangle$
in the place of Eq. \Eqref{condv}. 
Mathematically, the difference between the two expressions
is that weak values are generically complex whereas 
Eq. \Eqref{condv}
is real for hermitian operators. On the other hand, when 
$\hat{\cal O}$ and $\hat \Pi$ commute
they coincide since $\hat \Pi^2 = \hat \Pi$. 
To understand the physical relevance of these two versions is a 
subtle question:
which version should one use in a given context when some 
information concerning the final
outcome is known ? The reader interested by this type of question will consult
the original references. In inflationary cosmology, because of the high
occupation number, a simplification occurs: the differences
between the two versions
%R%versions of conditional values 
%associated with the detection of classical waves
%to a ``strong'' detection which implies the use of Eq. \Eqref{condv}. 
%Notice also that 
%in a cosmological context, thanks to the presence of 
%leads to large occupation numbers 
%guarantee that the differences
%betwen the two conditional values 
are subdominant (i.e. $O(1/n)$) since 
they arise from commutators.} 
The above projector is ``partial'' in that it
is unity in all sectors but the $R$-mode $\bold k$. 

The use of projectors which act only on the $R$ sector
is particularly interesting because it guarantees that 
the $L$-mode content will be {\it only} 
determined by the correlations which are present
in the state of the field. It is therefore the appropriate tool
to unravel the intrinsic properties of the $R-L$ correlations.
The next step is to determine how these 
correlations give rise to the spatial properties of conditional values.
This is done in Sections VI.
Notice finally that we shall no longer use 
projectors which specify the occupation number as in Eq. \Eqref{proj1}.
Instead we shall use projectors based on coherent states
because these properly characterize the various realizations 
of the ensemble, see footnote 3.

% we shall compute conditional values as in Eq. \Eqref{condv} 
%with some physically motivated projectors.
% which are more
% physically motivated than Eq. \Eqref{proj1}. 
%On the one hand, in order to describe classical waves, 
%we shall use coherent states
%and no longer Fock states.
%On the other hand, in order to describe localized waves in space, 
%we shall consider wave packets and no longer delta functions in $\bold k$ space.

\subsection{Coherent states and classical waves}

Because of the entanglement between $R$ and $L$ modes in the in vacuum, 
only half of the modes are independent.
Hence we shall use coherent states which are defined
in the right sector only. 
We also work with out modes because the detection of waves
is performed at late time, i.e. during the radiation or the matter 
dominated epoch.
In this Section, we consider coherent states which 
specify the amplitudes of
all $R$-modes:
\ba
\ket{{\cal V}, R} = \widetilde{\prod\limits_{\bf k}} \otimes \ket{v_{\bf k}, R} \, ,
\label{cohR}
\ea
where $\ket{v_{\bf k}, R}$ is a (one-mode) coherent state of complex amplitude 
$v_{\bf k}$, see Appendix B.
% for definitions and properties of coherent states.

The product  $\ket{{\cal V}, R}$ is an eigenstate of $ \hat \phi_{R}^{(+)}$,
the right-moving positive frequency part of the field operator :
\ba \label{defbarv}
 \hat \phi_{R}^{(+)}
 %(\eta, {\bf x}) 
 \ket{{\cal V}, R} &=& 
 % \tintk  e^{i \bold k \bold x} 
 \int\!\!\widetilde{d^3k}\,  
 %\tilde \phi_{\bold k}^{out} 
 \frac{e^{i \bold k \bold x}}{(2\pi)^{3/2}} 
  \phi_{k}^{out}(\eta)
 \, 
 \hat a_{\bold k}^{out}\ket{{\cal V}, R}   \nonumber \\
 &=& 
 %\tintk e^{i \bold k \bold x} \phi_k^{out} 
 \int\!\!\widetilde{d^3k} \, 
 \frac{e^{i \bold k \bold x}}{(2\pi)^{3/2}} 
  \phi_{k}^{out}(\eta)
  %\tilde \phi_{\bold k}^{out} \, 
 v_\bold k \ket{{\cal V}, R}     \nonumber \\
 &=& \bar \phi_{R}^{(+)}  \, \ket{{\cal V}, R} \, .
\ea
where the tilde on $d^3k$ means that one integrates over $R$ modes only,
i.e. $k_x$ is  integrated from $0$ to $\infty$. 
The function $\bar \phi_{R}^{(+)}$ is complex
because it contains only positive frequencies.
To get a real function, one should consider the expectation value 
of the observable $\hat \phi_{R}$, the field amplitude itself:
\ba \label{cohwave}
 \bar \phi_{R} = \bra{{\cal V}, R} \hat \phi_{R} \ket{{\cal V}, R} = 
 \bra{{\cal V}, R} (\hat \phi_{R}^{(+)}+ \hat \phi_{R}^{(-)})  
 \ket{{\cal V}, R} = 2 Re\left\{ \bar \phi_{R}^{(+)} \right\}
 \, .
\ea
The same can be done to the conjugate momentum of the field, 
see Eq. \Eqref{cltrajincohstate}.

We can now verify that the (dominant part of the) 
expectation values computed in the state $\ket{{\cal V}, R}$ 
can be expressed directly in terms of the mean value $ \bar \phi_{R}$,
in the same way that Eq. \Eqref{infl2ptf1}
was expressed in terms of the two sine functions. 
%R%

Using the normalisation of Eq. \Eqref{normv_k},
the amplitude $\bar v$  is related to the expectation
value of the occupation number
\ba
 \bra{{\cal V}, R}  \hat N_R \ket{{\cal V}, R} = 
  \int\!\!\widetilde{d^3k} \, \bra{{\cal V}, R}
  \hat a_{\bf k}^{R\, \dagger} \hat a^{R}_{\bf k}  \ket{{\cal V}, R} = 
  \bar v^2 \, .
\ea 
It is also related to the current of $\bar{\phi}_{R}^{(+)}$:
%D%
\ba
%\langle \bar{\phi}_{R}^{(+)} \, , \,  \bar{\phi}_{R}^{(+)} \rangle_{KG} = 
\int\!\!d^3x \,\, \bar{\phi}_{R}^{(-)} \, i \!\!\dpartialeta 
 \bar \phi_{R}^{(+)}
 = \bar v^2 \, .
\ea
Thus one has two alternative descriptions.
Either one uses the quantum description in terms of the
expectation value based on the counting operator or
the classical concept of 
%D% norm
current based on the mean wave $\bar{\phi}_{R}^{(+)}$. 
The same conclusion is valid for other 
%D% operators
quantities such as Green functions, see  Appendix B, or 
the 3-momentum.
In all cases, when $\bar v^2 \gg 1$, the classical expressions based 
of the mean field  $\bar{\phi}_{R}$ coincide with the corresponding
expectation values evaluated in the coherent state $\ket{{\cal V}, R}$. 
The reason is that
the ambiguities of operator ordering lead to differences
governed by commutators which are subdominant in the large
occupation number limit.

%%%%%%%%%%%%%%%%%%%%%%%%%%%%%%%%%%%%%%%%%%%%%%%%%%%%%%%%%%%%%%%%%%%%%%%%%%%%
%%%%%%%%%%%%%%%%%%%%%%%%%%%%%%%%%%%%%%%%%%%%%%%%%%%%%%%%%%%%%%%%%%%%%%%%%%%%

\subsection{Detections and random processes} \label{sec:ps,random}

In this subsection we show two important results. First, when the
(Heisenberg) state is the in-vacuum, the detection of the 
$R$-moving configurations $\bar \phi_R$  described 
by the coherent state $\ket{{\cal V}, R}$
is a stochastic Gaussian process. 
This result is exact: it requires no approximation 
and is valid even before applying any decoherence process.
Notice however that, because of the entanglement 
between the $R$ and $L$ sectors, the particle content of only half the
modes (e.g. the $R$-moving configurations) should be specified 
to get this result. Second, the detection of $\bar \phi_R$
fixes the $L$ modes to be also described by coherent states. This result
is also exact and follows from the Gaussianity of squeezed states
and the $R-L$ entanglement. 
These two results {\it entirely} determine the 
%modifications of the 
correlations 
%(with respect to their values in the in vacuum)
which result from the detection of semi-classical configurations
described by coherent states.

To determine the consequences of a detection,
it is appropriate to 
introduce the associated projector, see subsection
\ref{sec:EPR,two}. 
%acts only in the $R$ sector
%exhibe the space-time correlations materializing the EPR
%$correlations in the momentum space, we must perform a partial postselection 
%on the right sector only, and let Quantum Mechanics do the job for the 
%left sector.
%In a first attempt, 
%we define a normalized coherent state of the field that 
%contains only $k_x > 0$ components as in \Eqref{defbarv}, and 
%we introduce the projector
In the present case, it is 
\ba \label{ProjV}
 \hat \Pi_{{\cal V}_R} = \ket{{\cal V}, R} \bra{{\cal V}, R} \otimes 
 \bold{1}_{L} \, .
\ea 
It is non trivial in the $R$-sector only. 
The probability to detect the classical wave $\bar \phi_R$
is given by 
\ba \label{ProbaClWave}
 P^{in}_{{\cal V}_R} = \bra{0\,  in} \hat \Pi_ {{\cal V}_R} \ket{0\,  in} =  
 \tprod \abs{A^{in}_{\bold k}}^2 \, ,
%=  \left( \tprod \inv{\abs{\alpha_{k}}^2} \right)
% \exp\left(-\tintk \frac{|v_{\bold k}|^2}{|\alpha_{k}|^2}\right) \, ,
\ea
where the  amplitude for the $\bold k$-mode is
\ba
\label{ProbaClWaveA}
 A^{in}_{\bold k} &=&  
 \inv{|\alpha_{k}|} e^{-\frac{|v_{\bold k}|^2}{2|\alpha_{k}|^2}} \, ,
\ea
see Eq. \Eqref{Clproba} for the details.
The probability  Eq. \Eqref{ProbaClWave} defines 
a normalized gaussian distribution for each $R$-mode in every two-mode sectors.
The normalization follows from the ``density'' of coherent states,
see Eq. \Eqref{normP_z}.
As already mentioned, only $R$-modes have been so far specified.
Had we performed a projection on both right and left sectors,
the probability would have been exponentially smaller. 
Indeed, the ratio of the probabilities with or without double projection is, 
see Eqs. \Eqref{Impeq} and \Eqref{overlap},
\ba \label{Impeq-1}
 \frac{P^{in}_{{\cal V}_R\, , {\cal W}_L}}{P^{in}_{{\cal V}_R}} =   
 \tprod \exp\left(- \vert w_{\bf k} -
 z_k v_{\bf k}^* \vert^2 \right) \, , 
\ea 
where $w_{\bf k}$ is the amplitude of the coherent wave in the $L$-sector.

%This problem will be solved in the next section by the introduction of localized
%wave-packets. For the time being, we shall continue with this projector since it
%posess nice features that are worth analysing.
 
This exponentially suppression results 
from the entanglement between the left and right sectors.
Indeed, when applying the projector $\hat \Pi_{{\cal V}_R}$
on the in vacuum one gets
\ba
\label{Impeq}
 \hat \Pi_{{\cal V}_R} \ket{0, in} &=& 
 \tprod \, 
 A^{in}_{\bold k} \,  \ket{v_{\bold k},\, \bold k, R} \otimes 
 \ket{z_{\bold k}v_{\bold k}^*,\, \bold k, L} \, .
\ea
It is remarkable that in each two-mode sector, 
the $L$-state is also a coherent state.
This results from Eq. \Eqref{reducedstate}, see also Eq. \Eqref{vonz}. 
The $L$-mode amplitude is 
$ z_{k}v_{\bold k}^*$. It is fixed by the $R$-amplitude $v_{\bold k}$  and 
by the pair creation process which is 
governed by $ z_{k}$. The properties of the space time patterns we shall
later exhibit directly follow from this double origin.  

Taken together, Eqs. (\ref{ProbaClWave}-\ref{Impeq})
show that the notion of stochastic processes naturally
emerge when questioning the in vacuum
by making use of coherent out states.
More precisely we have the following.
Firstly, as one might have expected, the $R$-mode amplitude
$v_{\bf k}$ is a Gaussian stochastic variable of  variance 
equal to $\abs{\alpha_k}^2
= n_k + 1 \simeq n_k$. 
Secondly, the ${\bf k}$-th $L$-mode amplitude is ``slave driven''
by the detection of the $R$-mode
in that its probability is centered around  $ z_{k}v_{\bold k}^*$
with a spread equal to 1, see Eq. \Eqref{Impeq-1}. 
Therefore, in the large $n_k$ limit, 
this spread is negligible and one can consider that 
the  $L$-mode amplitude is equal to $ z_{k}v_{\bold k}^*$. 
%as it is equal to $ z_{k}v_{\bold k}^*$ and described by the same distribution.
%In Eq. \Eqref{Impeq} there is indeed one amplitude for both sectors. 
Thus, in the stochastic description as well,   $P^{in}_{{\cal V}_R}$ 
is a two-mode distribution, as clearly seen from Eq. \Eqref{Impeq}. 

These properties offer an alternative way to express
expectation values in the in vacuum. It suffices to apply
the following substitution: first, at the 
level of amplitudes $\hat a^{R}_{\bf k} \to v_{\bf k}$, 
$\hat a^{L}_{\bf k} \to z_{k}v_{\bold k}^*$,
and second at the level of the distributions, the quantum distribution 
$\hat \rho_{in}= \ket{0 in}\bra{0 in}$ should be replaced by 
$P^{in}_{{\cal V}_R}$ of Eq. \Eqref{ProbaClWave}. 
Notice that no dynamical assumption was needed, nor was 
it necessary to follow the time evolution of the modes.

The emergence of classicity rests on the high occupation number $n_k \gg 1$
and on the (restricted) set of questions formed by inquiring about the
coherent out state content of the in vacuum. 
Even though these conclusions are not new \cite{GuthPi85,Matacz94,PolStar96}, 
the derivation which makes  use of coherent states is particularly clear.
In particular, it disentangles the question of the late time description of the
in vacuum in the above stochastic terms from the more difficult question 
which concerns the evaluation of the time from which 
this stochastic description is valid. 
This time is determined by the efficiency of decoherence processes
in the early cosmology, a subject not addressed in the present paper 
\cite{Hu97,KieferPol98}.
%In doing so, we have not addressed the difficult question on when thi
%assumption  classical stochastic properties emerge. 
%how the statistical properties of the field treated as a 
%classical stochastic variable emerge. 

%%%%%%%%%%%%%%%%%%%%%%%%%%%%%%%%%%%%%%%%%%%%%%%%%%%%%%%%%%%%%%%%%%%%%%%%
%%%%%%%%%%%%%%%%%%%%%%%%%%%%%%%%%%%%%%%%%%%%%%%%%%%%%%%%%%%%%%%%%%%%%%%%

\subsection{1-point and 2-point functions} \label{sec:ps,two modes}

Besides the above substitution, one can consider the projector 
$\hat \Pi_{{\cal V}_R}$ as in subsection \ref{sec:EPR,two},
namely as defining a new ensemble of configurations with modified expectation
values given by Eq. \Eqref{condv}. It is of interest to present these
expectation values in some details. Starting with  
%The new statistical properties of the ensemble obtained after having detected
%the classical wave are enterely governed by the following 
%1-point and 2-point functions. Starting with the 
1-point functions, we have 
\ba \label{1pointcoh}
 \langle  \hat a_{\bold k,\, R}^{out} \rangle_{{\cal V}_R} = v_{\bold k} \, , 
 \quad \langle \hat a_{\bold k,\, L}^{out}  \rangle_{{\cal V}_R}
 %\, \hat \Pi_v    \rangle_{\rm{in}} 
 = z_{k} v_{\bold k}^{*} \, .
\ea
In the in vacuum we had $ \langle  \hat a_{\bold k,\, R}^{out} \rangle_{\rm in}
= \langle  \hat a_{\bold k,\, L}^{out} \rangle_{\rm in}=0$. The interpretation 
of the modification is clear: once we know that the classical wave 
$\bar \phi_R$
has been detected,
 the mean $R$-amplitudes of the $R$ ${\bold k}$-modes 
are those of that wave.
Moreover because of the EPR correlations in the in vacuum, 
the mean amplitudes of the associated $L$-modes are 
fixed by the detection of the $R$-wave $\bar \phi_R$ and $z_k$. 

For the 2-point functions we have, 
\sba \label{buildingblocks}
 &&\langle \hat a_{\bold k,\, R}^{\dagger \,  out}\, 
 \hat a_{\bold k',\, R}^{out}  \rangle_{{\cal V}_R} 
 = v^*_{\bold k} v_{\bold k'} \, , 
 \qquad \langle \hat a_{\bold k,\, L}^{\dagger \, out} \, 
 \hat a_{\bold k',\, L}^{out}   \rangle_{{\cal V}_R}    
 = z^*_{k} v_{\bold k} \, z_{k'} v^*_{\bold k'} \, , \\
 &&\langle \hat a_{\bold k,\, R}^{out}  \, 
 \hat a_{\bold k',\, L}^{out}  \rangle_{{\cal V}_R}  
 = v_{\bold k} \, z_{k'} v^*_{\bold k'}  \, , \\
 &&\langle \hat a_{\bold k,\, R}^{\dagger\, out} \, 
 \hat a_{\bold k',\, L}^{out}  \rangle_{{\cal V}_R}  
 = v^*_{\bold k} \,  z_{k'} v^*_{\bold k'} \,  .
\sea 
In the first line, the main modification with respect to in vacuum correlations
is the loss of the diagonal character in ${\bold k}$.
This 
%D25/11/03 dramatic %tu y tiens vraiment
radical
change follows from the strength 
of the projection induced by $\hat \Pi_{\cal V_{R}}$. 
Since all $R$-${\bold k}$ components are now described by coherent states, 
the above 2-point functions are entirely given by a disconnected contribution.
For these 2-point functions, the 
%D25/11/03
correspondence mentioned in section IV.B
%above mentioned correspondence 
is exact.
However, this is not the case in general because of non-vanishing 
commutators (consider for instance $ \langle \hat a_{\bold k,\, R}^{out}\, 
 \hat a_{\bold k',\, R}^{\dagger \,  out}  
\rangle_{{\cal V}_R} $). It is only in the large occupation number regime 
that the operator ordering gives subdominant corrections. 
In the second line, we see that the in vacuum 
correlations between $R$ and $L$ modes  have been 
replaced by the ``coherent state correlations'' described by Eq. \Eqref{Impeq}.
They fix $\langle \hat a_{\bold k,\, L}^{out}  \rangle_{{\cal V}_R}$ in terms
of $ \langle  \hat a_{\bold k,\, R}^{out} \rangle_{{\cal V}_R}$. 
Notice finally that the variances of $\hat a^{out}_{\bold k,\, R}$ and 
$\hat a_{\bold k',\, L}^{out}$
vanish, see Eq. \Eqref{nullvariance}.
The detection of a coherent state of the field can thus be
seen as 
%a snapshot in the sense that it 
providing one classical realization of the stochastic ensemble.

Before examining the correlations in space induced by the 
detection of $\bar \phi_R$, it is of value to determine to what extend 
one recovers (in vacuum) mean values from these
conditional expectation values and from the distribution $P^{in}_{{\cal V}_R}$. 
Using Eq. \Eqref{Impeq-1} and Eq. \Eqref{ProbaClWaveA},
one finds that the ensemble average is defined by
\ba \label{ensaverage}
% \langle \hat a_{\bold k,\, R}^{out}  \hat a_{\bold k',\, L}^{out} \rangle_{\rm{in}} &=& 
 \langle\langle v_{\bold k,\, R} \, w_{\bold k',\, L} \rangle \rangle_{\rm{in}}
 &=&  \frac{\int\!\!\widetilde{\cal D}v_{\bf p} \widetilde{\cal D}w_{\bf q} \,\,  
 v_{\bf k} w_{\bf k'}\,  P^{in}_{{\cal V}_R, \, {\cal W}_L}}
 {\int\!\!\widetilde{\cal D} \, v_{\bf p} \, 
 \widetilde{\cal D}w_{\bf q} \, P^{in}_{{\cal V}_R, \, {\cal W}_L}}
 \nonumber \\ 
 &=&  \frac{\int\!\!\widetilde{\cal D}v_{\bf p}  
 \,\, v_{\bf k} z_{k'} v_{\bf k'}^{*} \, P^{in}_{{\cal V}_R}}
 {\int\!\!\widetilde{\cal D}  v_{\bf p}  \, P^{in}_{{\cal V}_R}}
%frac{d^2v}{\pi} \, 
%      \bra{0\rm{in}} \hat a_{\bold k,\, R}^{\dagger \, out} 
%      \hat a_{\bold k,\, R}^{out} \, \hat \Pi_v \otimes {\bf{1}}_L \ket{0\rm{in}} 
% \,  \nonumber \\
% &=&\int\!\!\frac{d^2v}{\pi} \, \abs{z_{k}}^2 \abs{v_{\bold k}}^2 
% P_{\bar z}({\cal V}) \,  \nonumber \\ 
 = z_{k} \abs{\alpha_k}^2 \delta^3({\bf k} - {\bf k'}) \, .
\ea
%This is in agreement with the quantum result Eq. \Eqref{cross}.
The tilde over the functional integration 
is there to remind that the integration variables $v_{\bf k}$ ($w_{\bf k}$)
are defined only for $k_x >0 $, thus 
$\widetilde{\cal D} v_{\bf k} = \widetilde{\prod\limits_{\bf k}} dv_{\bf k} $.
The result of Eq. \Eqref{ensaverage} is in agreement with the quantum result 
Eq. \Eqref{cross}. 
For the diagonal $R$ term, the correspondence between the ensemble average 
and the quantum result Eq. \Eqref{occup} is not exact and the 
mean values differ by a factor equal to $\vert z_k \vert^2$. 
The origin of this discrepancy is that the 
$\hat a^\dagger_{{\bf k}, \, R} \hat a_{{\bf k'}, \, R}$
does not commute with the projector $\hat \Pi_{{\cal V}_R}$, see also 
the third footnote.
However, since the discrepancy rests on commutators, 
the two versions will agree when $n_k \to \infty$.
This agreement in the large occupation number limit
confirms that field configurations are effectively
characterized by a 
set of stochastic variables ${\cal V}$ with the two-mode probability 
distribution $P^{in}({\cal V})$. 

\section{Spatial correlations}

In the preceding subsection we gave the new expectation values 
%given the detection of 
when having detected the classical right moving configuration $\bar \phi_R$.
Here we shall see that this detection leads to specific correlations in space.
To exhibit these correlations it suffices to compute the modified
expectation value of the field amplitude: 
\ba \label{spacetimestruct}
\langle   \hat \phi (\eta, {\bf x}) \rangle_{{\cal V}_R}=  
 \bar \phi_{R, \, {\cal V}_R} +  \bar \phi_{L, \, {\cal V}_R}\, ,
 % \phi_R^{cl} + \phi_L^{cl} \, ,
\ea
where the $R$ and $L$ mean waves are 
\sba \label{condvalues2}
 \bar \phi_{R, \, {\cal V}_R}(\eta, {\bf x}) &=& 
 \int\!\!\widetilde{d^3 k} \, 
     \left( v_{\bold k} \, \frac{e^{i \bold k \bold x}}{(2\pi)^{3/2}} 
  \phi_{k}^{out}(\eta)
  %\tilde \phi_{\bold k}^{out}(\eta, {\bf x}) 
  + c.c.
     %v_{\bold k}^{*}\,  \tilde \phi_{\bold k}^{out \, *} (\eta,{\bf x}) 
     \right)
     \, ,\\
 \bar  \phi_{L, \,{\cal V}_R}(\eta, {\bf x}) &=& 
 \int\!\!\widetilde{d^3 k} \, 
     \left( z_{k} v_{\bold k}^{*} \, \, 
     \frac{e^{-i \bold k \bold x}}{(2\pi)^{3/2}} 
  \phi_{k}^{out}(\eta)
%     \tilde \phi_{-\bold k}^{out}(\eta, {\bf x}) 
+ c.c.
%     z_{k}^* v_{\bold k} \,     \tilde \phi_{-\bold k}^{out \, *}(\eta, {\bf x})
\right)
  \, .
\sea
%These are the central equations of this section.
Eq. (\ref{condvalues2}a) should cause no surprise. Since the 
projector $\hat \Pi_{{\cal V}_R}$ completely specifies $R$-configurations, 
expectation values in the $R$ sector are given by the coherent state 
expectation values as in subsection IV.A. 
%is mean 
Eq. (\ref{condvalues2}b) is more subtle as it arises both from 
this projector as well
as from the EPR correlations, Eq. \Eqref{Impeq}.
%reducedstate}.p
%coherence of the in vacuum. 
Because of the latter, 
the mean value of the $L$-part of the field operator, $\hat \phi_L$,
is also described by a {\it local} wave packet 
even though {\it nothing} has been specified about $L$-configurations.

The main lesson from these equations is that the simultaneous 
specification of the 
various $v_{\bf k}$ has introduced some {\it spatial coherence} 
by coupling modes which were so far independent.
To obtain these spatial properties,
we re-use the wave packet given in Eq. \Eqref{wp}.
The $R$-wave is then given by Eq. \Eqref{barphiR}
which is maximum along the 
classical light-like trajectory of Eq. (\ref{maxR}).
More interesting is the partner wave, the $L$ component.
Using Eq. \Eqref{z}, one has 
\ba\label{barphiL}
 \bar \phi_{L, \,{\cal V}_R} = - \,  
        \bar v N \int \widetilde{d^3k} \inv{\sqrt{2k}} 
 	e^{-\frac{(\bf k - \bar{\bf k})^2}{4\sigma^2}} 
	%e^{ik \eta} e^{-i\bf k\bf x}
 	\left( e^{- i\bf k(\bf x- \bf x_0)} e^{-i k (\eta +  \eta_0 - 4 \eta_r)}
	e^{-i\phi} + c.c \right) \, .
\ea
Two interesting properties should be discussed.
First, using the stationary phase condition one determines the 
partner's trajectory ${\bf x}_L$ defined in Eq. \Eqref{maxL}. 
As expected 
one verifies that the partner propagates in the opposite direction. 
More importantly, it is separated from the detected wave by
the `universal' distance given in Eq. \Eqref{maxL2}.
Secondly, the phases of the two waves are opposite 
when evaluated at the centers of the wave-packets, 
compare Eq. \Eqref{barphiR} and \Eqref{barphiL}. 
%D25/11/03
This phase opposition is particularly clear when working in 1 dimension.
In this case, for {\it any} right moving wave packet, 
%In this case, 
%Indeed, 
one has
% in the following sense :
\ba
\label{phaseop}
 \bar \phi_{L, \,{\cal V}_R}({\bf x} ; \phi) = 
 \bar \phi_{R, \,{\cal V}_R}({\bf x} -{\bf x}_R + {\bf x}_L  ; \phi + \pi) \, .
\ea
%In this equation, we have used the saddle-point approximations of 
%Eqs. \Eqref{barphiR} and \Eqref{barphiL}.
In Section IV, we 
%D25/11/03
%%%%%%%%%%%%%obtain explicitely this result 
have seen that a similar result obtains
in 3 dimensions when working at the saddle-point approximation.
%, the waves depend parametrically on the temporal through 
%Eqs. \Eqref{maxR} and \Eqref{maxL},
%together with a time-dependant spread.
%is $\pi + \phi$ whereas that of the $R$ wave is $\phi$,
%see Eq. \Eqref{barphiL} and Eq. \Eqref{barphiR}.
%R%sign in factor in Eq. \Eqref{barphiL} 
%is opposite to that in Eq. \Eqref{barphiR}. 
This phase opposition originates from the coherence in the in-vacuum,
see Eq. \Eqref{inoutvact} and Eq. \Eqref{z}. (Notice that it also 
follows from the neglect of the decaying mode). 
It has important physical consequences. It implies that the partner
of a local Newton (Bardeen) potential dip is a local hill. 
For adiabatic perturbations,
% ($\delta T/T \propto \Phi$) , 
it means that the partner of a hot region
%in the long wave-length approximation, the 
%temperature fluctuation $\delta T/T$ 
is a cold region.
It is to be emphasized that these correlations 
are valid for every configurations specified by Eq. (\ref{cohR})
and not only in the mean. 

%have the opposite sign.
%is proportional 
%to the Bardeen potential $\Phi_B$. Relation between Mukhanov field and Bardeen
%potential is complecated \cite{MFB}.......... In principle, the partner of a hot
%region is a cold region.......

%Therefore 
In brief, by having isolated from the in vacuum
the $(\bar {\bf k} , \bar v)$ $R$ configuration centered around 
$ {\bf x_0} $ at $\eta_0$, 
%with momentum $\bar {\bf k}$ and amplitude $\bar v$, 
we obtain a causally disconnected $L$ configuration which is centered 
around $ {\bf x}_0 - 2 (\eta_0 - 2\eta_r) \, {\bf 1}_{\bar{\bf k}}$, 
has momentum $-\bar {\bf k}$, and which has the same amplitude $\bar v$
and opposite phase.
 Notice that these results follow from the fact that
  $\langle   \hat \phi \rangle_{{\cal V}_R} $
given in Eq. \Eqref{spacetimestruct}
 is in fact a single wave packet of in modes. Notice finally
 that we have reached these results by making use of the complete
projector $\hat \Pi_{{\cal V}_R}$ which specified the amplitudes
of all $R$-modes. However these results can also be
%could be also
obtained when performing only a partial selection which leaves unspecified
the amplitudes of all $R$-wave packets orthogonal to the chosen one. 
The proof is given in the Appendix \ref{sec:wp}. It rest on the 
Gaussianity of the distribution.

%\begin{figure}[ht] \label{wptranform}
% \epsfxsize=8.0truecm
% \epsfysize=6.0truecm
% \centerline{{\epsfbox{wptransform.eps}}}
% \caption{The result of a wave-packet transform of the two-point function with
% respect to one of its arguments. The characteristics of the wave-packet are the
% same as in Fig. 1. The conformal time ranges from reheating to combination 
% at $\eta_{LSS} = 4$.}
%\end{figure}

%The general case where $\eta \neq \eta_{eq}$ has been represented on Figure XXX. 
%The Fig. XXX shows the result of such a wave-packet transformation. 
%As in Fig XXX, the two partners are perfectely anti-correlated to the central 
%temperature fluctuation. 
%wave. The relative heigh of these symmetric fluctuations with respect to
%the central one is $1/2$. This is because
%this three-fold structure is the superposition of two identical
%patterns, symmetric with respect to the the center ${\bf x}_R(\eta_{LSS})$ of
%the wave-packet $\Phi_{\lambda,R}^{out}$. The two patterns correspond to
%wave-packets of a same mean momentum $\bar k_{\lambda}$ but with opposite
%velocities.

\section{Conclusions and Discussions}

In the second part of this paper we have shown the following results.
When considering (long after the reheating)
the set of final configurations
of a quantum field in an inflationary model,
the detection of a coherent state describing half the modes
(e.g. a right moving configuration)
 is a random process, see Eq. \Eqref{ProbaClWave}. 
Second, the fact that modes have been amplified in pairs
implies that the ``reduced state'' 
which follows from this detection is 
%such that the partner (left moving) configurations are 
also a coherent state,
see Eq. \Eqref{Impeq}. Therefore, when
%when restricting the set of final configurations by 
specifying that some local waves have been detected,
% long after the reheating,
one introduces spatial correlations which possess definite properties. 
In particular these correlations
always have the same dipolar structure
%which is the signal that the 
since the amplification process
%modes have been amplified in a 
is scale invariant.
% manner.
Moreover, the two waves in each pair are
%%%R exactly
in phase opposition
because of Eq. \Eqref{z}
%the high occupation numberimplies that the
% we showed that the components of 
which tells us that
 the linear $k$ dependence of the
phase of $z_k$ will only induce a translation of the partner's
wave with respect to the chosen one. 
Indeed, this linear dependence implies that 
%leads to the fact that
the separation between the two waves 
%in any pair 
is always given by twice the Hubble time multiplied by the speed of the waves,
in a direction specified by their wave vector, see Eq. \Eqref{maxL}.
The phase opposition means that
%besides a question of spread,see Eqs. (\ref{spread}),
%D25/11/03
%in a saddle point approximation, 
the spatial profile of the partner wave 
is the symmetrical of that of the chosen wave, see Eq. \Eqref{phaseop},
%In the perpendicular directions
up to a question of the spreads in the perpendicular
directions, see Eqs. (\ref{spread}).
%the symmetric
It should be also pointed out that there also exists a strict correlation
in the amplitudes of the two waves in each pair. 
This correlation in amplitude cannot be seen from the simpler treatment
based on the two-point function since the mean has been taken
before having applied the wave packet transformation.
%In this paper we have shown the following results.
%When considering a quantum scalar field in inflation 
%and restricting the set of final configurations by specifying that some 
%local waves have been realized long after the reheating,
%one introduces new correlations which possess definite
%spatial properties. In particular these correlations 
%always have the same dipolar structure
%which is the signal that modes have been amplified in pairs 
%in a conformally invariant manner.
%Therefore the mean separation between the two waves in each pair is
%always given by twice the Hubble time multiplied by the speed of the waves,
%in a direction specified by their wave vector. 
%Moreover, the high occupation number limit
%implies that the 
%% we showed that the 
%components of the two waves in each pair are 
%%%%R exactly
%in phase opposition, besides the linear dependence of the
%phase of Eq. (\ref{z}) which is simply a shift in $\eta$. Therefore, 
%%besides a question of spread,see Eqs. (\ref{spread}), 
%%D25/11/03
%%in a saddle point approximation, 
%the profile of the partner wave is the symmetrical of that of the chosen wave,
%%In the perpendicular directions  
%up to a question of spread in the perpendicular 
%directions, see Eqs. (\ref{spread}).
%%the symmetric 
%In this we have completed the simpler 
%analysis based on a 
%We then show that these dipolar structures can also be found by 
%wave packet transformation of the two point function.
% with respect to one of its arguments. More precisely,
In addition to this discrepancy,
when using a product which 
is insensitive to the sign of the wave velocity, as in Eq. \Eqref{Theta},
%does not distinguish the velocity 
%D25/11/03 of a plane wave,
%for a given wave vector, 
we obtain the three folded structure 
since the contribution of two pairs are isolated 
by computing the spatial overlap, see Figure 1.

One might finally question if it would be possible to 
%put these structures into observational evidence.
observe these structures.
Let us briefly mention 
the different aspects which 
should be confronted.

One should first analyze the statistical basis for the
identification. We first notice that
Eq. \Eqref{Gbar} results from having taken an
ensemble average 
(which could be thought to be either quantum or stochastic). 
This mean value could be reached observationally if
a sufficiently large number of 
(independent) 
pairs are considered. In this, 
one should exploit the isotropy and the homogeneity of the 
distribution, as for the temperature anisotropy multipoles. 
When dealing with modes with sufficiently high wave vectors
(corresponding to angles smaller than a degree), this condition can 
probably be met.

%In addition to these aspects, one should also confront the
%following difficulties.

Second, we have access only to a portion of the configurations 
at recombination time, namely the Last Scattering Surface 
${\cal{S}}$, the intersection of 
$\eta=\eta_{rec}$ with our past light cone, see Fig. 3.
Since all pairs propagate on the particle horizon of the locus of birth, at
recombination, they fall into three classes.
First there exist pairs which do not intercept the LSS,
such as the pair $1$. These do not contribute at all
to the temperature anisotropies. 
Second there exist pairs such as $2$ for which only one 
member crosses the LSS. These contribute incoherently 
(i.e. as if $c_k = 0$)
to the temperature anisotropies. 
Third one finds the pairs such that both members
live on the LSS. They contribute coherently
to the anisotropies. Hence only these are responsible for the 
dip in the function $C(\theta)$ mentioned in \cite{BB}.
%In order to have both members on the LSS, t
These pairs have their wave vector tangent to ${\cal{S}}$. 
Their number is therefore limited by these geometrical constraints.
To quantify the percentage of such pairs, 
one must consider the depth of the LSS. 
%Notice that this percentage determines the relative value of the dip.
These aspects will be 
presented in a forthcoming publication. 

%the is not orthogonal to the
%radius, hence the two parners can never be embraced within a $2$-sphere centered
%on us. 

%unless one can use wave length much shorter than the 
%Hubble radius at recombination. 
%much smaller than the accual number of pairs. 
%For an estimation of their nurmber, recall  t
%The density of modes per unit conformal volume per unit conformal momentum 
%is $1$. Since the LSS has a finite width of redshift $z_w \sim 8$. Thus, the density 
%of pairs 
%of modes accessible with momentum $k$ is $dN/dk \sim k^2 z_w \eta_{LSS} R_H^2$ with 
%$R_H$ the Hubble radius at recombination. 
%0can be obt$N \sim ...$ where $z_w \sim 8$ is the 
%finite width of the last scattering surface, $R_H$ the
%Hubble radius at recombination, and $\abs{\beta_k}^2$ the
%density of pairs per unit conformal volume per unit momentum.
%is To estimate the former, we take into account the finite width of the last
%scattering surface $z_w\sim 8$. With a density of pairs per unit conformal volume 
%per unit momentum proportional to $\abs{\beta_k}^2$, one gets $N \sim ...$
%which is a number of observable pair large enough.

Moreover, the temperature anisotropies do not arise solely from the 
density fluctuations on the LSS. For a description of the 
various contributions, we refer to \cite{priorM,Mukhanov03}.
Notice that the Doppler
effect does not affect the temperature fluctuations 
which propagate longitudinally with respect
to the LSS. 
%the contribution which is due to the Doppler
%effect should  still be coherent as it arises from the derivative of the
%adiabatic density fluctuations. This could be exploited to 
%enhance the detectability of the correlations. 
Instead the contribution of secondary anisotropies
will lower the level of coherence of temperature anisotropies.
% so as to prevent a clear pair identification.

Finally, 
%there is little hope that the velocity field be 
%accessible since the LSS is very thin \cite{wmap}.  It 
would be very interesting to have access to the velocity field
on the LSS in order to be able to suppress the
doubling of the partners. 
Maybe a clever use of the polarization spectra might allow to 
reconstruct this field. 
%egeneracy 
Before trying to do so one might first
look for a statistical identification
of three folded structure. We re-emphasize the statistical 
character of this structure  which results from
an averaging procedure over pair creation events which individually
form local dipoles. 
% knowledge of the polarisation 
%field might however be used to distinguish the various contributions,
%thereby increasing the possibility of pair identification.

\begin{figure}[ht] \label{magiccone}
 \epsfxsize=10.0truecm
 \epsfysize=7.5truecm
 \centerline{{\epsfbox{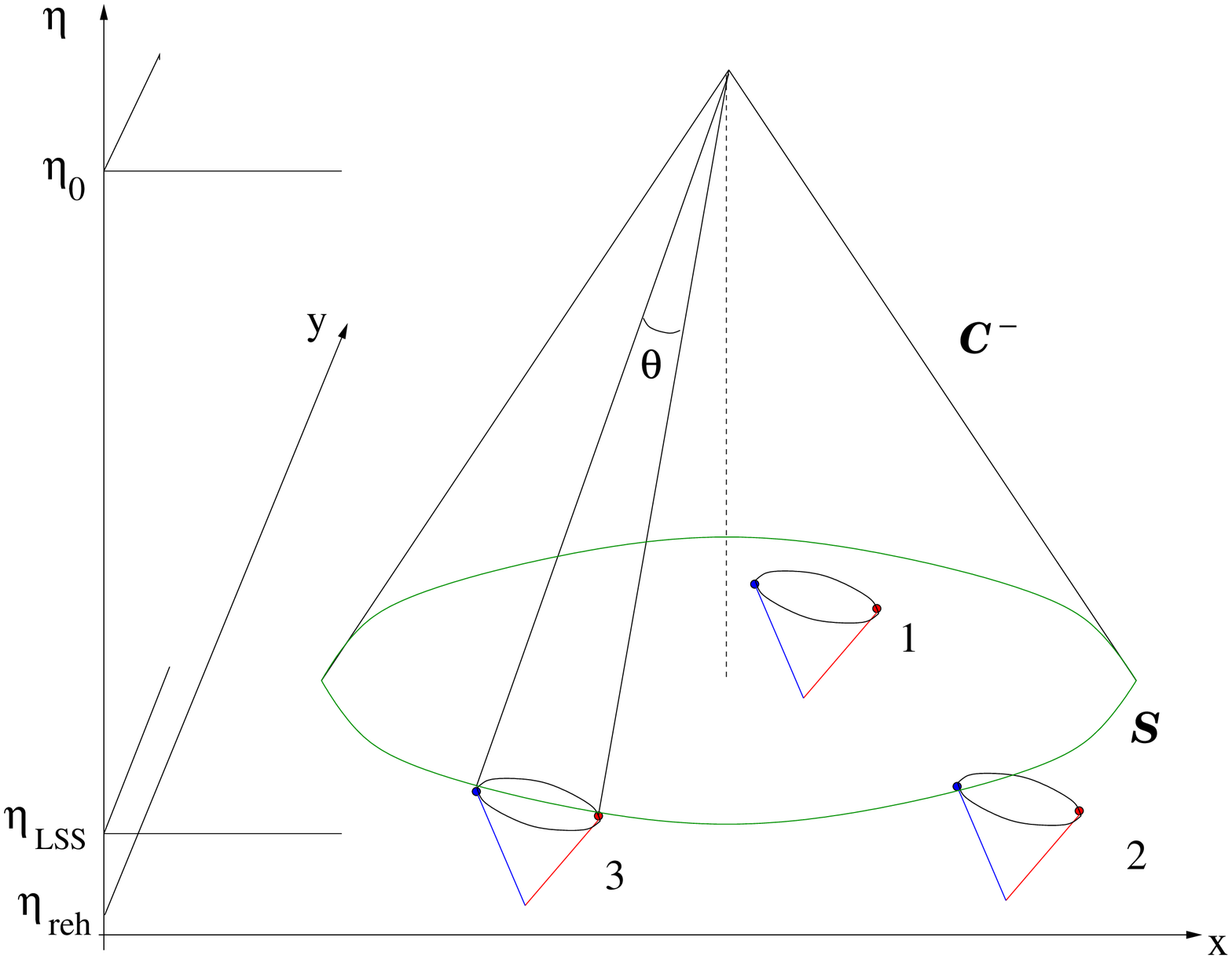}}}
 \caption{A $2+1$ dimensional space-time diagram in conformal coordinates
 $(\eta,x,y)$. 
 The Last Scattering Surface ${\cal{S}}$ is here represented by a circle 
 defined by the
 intersection of our past light-cone ${\cal{C}^{-}}$ with the 
 2d-spacelike surface $z=cte, \eta = \eta_{LSS}$. 
%We work in the approximation that  the recombination in instantaneous.
% $3$-spacelike 
% surface at $\eta_{LSS}$
% the $2$-sphere defined by the
% intersection of our past light-cone ${\cal{C}^{-}}$ with the $3$-spacelike 
% surface at $\eta_{LSS}$. 
 All pairs are created at reheating $\eta_{reh}$ and propagate on 
 light-cones. Given that the recombination is 
 almost instantaneous, only few pairs, such as $3$, 
 are such that both particles intercept ${\cal{S}}$. 
%the Last Scattering Surface. 
%D25/11/03 
% They are seen by us
% with an angle $\theta = 2c \Delta \eta \simeq 1 \deg$.
}
\end{figure}

\vskip .3 truecm
%\newpage

${\bf{Acknowledgements}}$
\newline
We would like to thank Ted Jacobson, Serge Massar, Simon Prunet, and 
Jean-Philippe Uzan for useful remarks.
%for interesting comments concerning the first version of this paper.

%%%%%%%%%%%%%%%%%%%%%%%%%%%%%%%%%%%%%%%%%%%%%%%%%%%%%%%%%%%%%%%%%%%%%%%%%%%%
%%%%%%%%%%%%%%%%%%%%%%%%%%%%%%%%%%%%%%%%%%%%%%%%%%%%%%%%%%%%%%%%%%%%%%%%%%%%
%%%%%%%%%%%%%%%%%%%%%%%%%%%%%%%%%%%%%%%%%%%%%%%%%%%%%%%%%%%%%%%%%%%%%%%%%%%%
%%%%%%%%%%%%%%%%%%%%%%%%%%%%%%%%%%%%%%%%%%%%%%%%%%%%%%%%%%%%%%%%%%%%%%%%%%%%
%%%%%%%%%%%%%%%%%%%%%%%%%%%%%%%%%%%%%%%%%%%%%%%%%%%%%%%%%%%%%%%%%%%%%%%%%%%%
%%%%%%%%%%%%%%%%%%%%%%%%%%%%%%%%%%%%%%%%%%%%%%%%%%%%%%%%%%%%%%%%%%%%%%%%%%%%
%%%%%%%%%%%%%%%%%%%%%%%%%%%%%%%%%%%%%%%%%%%%%%%%%%%%%%%%%%%%%%%%%%%%%%%%%%%%
%%%%%%%%%%%%%%%%%%%%%%%%%%%%%%%%%%%%%%%%%%%%%%%%%%%%%%%%%%%%%%%%%%%%%%%%%%%%

\begin{appendix}

\section{The Bogoliubov transformation from inflation to the adiabatic era} 
\label{sec:settings,bogol}

Since we use quantum settings, we need 
the Bogoliubov coefficients relating positive frequency  
modes during inflation (taken here to be for simplicity
a de Sitter period) to 
%positive frequency  
modes during the radiation and matter dominated eras.

In these three periods,
the scale factor is given by 
\sba \label{aeta}
 a(\eta) &=& - \frac{1}{H\eta} \, , \qquad 
 {\mbox{for\, \, }} -\infty < \eta < \eta_r < 0  \, , \\
 a(\eta) &=& \frac{1}{H\eta_r^2} (\eta -  2 \eta_r) \, , \qquad 
  {\mbox{for\, \, }}  \eta_r < \eta  < \eta_{eq} \, , \\
  a(\eta) &=&  \frac{1}{4H\eta_r^2} 
  \frac{(\eta-4\eta_r+\eta_{eq})^2}{\eta_{eq}-2\eta_r} \, , \qquad 
 {\mbox{for\, \, }} \eta_{eq} < \eta \, ,
\sea
where $\eta_r$ and $\eta_{eq}$ designate respectively 
the end of inflation 
and the time of equilibrium between radiation and matter. The
shifts in the parenthesis in the second and third lines are 
necessary to parametrize the three periods by a single conformal time $\eta$.
The sudden pasting of these periods are such that
the scale factor and the Hubble parameter are
continuous functions.
% of conformal time. The phase transition between two
%subsequent expansion epochs are approximated to be instantaneous \cite{Allen94}. 

The positive-frequency solutions of Eq. \Eqref{PropPhi_k} 
(corresponding to
gravitational waves, the modes for density fluctuations having a different 
dispersion relation, see the second footnote) are 
\sba \label{modes}
 \phi_k^{in}(\eta) &=& \frac{1}
%e^{i\phi_{in}}}
{\sqrt{2k}} 
 	\left(1-\frac{i}{k\eta}\right)e^{-ik\eta} \, , \quad 
 \mbox{for\, \,} -\infty < \eta < \eta_r   \, , \\
 \phi_k^{rad}(\eta) &=& 
 	\frac{1}
%e^{i\phi_{rad}}}
{\sqrt{2k}} e^{-ik\eta} \, ,\quad 
  \mbox{for\, \, }  \eta_r < \eta  < \eta_{eq} \, , \\
 \phi_k^{mat}(\eta) &=& \frac{1}
%e^{i\phi_{mat}}}
{\sqrt{2k}} 
 	\left(1-\frac{i}{k(\eta - \eta_m)}\right)e^{-ik\eta} \, , \quad 
 \mbox{for\, \,} \eta_{eq} < \eta \, ,
\sea
where we note $\eta_m= 4\eta_r - \eta_{eq}$.
The Bogoliubov coefficients between these
modes are
\sba \label{defBogol}
 \phi_{k}^{in}(\eta) &=& \alpha_{k}^{in-rad} \ \phi_{k}^{rad}(\eta)
  + \beta_{k}^{in-rad\, *} \ \phi_{k}^{rad \, *}(\eta) \, , \\
 \label{defB2}
 \phi_{k}^{rad}(\eta) &=& \alpha_{k}^{rad-mat} \ \phi_{k}^{mat}(\eta)
  + \beta_{k}^{rad-mat\, *} \ \phi_{k}^{mat \, *}(\eta) \, . 
\sea 

The coefficients $\alpha_{k}$ and $\beta_{k}$ of the first line are 
given by the Wronskians
\ba \label{abcoef}
  \alpha_{k}^{in-rad} = \left( \phi_{k}^{R} , \phi_{k}^{in} \right)
  \mbox{, \ \ }
  \beta_{k}^{in-rad\, *} = -  \left( \phi_{k}^{R\, *} , \phi_{k}^{in} \right) \, ,
\ea 
evaluated at transition time $\eta_r$ since modes satisfy different equations
in each era. Similar expressions evaluated at the equilibrium time $\eta_{eq}$
hold for the coefficients between $\phi^{rad}_k$ and
$\phi^{mat}_k$. 
One gets
\sba \label{bogolincosmo}
 \alpha_k^{in-rad} &=& 1-\frac{i}{k\eta_r}-\frac{1}{2k^2\eta_r^2} \, , \\
 \left(\beta_k^{in-R}\right)^{*} &=& 
 	\frac{e^{-2i k \eta_r}}{2k^2\eta_r^2} \, , \\
 \alpha_k^{rad-mat} &=& 1+\frac{i}{k(\eta_{eq}-\eta_m)} - 
 	\frac{1}{2k^2(\eta_{eq}- \eta_m)^2}   \, , \\
 \left(\beta_k^{rad-mat}\right)^{*} &=& 
 	\frac{-e^{-2ik \eta_{eq}}}{2k^2(\eta_{eq}-\eta_m)^2} \, . 
\sea
These Bogoliubov coefficients have been calculated for gravitational waves 
in \cite{Allen88,Allen94}. Their results agree with ours up to constant phases
which can be gauged away by a mode redefinition.
%of the arbitrary phase of the modes \cite{CP1}.
For density fluctuations similar expressions are obtained when using
the appropriate frequency.

For relevant modes, i.e. modes which contribute to visible anisotropies in 
the CMB,
%today  that are well outside the 
%In the large wave-length limit, that is for modes that are well 
%outside the Hubble radius,
% at reheating, 
their wave length obeys 
%D25/11/03
$k/Ha(\eta_r) = k \vert \eta_r \vert \ll 1$. 
In the limiting case, to order 
%D% $(H k \abs{\eta_r} )^3$
$(k \abs{\eta_r} )^3$,
one has 
%We have repalced the squeezing parameter $z_k$ by its value
%which reads, in the conventions of Appendix A,
%the complex function $z_k$ is
\ba
\label{z}
\beta_k/\alpha_k^* = z_k = - e^{4 i k \eta_r}\, .
\ea
%$\alpha^{in-R}_k = - e^{2ik\eta_r}/2k^2 \eta_k^2$. 
Thus the in modes during the radiation dominated era read
%this amounts 
%to neglect the decaying mode. 
%Explicitely, putting all the arbitrary phases to zero, for modes such that
%$k\eta_r,  \ll 1$ (which is verified for the modes of interest, see for 
%instance \cite{Allen99a}), one has during the radiation dominated epoch: 
\ba \label{neglectdecay}
 \phi_k^{in}(\eta) 
 %&=& -\frac{1}{2k^2\eta_r^2} 
 %\left(\phi_{k}^{out} r \phi_{k}^{out\, *} + O(k\eta_r)\right) \nonumber \\
 &=& \frac{i}{\sqrt{2}k^{3/2}\eta_r^2} \left(\frac{\sin(k(\eta- 2 \eta_r) )}{k} 
      + O((k\eta_r)^3) \right)   \, , \qquad  \eta \gg \eta_r \, .
\ea
One then verifies that the physical modes 
$\xi_k = \phi_k^{in}/ a$ are 
constant (and $=H/(2k^{3/2})$ 
%D25/11/03
until they re-enter the Hubble radius, i.e. when $k(\eta- 2\eta_r)$
approaches $1$. This guarantees a scale invariant spectrum 
%$\frac{k^3}{2\pi^2} \frac{32\pi}{m_{Pl}^2} \abs{\xi_k}^2 = 
%\frac{16}{\pi m_{Pl}^2}
$k^3 \abs{\xi_k}^2 = H^2$ 
%until they re-enter the Hubble radius, i.e. when $k(\eta- 2\eta_r)$
%approaches $1$. 
Finally, since $\xi_k$
%%%R = \phi_k^{in}/ a$ 
%D25/11/03
is
%are 
constant 
%when starting with the exact expression,
the approximation which consists in dropping terms of $O((k\eta_r)^3)$ in
Eq. \Eqref{neglectdecay} corresponds to 
%can be reached by 
the neglect of the decaying mode, see the discussion after 
Eq. (\Eqref{infl2ptf1}).

%%%%%%%%%%%%%%%%%%%%%%%%%%%%%%%%%%%%%%%%%%%%%%%%%%%%%%%%%%%%%%%%%%%%%%%%%%
%%%%%%%%%%%%%%%%%%%%%%%%%%%%%%%%%%%%%%%%%%%%%%%%%%%%%%%%%%%%%%%%%%%%%%%%%%
%%%%%%%%%%%%%%%%%%%%%%%%%%%%%%%%%%%%%%%%%%%%%%%%%%%%%%%%%%%%%%%%%%%%%%%%%%

\section{Coherent states for a real harmonic oscillator}

This appendix aims to give a self-contained presentation of coherent states,
with special emphasis on properties which 
shall be used 
in the body of the manuscript. 
For more details, we refer to \cite{Glauber63a,Glauber63b,Zhang90}.

They are several equivalent ways to define coherent states.
The definition we adopt \cite{Glauber63b} 
is as an eigenstate of the annihilation operator:
\ba \label{DefCoh1}
 \hat a \ket{v} = v \ket{v} \, ,
\ea
where $v$ is a complex number.
The development on this state
%The components of this state in the
%occupation number representation $\{ \ket{n} \} $ are
on Fock basis is
\ba \label{DefCoh2}
 \ket{v} = e^{-\frac{\vert v \vert^2}{2}}  \sum_{n=0}^{\infty}
 \frac{v^n}{\sqrt{n!}} \ket{n} \, ,
\ea  
where the exponential prefactor guarantees that the state is normalized 
to unity $\langle v \vert v \rangle=1$.

The first interesting property of coherent states is that they correspond 
to states with a well defined complex amplitude $v$. 
Indeed, by definition \Eqref{DefCoh1}, 
the expectation values of the annihilation and creation operators are
\ba
 \bra{v} \hat a \ket{v} = v \, , \quad \bra{v} \hat a^{\dagger} \ket{v} = v^{*} 
 \, .
\ea
It is to be stressed that the variances vanish:
\ba \label{nullvariance}
 \Delta \hat a^2 = \bra{v} \hat a^2 \ket{v} - \bra{v} \hat a \ket{v}^2 = 0 \, , 
 \quad \Delta \hat a^{\dagger \, 2} = \bra{v} \hat a^{\dagger \, 2} \ket{v} - 
 \bra{v} \hat a^{\dagger} \ket{v}^2 =  0 \, .
\ea
Moreover the mean occupation number is
\ba \label{occupnumber}
 \bra{v} \hat a^{\dagger} \hat a \ket{v} = \abs{v}^2 \, ,
\ea 
in agreement with the mean value given by the Poisson distribution
\Eqref{DefCoh2}.

>From these properties one sees that the expectation values
of the position and momentum operators (in the Heisenberg picture, 
with $\hbar = 1$)
\ba 
 \hat q(t)=\frac{\hat a e^{-i \omega t} + 
 		\hat a^{\dagger} e^{i \omega t}}{\sqrt{2\omega}} \, , \qquad 
 \hat p(t)=-i\sqrt\frac{\omega}{2}(\hat a e^{-i \omega t} - 
 				\hat a^{\dagger}e^{i \omega t}) 
 \, , \nonumber
\ea
are
\ba \label{cltrajincohstate}
 \bar q(t) &=& \bra{v} \hat q(t) \ket{v} = 
 \inv{\sqrt{2\omega}} ( v e^{-i \omega t} + v^* e^{i \omega t}) 
            = \sqrt{\frac{2}{\omega}} \abs{v} \cos(\omega t - \phi_v) 
 \, , \nonumber \\ 
%D25/11/03
 \bar p(t) &=& \bra{v} \hat p(t) \ket{v} = 
  -i\sqrt{\frac{\omega}{2}} ( v e^{-i \omega t} - v^* e^{i \omega t}) 
            = -\sqrt{2\omega} \abs{v} \sin(\omega t - \phi_v) 
	    = \partial_t \bar q(t) \, .
\ea
We have used the polar decomposition $v=\abs{v} e^{i \phi_v}$.
These expectation values have a well defined amplitude and phase and 
follow a classical trajectory of the oscillator. 
This is due to the ``stability'' of coherent states which is better seen
in the Sch\"{o}dinger picture. If the state is a
coherent state $\ket{v}$ at a time $t_0$, one immediately gets from 
\Eqref{DefCoh2} that at a later time $t$, the state is a coherent state
given by $\ket{v(t)} = \ket{ve^{-i\omega (t-t_0)}}$.
Notice that the variances of the position and the momentum are 
\ba
 \Delta \hat q^2 = \frac{1}{2\omega} \, , \qquad  
 \Delta \hat p^2  = \frac{\omega}{2} \, .
\ea
%when $\hbar = 1$.
They minimize the Heisenberg uncertainty relations and are time-independent.
Hence, in the phase space $(q,p)$, a coherent state can be
considered as a unit quantum cell $2\pi \hbar$ in physical units 
(see also \Eqref{normP_z} for the measure of
integration over phase space) centered on the classical position and momentum of
the harmonic oscillator $(\bar q(t),\bar p(t))$. 
In the large occupation number limit $\abs{v}\gg 1$,
coherent states can therefore be
interpreted as classical states since 
$\Delta \hat q / \bar q = \Delta \hat p / \bar p \propto 1/ \abs{v}$.
%This is 
%one reason 
%why they are considered as the quantum counterparts of
%classical points in phase space. 
This is a special application of the fact that coherent states
can in general be used to define the classical limit 
of a quantum theory, see \cite{Zhang90} and references therein.

One advantage of coherent  states \cite{Glauber63a} is that the 
calculations of Green functions resembles closely to those of the 
corresponding classical theory (i.e. treating the fields not as operators 
but as c-numbers) provided either one uses normal ordering,
or one considers only the dominant contribution when $\abs{v}\gg 1$.
In preparation for the calculations with field
we compute the 
%D25/11/03
Wightman
%two-point Green  
function in the coherent state $\ket{v}$
%D25/11/03
\ba
\widetilde G_v(t,t') &=& \bra{v} \hat q(t) \hat q(t') \ket{v} \nonumber \\
           &=& \langle :\hat q(t)\hat q(t') : \rangle_v + 
	   \frac{1}{2\omega} e^{i\omega(t-t')}
	   %\langle [\hat q(t),\hat q(t')] \rangle_v \,  ,
\ea
where we have 
%split the correlator into its real an imaginary parts. 
isolated the contribution of the vacuum.
%The commutator is state independent and order 1:
%\ba
% [\hat q(t),\hat q(t') ] = \frac{i}{\omega} \sin(\omega (t-t')) \, .
%\ea
%It can be removed by normal ordering.
The normal ordered correlator is order $\abs{v}^2$: 
\ba \label{cohgreen}
 \langle :\hat q(t)\hat q(t') : \rangle_v &=& 
 \inv{\omega} \, Re\left[\langle \hat a^2 \rangle_v e^{-i\omega(t+t')} + 
 % (\hat a^{\dagger})^2 e^{+i\omega(t+t')} + 
    \langle  \hat a^{\dagger} \hat a \rangle_v e^{i\omega(t-t')} \right]
    \nonumber  \\ 
% &=& \frac{1}{2\omega} \, \left[  \vert v \vert^2 \cos(\omega(t-t'))
%+ 2 Re \left\{ v^2 e^{-i\omega(t+t')}\right\} 
%% + c.c. ) +      \vert v \vert^2 \cos(\omega(t-t'))	
%\right]   \nonumber \\
% &=& \vert v \vert^2 
% \left[ \cos(\omega (t+t') -2\phi_v) + \cos(\omega(t-t')) \right] \nonumber \\
 &=& \frac{2}{\omega} \vert v \vert^2 \cos(\omega t - \phi_v) 
     \cos(\omega t' -\phi_v)   = \bar q(t) \,  \bar q(t') \, .
\ea
We see that the perfect coherence of the state, namely 
$\abs{\langle \hat a  \hat a  \rangle_v } 
= \langle \hat a^\dagger  \hat a  \rangle_v  
%(1 + O(1/\abs{v})
$ 
is necessary to 
combine the contributions of the diagonal and the interfering term so as to
bring the time-dependent classical position $ \bar q(t)$ in 
Eq. \Eqref{cohgreen}.

The wave-function of a coherent state in the coordinate
representation is given by 
\ba \label{CohWaveFunct}
  \psi_{v}(q) = \left(\frac{\omega}{\pi}\right)^{1/4} 
 e^{-\frac{\omega}{2}(q-\bar q)^2}e^{i \bar p q} \, ,
\ea
%D25/11/03
where $v=(\bar q + i \bar p)/\sqrt{2\omega}$.
This follows from the definition 
$\bra{q} \hat a \ket{v} = v \langle q \vert v \rangle$. 
>From this equation one notes that two coherent states are not orthogonal.
The overlap between two coherent states is
\ba \label{overlap}
 \langle v \vert w \rangle = \exp\left(v^{*} w - \inv{2}\abs{v}^2 - 
 \inv{2}\abs{w}^2 \right)  \, .
\ea
Nevertheless they 
form an (over)complete
basis of the Hilbert space in that 
the identity operator in the
coherent state representation $\{ \ket{v} \}$ reads
\ba
\label{coh1}
 {\bf 1} = \int\!\!\frac{d^2v}{\pi} \, \ket{v} \bra{v} \, .
\ea 
The measure is $d^2v=d(Re v) d(Im v)$.
(This identity can be established by calculating 
the matrix elements of both sides of
the equality in the coordinate representation $\{ \ket{q} \}$, 
with the help of \Eqref{CohWaveFunct}.) 
%This writing permits to expand states and 
%operators on the coherent state basis, leading to interesting 
%analytical properties 
%\cite{Glauber63b}, but this is out of the main scope of this paper.
%For us the importance of Eq. \Eqref{coh1} is to use coherent states
%to represent the detection of classical configurations in quantum terms.

%%%%%%%%%%%%%%%%%%%%%%%%%%%%%%%%%%%%%%%%%%%%%%%%%%%%%%%%%%%%%%%%%%%%%%%%%%%%

\section{Two-mode squeezed states and coherent states} \label{app:sqandcoh}

Since one can view $\hat \phi_{\bf k}(\eta)$ of Eq. \Eqref{hatphik} as the
position of a complex harmonic oscillator, it is useful to analyze the 
two-mode coherent states and the two-mode squeezed states of a complex
harmonic oscillator.
%Since pair creation in cosmology leads to two-mode squeezed states,
%it is usefull to analyse the links between the latter and the coherent 
%states of a complex harmonic oscillator. Indeed, we can write the real field
%$\hat \phi$ as the real part of a complex field $\hat \psi$ :
%\ba \label{phiintopsi}
% \hat \phi &=& \hat \psi + \hat \psi^{\dagger} \, , \nonumber \\
% \hat \psi(\eta, {\bf k}) &=& \int\!\!\widetilde{d^3k}\, e^{i {\bf k}{\bf x}} 
% 	\left( \hat a_{{\bf k},\, R}  \phi_k(\eta) + 
% 	\hat a_{{\bf k},\, L}^{\dagger}  \phi_k^*(\eta) \right) \, .
%\ea
%The new field $\hat \psi$ is a collection of complex harmonic oscillators. We
%restrict toone of these two-mode.
Its position and momentum operators are
\ba
\hat q(t)=\frac{1}{\sqrt{2 \omega}}(\hat a_R e^{-i\omega t} + 
	\hat a_L^{\dagger} e^{i\omega t}) \, , \qquad
\hat p(t)=-i\sqrt{\frac{ \omega }{2}}(\hat a_R e^{-i\omega t} - 
	\hat a_L^{\dagger} e^{i\omega t}) \, .
\ea
%As in the main body of text, we use the notation $R,L$ to distinguish 
%the two independent modes. 
Two-mode coherent states obey:
\ba
 \hat a_R \ket{v,\, R} \ket{w,\ L} = v  \ket{v,\, R} \ket{w,\ L} 
 \, , \qquad
 \hat a_L \ket{v,\, R} \ket{w,\ L} = w  \ket{v,\, R} \ket{w,\ L} \, .
\ea
Therefore the expectation values of the position and momentum in the state 
$\ket{v,\, R} \ket{w,\ L}$ are
\ba
 \bar q(t) = \inv{\sqrt{2 \omega }} (v e^{-i\omega t} + 
 w^{*} e^{i\omega t})  \, , \quad
 \bar p(t) = -i\sqrt{\frac{\omega }{2}} (v e^{-i\omega t} - 
 w^{*} e^{i\omega t}) \, .
\ea
As for a real oscillator, the normal ordered two-point function 
is given by the product of the mean values:
\ba \label{greenforcomplexocs}
 \bra{v,\, R} \bra{w, \, L} :\hat q(t) 
      \hat q^{\dagger}(t'): \ket{v,\, R} \ket{w, \, L} = 
 \bar q(t)\, \bar q(t')^{*} \, . 
\ea

%%%%%%%%%%%%%%%%%%%%%%%%%%%%%%%%%%%%%%%%%%%%%%%%%%%%%%%%%%%%%%%%%%%%%%%%%%%%
%%%%
%%%%%%%%%%%%%%%%%%%%%%%%%%%%%%%%%%%%%%%%%%%%%%%%%%%%%%%%%%%%%%%%%%%%%%%%%%%%

A two-mode squeezed state $\kett{z}$ of this system is defined
by the action of the following operator on the two-mode vacuum,
$\kett{0} = \ket{0,\, R}\ket{0,\, L}$: 
\ba \label{squeezedvac}
 \kett{z} &=& S(r,\phi) \kett{0} = 
 \exp\left[r \left( e^{-i2\phi} \hat a_{R} \hat a_{L} - h.c. \right)\right]
 \, \kett{0}  \nonumber \\
 	&=& \inv{\ch r} \, \exp\left(-e^{+i 2 \phi} \th r
 \, \hat a_{R}^{\dagger} \hat a_{L}^{\dagger}\right) \, \kett{0} \nonumber \\
 	&=& (\sqrt{1-\abs{z}^2}) 
\exp\left( z\hat a_{R}^{\dagger} \hat a_{L}^{\dagger}\right) \,
	\kett{0} \nonumber \\
 &=& (\sqrt{1-\abs{z}^2}) \sum_{n=0}^{\infty} z^n \ket{n,\, R}\ket{n, \, L}
\ea
where we have introduced $z=-e^{-i2\phi} \th r $. 
%The reduced form in the second
%line is made possible by the fact that the operator acts on the vacuum state
%\cite{CP1}. The parameter $\zeta$ is called the squeezing parameter. 
The complex parameter $z$ fully specifies the two-mode squeezed state.
The correspondence with the Bogoliubov coefficients
is made by $z=\beta/\alpha^{*}$, see
Eq. (B13) and Appendix B in \cite{CP1}.

It is interesting to compute the projection of a two-mode squeezed state on 
a one-mode coherent state in the right sector $\ket{v, \, R}$.
Using Eqs. \Eqref{DefCoh2} and \Eqref{squeezedvac}, one gets
\ba \label{vonz}
 \langle{v, \, R} \vert z \rangle_{2} 
&=& 
% \inv{\abs{\alpha}} e^{-\frac{1}{2}\abs{v}^2}
% \sum_{n,n'=0}^{+\infty} \frac{v^{*\, n}}{\sqrt{n!}} z^{n'} 
% \langle{n, \, R} \vert n', \, R \rangle \ket{n', \, L} \nonumber \\ &=& 
 \inv{\abs{\alpha}} e^{-\frac{1}{2}\abs{v}^2}
 \sum_{n=0}^{+\infty}\frac{(v^{*}z)^{n}}{\sqrt{n!}} \ket{n, \, L} 
  \nonumber \\ &=& 
 \inv{\abs{\alpha}} 
%e^{-\frac{1}{2}\abs{v}^2} 
 e^{-\frac{1}{2}\abs{v}^2 \left(1-\abs{z}^2 \right)}\ket{zv^*, \, L}
 \nonumber \\
 &=& \inv{\abs{\alpha}}
 \exp{\left(-\frac{\abs{v}^2}{2\abs{\alpha}^2} \right)} \ket{zv^*, \, L} \, .
\ea
In the last line we have used $z=\beta/\alpha^{*}$ to 
write $(1-\abs{z}^2)$ as $1/\abs{\alpha}^2$.
The reduced state is
also a coherent state. This follows from the EPR correlations
in the two-mode squeezed state. 
The normalization factor can be interpreted easily. 
Its norm squared gives
the probability that the system, initially in a squeezed state, 
is found in the coherent state in the right sector, irrespectively of the state
in the left sector: 
\ba \label{Clproba}
 P_z(v) 
   %&=& \inv{\abs{\alpha}^2} \sum_{n,n'} z^{*n} z^{n'} \bra{n,\, R}\bra{n,\, L} 
   % \left(\ket{v, \, R}\bra{v, \, R} \otimes {\bf 1}_L\right) 
   % \ket{n',\, R} \ket{n',\, L} \nonumber \\
   &=& _2 \bra{z} \left(\ket{v, \, R}\bra{v, \, R} 
   \otimes {\bf 1}_L \right) \ket{z}_2 \nonumber \\
   &=& \inv{\abs{\alpha}^2} \sum_{n} \abs{z}^{2n} 
   \abs{\langle n,\, R \vert v, \, R \rangle}^2  \nonumber \\
   &=& \inv{\abs{\alpha}^2} e^{-\frac{\abs{v}^2}{\abs{\alpha}^2}} \, .
\ea
This gaussian probability is centered and naturally normalized to unity owing to
%the completeness of the coherent state basis:
the representation of unity in the coherent state basis Eq. \Eqref{coh1}
%D25/11/03
\ba \label{normP_z}
 \int\!\!\frac{d^2v}{\pi} \, P_z(v) = 
 \int_{-\infty}^{+\infty}\!\!\frac{d\bar q d\bar p}{2\pi\omega} \, 
 \tilde P_z(\bar q,\, \bar p ) = 1 \, .
\ea
We have used the decomposition $v=(\bar q + i\bar p)/\sqrt{2\omega}$ 
and the measure is $d^2v=d(Re v) d(Im v)$. 
One can see $v$ as a stochastic variable characterized by the probability
distribution $P_z(v)$. 
The variance of $v$ is given by $\abs{\alpha}^2$. In the large occupation number
limit $\abs{\alpha}^2 \gg 1$ 
the dominant contributions of expectation values in the squeezed state
$\ket{z}_2$ can be all obtained by making use of $v$ and the distribution $P_z$.

%to the occupation number $n=$
%operator in the squeezed state by:
%\ba
%  \bra{z} \hat a^{\dagger} \hat a \ket{z} = 
%  ???=\bra{z} \frac{\hat q^2 + \hat p^2}{2} \ket{z} =  
%  \int_{-\infty}^{+\infty}\!\!\frac{d\bar q d\bar p}{2\pi} \, 
%  \tilde P_z(\bar q,\, \bar p ) \frac{\bar q^2+\bar p^2}{2} = 
%  \abs{\alpha}^2 - 1 \, .  
%\ea 
%It may be surprizing that the variance is $\abs{\alpha_k}^2$ and not 
%the mean occupation number $\abs{\beta}^2$. 
%This is actually quite normal, since the projection made
%here which leads to compute diagonal expectation values of the density operator.
%It is therefor linked with the "$Q$" representation of the density matrix (see
%\cite{Glauber69} for further details). 

%%%%%%%%%%%%%%%%%%%%%%%%%%%%%%%%%%%%%%%%%%%%%%%%%%%%%%%%%%%%%%%%%%%%%%%%%%%%
%%%%%%%%%%%%%%%%%%%%%%%%%%%%%%%%%%%%%%%%%%%%%%%%%%%%%%%%%%%%%%%%%%%%%%%%%%%%
%%%%%%%%%%%%%%%%%%%%%%%%%%%%%%%%%%%%%%%%%%%%%%%%%%%%%%%%%%%%%%%%%%%%%%%%%%%%
%%%%%%%%%%%%%%%%%%%%%%%%%%%%%%%%%%%%%%%%%%%%%%%%%%%%%%%%%%%%%%%%%%%%%%%%%%%%
%%%%%%%%%%%%%%%%%%%%%%%%%%%%%%%%%%%%%%%%%%%%%%%%%%%%%%%%%%%%%%%%%%%%%%%%%%%%
%%%%%%%%%%%%%%%%%%%%%%%%%%%%%%%%%%%%%%%%%%%%%%%%%%%%%%%%%%%%%%%%%%%%%%%%%%%%

\section{The detection of a single wave packet} \label{sec:wp}
%D%
%In the preceeding Section, the projection on the coherent states
%specified all $R$-modes, hence covering the full sky.
%%we have performed a postselection with the help of
%%coherent states in a naive way: half of the modes where post-selected, yielding
%%exponantially small probabilities, hence covering the full sky.
%%The cure is 

%In this Appendix we show that the results of Sections IV and V are robust
%when only a {\it local} realization of the Gaussian ensemble has been detected.
%In Sections IV and V we annalysed the space-time correlations for a complete
%selection. Since it requires the knowledge of the whole sky, 
%%% since it requires the knowledge of the whole sky. 
%%and this situation therefore
%%unrealistic for an implementation
%Rather it is meaningful to consider local wave-packets, leaving the state of the
%rest of the map unspecified.

In this appendix we consider a partial projection which concern only a subset
of modes. In fact, we shall consider the projection operator which concerns only
one $R$-mode and which acts as unity for 
%D25/11/03
all modes orthogonal to it. %Misleading ? -> all modes k ortho to \bar k

%In this Appendix we show that the results of Sections ... and .... are robust
%when considering coherent states of local wave packets. 
%The reason for this study is that the ``complete'' selections there performed
%require the knowledge of the whole sky which is irrealistic for 
%the treatment of observational data. Rather, one considers local wave packets.
%Therefore, we perform a
%%we introduce a more physically motivated 
%projection corresponding to the detection of 
%one local (on the map) wave packet of $R$-modes, unspecifying the state of the
%CMB on the rest of the map.
%%which contains no 
%%information for the modes orthogonal to it. 
%%To implement this projection, one needs to 
%%introduce a familly of orthogonal wave-packets.
%%% that will permit to select only
%%%a part of the sky without specifying the state of the rest.
%%%As a bonus, it provides a neat way to understand the classical
%%%nature of the fluctuations of the field in the high-sqeezing limit.

%%%%%%%%%%%%%%%%%%%%%%%%%%%%%%%%%%%%%%%%%%%%%%%%%%%%%%%%%%%%%%%%%%%%%%%%%%%%%%%
%%%%%%%%%%%%%%%%%%%%%%%%%%%%%%%%%%%%%%%%%%%%%%%%%%%%%%%%%%%%%%%%%%%%%%%%%%%%%%%

\subsection{Family of wave-packets} \label{sec:wp,familly}

%We suppose 
%%D% to me it is still a loose assumption
%that we have a discrete family of positive frequency 
Let us consider a family of positive frequency 
right-moving wave-packets $\Phi_{\lambda}^{R}(\eta, {\bf x})/a(\eta)$ 
solutions of \Eqref{PropPhi}. 
%DD%(as a concrete exemple we use wavelets \cite{waveletsbook}). 
Their Fourier contend is written 
%Fourier transform they are given by
\ba \label{DefWP}
 \Phi_{\lambda}^{R} = 
 \int \!\!\widetilde{d^3k}\,  \gamma_{\lambda,\, \bold k}^{R\,*} 
 {\frac{e^{i \bf k \bf x}}{(2 \pi)^{3/2}}} \phi^{out}_{k}(\eta) \, .
\ea
%where $\tilde \phi_{\bold k}$ are plane waves (out) modes of Eq. \Eqref{stmodes}.
The functions $\gamma^R_{\lambda,\, \bold k}$ 
are parametrized by six integers, designated generically by $\lambda$.
Three of them fix the mean momentum 
${\bold{\bar k_j}} = {\bf{j}} \,  \bar k$ 
where the vector 
${\bf{j}}=(j_x,j_y,j_z) \in \bold Z_+ \times \bold Z^2$ since $j_x \geq 0$.
The other specify the mean position  
${\bf{\bar x_n}} =  {\bf{n}} \,  \bar x$, $ {\bf{n}} \in \bold Z^3$ at $\eta_0$.
As a concrete example, one can work in a box. Then the 
functions $\gamma_{\lambda,\, \bold k}$ are matrices.
%momentum $\bar k_j^{Max} = j \times \bar k, \, j \geq 0$, where $\bar k_{x} > 0$
%and $\bar k_{y}=\bar k_{z}=0$ (QUESTION: does it cover the full Hilbert space?).
%The width in momentum is $\sigma$. 
%The resulting wave function 
%$\Phi_{j,n}^{R,L}$ is maximum at $\bar U_n = n \times \bar U, \, n \in
%\bold Z$. 
%We demand that this wavepackets satisfy 
%$\bar \omega \sim \sigma$ and $\bar U \sim \sigma^{-1}$ {\large{NECESSARY?}}.
%\newline

We assume that the $\Phi_{\lambda}^{R}(\eta, {\bf x})$  are
orthonormal with respect to the Klein-Gordon scalar product
\ba
\label{orthoWPR}
% \av{\Phi_{\lambda}^{R},\Phi_{\lambda'}^{R}}_{KG} = 
 \int \!\!{d^3x}\, \Phi_{\lambda}^{R\, *}\, i\!\dpartialeta \, 
 \Phi_{\lambda'}^{R}
 = \int \!\!\widetilde{d^3k}\,  \gamma_{\lambda,\,\bold k }^{R} \, 
 \gamma_{\lambda', \, \bold k}^{R\, *} =
 \delta_{\lambda \lambda'}\, .
\ea
This implies that the matrices $\gamma^R$ are invertible with inverse 
$\gamma^{R\, \dagger}$.
We also assume that the family is complete:
\ba \label{complete}
 \sum_{\lambda} \gamma^{R \, *}_{\lambda,\,\bold k } \, 
 \gamma_{\lambda,\,\bold k' }^{R} 
 = \delta^3( \bold k - \bold k' )\, .
\ea

We then introduce a family of positive frequency,
left-moving, wave-packets:
\ba
 \Phi_{\lambda}^{L} =  \int \!\!\widetilde{d^3k}\, 
 \gamma_{\lambda,\, \bold k}^{L\, *} 
 {\frac{e^{-i \bf k \bf x}}{(2 \pi)^{3/2}}} \phi^{out}_{k} \, .
\ea
For reasons which shall become clear in the sequel (see Eq.
\Eqref{Bll'}, \Eqref{implicitdefL} and discussion below),
we relate  $\gamma^{L}$ and $\gamma^{R}$ by
\ba \label{defL}
 \gamma_{\lambda,\,\bold k}^{L} = -
 e^{-2 i\psi_k}\gamma_{\lambda,\, \bold k}^{R\, *} \, ,
\ea
where $\psi_k$ is the phase of the squeezing parameter 
$z_k= - \abs{z_k} e^{i2 \psi_k}$.
It follows that the matrices $\gamma^{L}$ are invertible as well, and that the 
$\Phi_{\lambda}^{L}$ are orthonormal.
%and completeness of the left-movers 
%$\Phi_{\lambda}^{L}$ follows
%from the orthonormality and completeness of the right-movers.
Since $R$ and $L$ wave-packets are orthonormal, the family 
$\{ \left(\Phi_{\lambda}^{R}, \Phi_{\lambda}^{L} \right) \}$ 
forms a complete orthonormal basis of the solutions of the field equation.
 
%If we note $\Gamma^{R}$ its inverse, one has by definition \ba
% \int \!\!\widetilde{d^3k}\, \gamma_{\lambda,\, \bold k}^{R} 
%\Gamma_{\bold k,\, \lambda'}^{R} =
% \delta_{\lambda\lambda'} \, ,
%\ea
%from which it follows, by identification with \Eqref{orthoWPR}
%\ba
% \Gamma_{\bold k,\, \lambda}^{R} = \gamma_{\lambda,\, \bold k}^{R\, *} \, ,
%\ea
%which means that the matrix $\gamma^{R}$ is unitary.
%Then, by definition, the $\gamma^{L}$ are invertible. 
%If we note $\Gamma^{L}$ their inverse, we have a similar expression:
%\ba
% \Gamma_{\bold k,\, \lambda}^{L} = \gamma_{\lambda,\, \bold k}^{L\, *} \, .
%\ea
%

%Since the double family of wave-packets is orthonormal and complete, 
Hence, the field can be decomposed as
\ba
 \hat \phi = \sum\limits_{\lambda} \left( 
 \hat a_{\lambda}^{R} \Phi_{\lambda}^{R} +  \hat a_{\lambda}^{L} 
 \Phi_{\lambda}^{L}  + h.c. \right) \, .
\ea 
The annihilation (creation) operators are 
%defined as the
%overlaps betwenn the field and the positive (negative) frequency 
%$R$ and $L$ wave-packets  
given by
\ba \label{defa_lambda}
 \hat a_{\lambda}^{R} %\langle \Phi_{\lambda}^{R},\, \hat \phi \rangle 
 &=& \int \!\!\widetilde{d^3k}\,  
 \gamma_{\lambda,\, \bold k}^{R} \, \hat a_{\bold k}^{R}
 \,\, , \quad  
 \hat a_{\lambda}^{L} %\langle \Phi_{\lambda}^{L},\, \hat \phi \rangle 
 = \int \!\!\widetilde{d^3k}\, 
 \gamma_{\lambda,\, \bold k}^{L} \, \hat a_{\bold k}^{L}  \, . 
%\nonumber \\
% \hat a_{\lambda}^{R \, \dagger} 
% = - \langle \Phi_{\lambda}^{R \, *},\, \hat \phi \rangle
% &=& 
% \int \!\!\widetilde{d^3k}\,  \gamma_{\lambda,\, \bold k}^{R\, *} \, \hat a_{\bold k}^{\dagger} 
% \,\, , \quad  
% \hat a_{\lambda}^{L \, \dagger} 
% = - \langle \Phi_{\lambda}^{L \, *},\, \hat \phi \rangle
% = \int \!\!\widetilde{d^3k}\, \gamma_{\lambda,\, -\bold k}^{L\, *} \, \hat a_{-\bold k}^{\dagger}  
%  \, . \nonumber 
\ea
and satisfy the commutation relations 
\ba
 \left[\hat a_{\lambda}^{\mu},\, \hat a_{\lambda'}^{\nu \, \dagger}\right] =
 \delta_{\lambda\lambda'} \, \delta^{\mu \nu} \, ,
\ea
where $\mu,\nu$ stand for $R, L$. 
%Existense of a rotation operator that relates the operators
%\ba
% \hat a_{\lambda} = \hat C_{\lambda,\bf k} \hat a_{\bf k } 
% 		    \hat C_{\lambda,\bf k}^{\dagger} \, .\ea

The vacuum is the tensorial product
\ba
 \ket{0} = \tprodl \otimes \ket{0,\, \lambda}_2  \, ,
\ea
where each two-mode vacuum $\ket{0,\, \lambda}_2$ state is defined by  
\ba
% \ket{0,\, \lambda}_2 = \ket{0,\, \lambda, R} \otimes \ket{0,\, \lambda, L}
% \, , \\
 \hat a_{\lambda}^{R} \kett{0,\, \lambda} = 
 \hat a_{\lambda}^{L} \kett{0,\, \lambda} = 0 \, .
\ea
We have introduced the ``tilde'' tensorial product $\tprodl$ to indicate that it
takes into account the indexes $\lambda$ which belong
to ${ \bf Z}_+ \times { \bf Z}^2 \times { \bf Z}^3$. 
%since the summation is carried over half the wave-vector space. 
%We have thus constructed a new basis of the Fock space 
%(based on $\Phi_{\lambda,R}$ and $\Phi_{\lambda,L}$)
%which is unitary related to the standard one based on plane waves. 

It should be stressed that the above expression of the
vacuum in terms of two-mode states is somehow artificial since the 
Bogoliubov transformation between in and out wave-packets 
will, in general, be non diagonal in $\lambda$.
%since they are not $\bf k$-eigenstates.
%Indeed the Bogoliubov relation is 
%The Bogoliubov coefficients between in and out $\hat a_\lambda$ operators is not
%diagonal. 
%The non-diagonal character arises from the fact that the wave packets 
%$\Phi_{\lambda}$ are non-stationarity. 
Indeed, using Eq. \Eqref{inoutoperator} and \Eqref{defa_lambda}, one has:
\ba \label{inoutlambdaop}
  \hat a_{\lambda,\, R}^{in } = 
 %\sum\limits_{\lambda'} \,  \left( \left\{ \int\!\!\widetilde{d^3k}\, 
 %\gamma_{\lambda, \, \bold k}^{R}  \alpha_{k}^{*} 
 %\gamma^{R\, *}_{\lambda', \, \bold k} \right\} \, \hat a_{\lambda',\, R}^{out} -
 %\left\{ \int\!\!\widetilde{d^3k}\,  \gamma_{\lambda, \, \bold k}^{R} \beta_{k} 
 %\gamma^{L}_{\lambda',\, -\bold k} \right\}  \,  \hat a_{\lambda',\, L}^{out \dagger}  \right)
 %\, \nonumber \\
 %&=& 
 \sum\limits_{\lambda'} 
 \left( \alpha_{\lambda \lambda'}^{*} \,  \hat a_{\lambda',\, R}^{out} - 
 \beta_{\lambda \lambda'} \,  \hat a_{\lambda',\, L}^{out \dagger} \right) \, .
\ea
where
\ba
 \alpha_{\lambda \lambda'}^{*} = \int\!\!\widetilde{d^3k}\, 
 \gamma_{\lambda, \, \bold k}^{R}  \alpha_{k}^{*} 
 \gamma^{R\, *}_{\lambda', \, \bold k} \, , \qquad
 \beta_{\lambda \lambda'} = \int\!\!\widetilde{d^3k}\,  
 \gamma_{\lambda, \, \bold k}^{R} \beta_{k} 
 \gamma^{L}_{\lambda',\, -\bold k} \, ,
\ea
are not diagonal.
%D%
%There is no reason that 
%$\alpha_{\lambda \lambda'}$ and $ \beta_{\lambda \lambda'}$
%be diagonal in $\lambda,  \lambda'$. 
%Hence in general $\Phi_{\lambda,L}$ will not be 
%the partner wave of $\Phi_{\lambda,R}$. 

%However, in the limiting case of high occupation number, as is the case in
%inflationary cosmology, Eq. \Eqref{inoutlambdaop} is diagonal, see the
%discussion in Appendix \ref{sec:wp,vacuum} for more details.

%D%
%In inflationary cosmology however, thanks to the very high occupation
%number, it is possible to diagonalize the ``squeezing operator'' (wrong) in the
%$\lambda$ space. This requires a fine tuning of the $L$ modes $\Phi_{\lambda,L}$
%given the $\Phi_{\lambda,R}$ and the phase $\psi_k$.
%%s incoding the pair creation processes. 
%In orther words one can choose the $\Phi_{\lambda,L}$ 
%so as to let them be the partner waves of the $\Phi_{\lambda,R}$. 
%As we shall see below this leads to Eq. \Eqref{defL}.
%%Notice finaly that it might be not possible to diagonalize both 
%%$\alpha_{\lambda \lambda'}$ and $ \beta_{\lambda \lambda'}$. 

\subsection{In and out vacuum states} \label{sec:wp,vacuum}

To obtain the relation between in and out vacua, 
we start with Eq.\Eqref{inoutvact}, the expression in terms of the
operators $\hat a_{\bf k}$ which diagonalize the Bogoliubov transformation.
We then express the  $\hat a_{\bf k}$ in terms of $\hat a_{\lambda}$
so as to get
%ee Eq.\Eqref{inoutvact},
\ba \label{complicatedinvac}
 \ket{0 in} &=& 
 \left(\tprod \frac{1}{\vert \alpha_k \vert} \right) 
 \exp\left( \frac{L^3}{(2\pi)^3}  \int \!\!\widetilde{d^3k}\, z_k
 \hat a_{\bf k}^{R, \, out\, \dagger} \hat a_{\bf k}^{L, \, out\, \dagger}
 \right) \ket{0 out} \, \nonumber \\
  &=& 
% Z^{-1}
 \left(\tprod \frac{1}{\vert \alpha_k \vert} \right) 
 \exp\left( \sum\limits_{\lambda\lambda'} 
%\frac{L^3}{(2\pi)^3 } \int \!\!\widetilde{d^3k} \, z_k \, 
% \gamma^{R}_{\lambda,\, \bold k} \gamma^{L}_{ \lambda,\,-\bold k} 
B_{\lambda \lambda'} \, 
\hat a_{\lambda,\, R}^{out\,  \dagger} 
 \hat a_{\lambda',\, L}^{out\, \dagger} \right) \ket{0 out}  
 \, .
\ea
The factor $ {L^3}/{(2\pi)^3}$ gives the density of states in a cube 
of size $L$. 
The matrix $ B_{\lambda\lambda'} $ 
%to create the pair $(\lambda,\, R), (\lambda',\, L)$  
is given by 
\ba \label{Bll'}
 B_{\lambda\lambda'} &=&  \frac{L^3}{(2\pi)^3} 
 \int\!\!\widetilde{d^3k}\, z_k \, 
 \gamma^{R}_{ \lambda,\, \bold k,} 
 \gamma^{L}_{ \lambda',\, \bold k} \, .
\ea
%It corresponds the 
%%The terms contained in the triple summation in the exponential are 
%probability amplitude to create the $\lambda \lambda'$  pair
%\ba
%\frac{ \bra{0 \, out} \hat a_{\lambda}^{R \, out} 
% \hat a_{\lambda'}^{L\,out} \ket{0 \, in}}{  \bra{0 \, out} 0 \, in \rangle}  = 
% B_{\lambda \lambda'} \, ,
%\ea
%The above expressions hold for every complete family of wave-packets
%$\Phi_{\lambda}^{R}$ and $\Phi_{\lambda}^{L}$. 
 
In inflationary cosmology, 
one can fine-tune the family $\Phi_{\lambda}^{L}$ so as to get 
a diagonal matrix:
%that
\ba \label{implicitdefL}
 B_{\lambda\lambda'} =  z_\lambda \, \delta_{\lambda\lambda'} \, ,
\ea
where $z_\lambda$ is real and positive. 
One has $z_\lambda = \abs{z_{\bar k}}$ 
where $\bar k_\lambda$ is the mean momentum in the wave packet $\Phi_\lambda^R$.
Hence $z_\lambda - 1 \ll 1$ and 
%Thus $z_\lambda $ fixes the mean occupation number to be 
$n_\lambda = n_{k_\lambda} =1/(1-z_\lambda^2) \gg 1$. 
To obtain a diagonal matrix, one first 
%More precisely, by 
fine-tunes the $k$ dependent phase of the $\gamma_{\lambda}^{L}$,
as specified in Eq. \Eqref{defL}. Thus
one gets rid of the phase of $z_k$ in the integrand of Eq. \Eqref{Bll'}.
Second one uses the fact that the modulus $\abs{z_k}$ is a 
slowly varying function of $k$ in the 
large occupation number limit when dealing with narrow wave packets in $k$. 
Then
one can use Eq. \Eqref{orthoWPR} to show that $B_{\lambda\lambda'}$
is diagonal. 
Let us stress that the slowly varying character of $\abs{z_k}$ follows from the 
large occupation number limit.
Indeed, one has 
$k \delta_{k}  \simeq 1/ n_k \ll 1 $ where  
$\delta_{k}= \partial_k z_k \simeq (1/ n_k) \partial_k \ln n_k $
and where we have supposed that $n_k$ is a power law.

%Therefore, the diagonality of the $B_{\lambda \lambda'}$ matrix 
%is valid  in the limiting case $\abs{z_k}^2 = n_k /(n_k + 1) \to 1$.
Using the fine-tuned $L$ modes, 
the in vacuum factorizes as a product of two-mode out states,
as in the ${\bf k}$-basis 
\ba
 \ket{0 in} = \widetilde{\prod\limits_{\lambda}} 
\left( \inv{(n_{\lambda}+1)^{1/2}} 
	    \exp\left( z_\lambda \, \hat a_{\lambda,\, R}^{out\,  \dagger} \, 
 	                  \hat a_{\lambda,\, L}^{out\, \dagger}\right) 
                  \kett{0,\, \lambda, out} \right) \, .
\ea
The terms in the exponentials should be understood as a 
first order approximation in $\delta_k$. 
%Subleading terms of order $\delta_k$ describing
%multiple pair-creations have been omited. 
In this first order approximation, one can also
treat the Bogoliubov coefficients $\alpha_{\lambda\lambda'}$ and 
$\beta_{\lambda\lambda'}$ as diagonal. In the sequel we shall work in this
limit, and all equations should be understood as giving the leading behavior
when $\delta_k \ll 1$.

%%%%%%%%%%%%%%%%%%%%%%%%%%%%%%%%%%%%%%%%%%%%%%%%%%%%%%%%%%%%%%%%%%%%%%%%%%%%%%%
%%%%%%%%%%%%%%%%%%%%%%%%%%%%%%%%%%%%%%%%%%%%%%%%%%%%%%%%%%%%%%%%%%%%%%%%%%%%%%%

\subsection{Modified correlations} \label{sec:wp,ps}

The projector on a coherent state of a out wave-packet $\Phi_\lambda^R$ 
of amplitude $v$ is, in total analogy with Eq. \Eqref{proj1}, 
\ba
 \hat \Pi_{v_{\lambda}^R} = \ket{v,\, \lambda, R, out} \bra{v,\,\lambda, R, out} 
 \otimes {\bf 1}_{\lambda, L} 
 \widetilde{\prod\limits_{\lambda' \neq \lambda}} \otimes{\bf 1}_{\lambda'} \, .
\ea
Unlike what we had with Eq. \Eqref{ProjV}, this operator is unity in all 
sectors orthogonal to that defined by $\Phi^R_\lambda$.

Since the relation between in and out vacua is diagonal, 
we can collect the
results from Section IV.B and V and re-express them with the $\lambda$ basis.
The action of the projector on the in-vacuum is
\ba
 \hat \Pi_{v_\lambda^R} \ket{0 \, in}  = 
%Z_{\lambda}^{-1}\alpha_{\lambda}^{-1}
(n_\lambda + 1)^{-1}
 e^{-\frac{\abs{v}^2}
%{(1-\abs{B_{\lambda}}^2)/2}
{2(n_\lambda + 1)}}
\ket{v,\,\lambda, R, out} \otimes \ket{(z_{\lambda}v^{*}),\, \lambda, L, out} 
 \widetilde{\prod\limits_{\lambda' \neq \lambda}} 
 \otimes \kett{0,\, \lambda', in}
%\bigotimes_{\lambda' \neq \lambda} \ket{0,\, \lambda', in} \, .
\ea
It displays coherent state correlations in that the $\lambda$-th 
$L$-component is a coherent state of amplitude $(z_{\lambda}v^{*})$.
%phase-space and in occupation number as well.
%The interpretation of factor $B_{\lambda}$ is given at \Eqref{Bll'} and is
%consistent with the result of Section \ref{sec:ps}.
The probability to find the $\lambda$ wave-packet is given by
\ba
 P_{v_\lambda^R}^{\rm in}(v,\, \lambda) = 
 \bra{0 \, in}  \hat \Pi_{v_\lambda^R} \ket{0 \, in} = 
 \frac{1}{n_\lambda + 1} e^{-\frac{\abs{v}^2}{2(n_\lambda + 1)}} \, .
 %\inv{Z_{\lambda}^2} e^{-\abs{v}^2(1-\abs{B_{\lambda}^2})}  \, .
\ea
Notice that this amplitude is much larger than $P^{in}_{{\cal V}_R}$ of Eq.
\Eqref{ProjV} since the coherent state projection concerns one mode only.
(However the probability $P_{v_\lambda^R}^{\rm in}$ 
is still small because the resolution of coherent
states is very high with respect to the occupations number 
%D% $n_k \gg 1$).
$n_{\lambda} \gg 1$).
The modified ensemble obtained by having detected $\Phi_\lambda$ with
amplitude $v$ is thus much closer to the mean than that obtained by the
projector $\hat \Pi_{{\cal V}_R}$. 
%The matrix element are identicle to that of Section \ref{sec:ps} 
%with $v_{\bold  k}$ replaced by $v$ and $z_k$ by $B_{\lambda}$:
This is clearly seen by computing the modified expectation values Eq. 
\Eqref{condv}.
The 1-point functions are 
%are identicle to that of Section \ref{sec:ps} 
%with $v_{\bold  k}$ replaced by $v$ and $z_k$ by $B_{\lambda}$:
\ba \label{1pointlambda}
 \label{Pi2ampR} 
 &&\langle \hat a_{\lambda',\, R}^{out} \rangle_{v_\lambda^R} = 
 v\, \delta_{\lambda, \lambda'}, \quad \quad 
 \langle\, \hat a_{\lambda,\, L}^{out} \rangle_{v_\lambda^R} =  
 z_\lambda v^{*} \, \delta_{\lambda, \lambda'} \, , 
\ea
or equivalently
\ba \label{1pointk}
 &&\langle \hat a_{{\bf k},\, R}^{out} \rangle_{v_\lambda^R} = 
 v\, \gamma_{\lambda,{\bf k}}^{R\, *}, \quad \quad 
 \langle\, \hat a_{{\bf k},\, L}^{out} \rangle_{v_\lambda^R} =  
 z_\lambda v^{*} \, \gamma_{\lambda,{\bf k}}^{L\, *} \, .
\ea
%\sba
% \label{Pi2ampR} 
% &&\langle \hat \Pi_{\lambda}^{v} \, \hat a_{\lambda,\, R}^{out} \rangle_{\rm{in}} = 
% \abs{B_{\lambda}}^2 v \, , \qquad 
% \langle \hat a_{\lambda,\, R}^{out} \,  \hat \Pi_{\lambda}^{v} \rangle_{\rm{in}} = 
% v \, , \\
%\label{Pi2ampL}
% &&\langle \hat \Pi_{\lambda}^{v} \, \hat a_{\lambda,\, L}^{out} \rangle_{\rm{in}} = 
% \langle \hat  a_{\lambda,\, L}^{out} \, \hat \Pi_{\lambda}^{v} \rangle_{\rm{in}} = 
%    B_{\lambda} v^{*} \, , \\
%\label{Pi2cros}
% &&\langle \hat \Pi_{\lambda}^{v} \, \hat a_{\lambda,\, R}^{out} 
% 	   \hat a_{\lambda,\, L}^{out} \rangle_{\rm{in}} =
%	   B_{\lambda}\left( 1 + \abs{B_{\lambda}}^2 \abs{v}^2 \right) \, , \qquad 
% \langle \hat a_{\lambda,\, R}^{out}
% 	   \hat a_{\lambda,\, L}^{out} \, \hat \Pi_{\lambda}^{v} \rangle_{\rm{in}} =
%	   B_{\lambda}\abs{v}^2 \, , \\
%\label{Pi2occup} 
% && \langle \hat \Pi_{\lambda}^{v} \, \hat a_{\lambda,\, R}^{\dagger \, out} 
% 	   \hat a_{\lambda,\, R}^{out} \rangle_{\rm{in}} =
%    \langle \hat \Pi_{\lambda}^{v} \, \hat a_{\lambda,\, L}^{\dagger \, out} 
% 	   \hat a_{\lambda,\, L}^{out} \rangle_{\rm{in}} = 
%	   \abs{B_{\lambda}}^2 \abs{v}^2 \, .
%\sea
Eqs.\Eqref{1pointk} coincide with Eqs.\Eqref{1pointcoh} with 
$v_{\bf k} = v \gamma_{\lambda,{\bf k}}^{R\, *}$ even though the projection
enforced by $\hat \Pi_{v_\lambda^R}$ is much weaker than that of 
$\hat \Pi_{{\cal V}_R}$. The reason of the agreement is that the mean value
of the modes orthogonal to $\Phi_{\lambda}^{R}$ 
vanishes because we are in vacuum whereas it vanished in Eqs.\Eqref{1pointcoh} 
because we set all amplitudes to zero by the complete projection. 
For the two-point functions we have
%We can finally give the conditionnal value of the field in the post-selected
%ensemble:
\ba \label{Pi2occup} 
 && \langle \hat a_{\lambda_1,\, R}^{\dagger \, out} 
 	    \hat a_{\lambda_2,\, R}^{out} \rangle_{\rm{in}} =
    \langle \hat a_{\lambda_1,\, L}^{\dagger \, out} 
 	   \hat a_{\lambda_2,\, L}^{out} \rangle_{\rm{in}} = 
	   \delta_{\lambda_1,\lambda_2} \left[ 
	   (1-\delta_{\lambda,\lambda_1}) \beta_{\lambda}^{*}  \beta_{\lambda} 
	   + \delta_{\lambda,\lambda_1} \abs{v}^2 \right]  \, . \\
	   %\sum\limits_{\lambda'}
	   %\beta_{\lambda,\lambda'}^*  \beta_{\lambda,\lambda'} \right]  \, .
 \label{Pi2cros} 
 &&\langle  \hat a_{\lambda_1\, R}^{out} 
 	   \hat a_{\lambda_2\, L}^{out} \rangle_{v^R_\lambda} = 
%	   B_{\lambda}\left( 1 + \abs{B_{\lambda}}^2 \abs{v}^2 \right) \, , \qquad 
% \langle \hat a_{\lambda,\, R}^{out}
% 	   \hat a_{\lambda,\, L}^{out} \, \hat \Pi_{\lambda}^{v} \rangle_{\rm{in}} =
	  \delta_{\lambda_1,\lambda_2} \left[  
	  (1-\delta_{\lambda,\lambda_1})  \alpha_{\lambda}  \beta_{\lambda}
	  + \delta_{\lambda,\lambda_1} B_{\lambda} \abs{v}^2 \right]  \, .	\\  
	  %\sum\limits_{\lambda'}
	  %\alpha_{\lambda,\lambda'}  \beta_{\lambda,\lambda'} \right] \, ,\\
\ea
We clearly see that the 2-point functions split into two contributions. 
First, one finds the usual in vacuum 
expectation values in the two-mode sectors orthogonal to the chosen mode $
\Phi^R_{\lambda}$.
Second, there is
the coherent state 2-point function of amplitude $v$ in this 2-mode sector.

Given these results, the conditional value
of the field are 
%space time correlations obtained by using the partial projector 
%$ \hat \Pi_{v_\lambda^R}$ are 
%as in Section \ref{sec:ps,one-point}
\sba
 \bar \phi_{R,v_\lambda^R} &=& \langle  \hat \phi_R  \rangle_{v^R_\lambda} = 
 \left(v \Phi_{\lambda, \, R}^{out} + 
 v^{*} \Phi_{\lambda \, R}^{out \, *}\right)  \, ,\\
 \bar \phi_{L, v^R_\lambda} &=& 
 \langle  \hat \phi_{L,v_\lambda^R}  \rangle
 = \left(z_{\lambda} v^{*} \, \Phi_{\lambda\, L}^{out} + 
 z_{\lambda}^{*} v \, \Phi_{\lambda\, L}^{out \, *} \right)   \, .
\sea
%D%
%The conditional value of the right sector describes a field in a coherent state
%with mean momentum and position given by $\lambda$.
%For the left sector, due to the definition of the corresponding wave-packets
%\ref{defL}, the phase of the wave packet is shifted. Recall that 
%this additional term comes from the squeezing variables and is thus the
%bookkeeper of the quantum origin of the fluctuations.
They agree with the expressions of Section V.D.
%\ref{sec:ps,one-point}
because, in the mean, the expectation values of the unspecified modes
all vanish in the vacuum.

\end{appendix}
%%%%%%%%%%%%%%%%%%%%%%%%%%%%%%%%%%%%%%%%%%%%%%%%%%%%%%%%%%%%%%%%%%%%%%%%%%%%
%%%%%%%%%%%%%%%%%%%%%%%%%%%%%%%%%%%%%%%%%%%%%%%%%%%%%%%%%%%%%%%%%%%%%%%%%%%%
%%%%%%%%%%%%%%%%%%%%%%%%%%%%%%%%%%%%%%%%%%%%%%%%%%%%%%%%%%%%%%%%%%%%%%%%%%%%
%%%%%%%%%%%%%%%%%%%%%%%%%%%%%%%%%%%%%%%%%%%%%%%%%%%%%%%%%%%%%%%%%%%%%%%%%%%%
%%%%%%%%%%%%%%%%%%%%%%%%%%%%%%%%%%%%%%%%%%%%%%%%%%%%%%%%%%%%%%%%%%%%%%%%%%%%
%%%%%%%%%%%%%%%%%%%%%%%%%%%%%%%%%%%%%%%%%%%%%%%%%%%%%%%%%%%%%%%%%%%%%%%%%%%%
%%%%%%%%%%%%%%%%%%%%%%%%%%%%%%%%%%%%%%%%%%%%%%%%%%%%%%%%%%%%%%%%%%%%%%%%%%%%
%%%%%%%%%%%%%%%%%%%%%%%%%%%%%%%%%%%%%%%%%%%%%%%%%%%%%%%%%%%%%%%%%%%%%%%%%%%%

\end{document}